\documentclass[%
 reprint,
 amsmath,amssymb,
 aps,prl
]{revtex4-2}

\usepackage{graphicx}
\usepackage{dcolumn}
\usepackage{bm}

\usepackage{amsmath}
\usepackage{amssymb}
\usepackage{verbatim}
\usepackage{graphicx}
\usepackage{psfrag}
\usepackage[british]{babel}
\usepackage{tikz} 
\usetikzlibrary{shapes,arrows.meta}
\usepackage{textcomp}
\usepackage{mathtools}
\usepackage{capt-of}
\usepackage{textgreek}
\usepackage{setspace}
\usepackage{cases}
\usepackage{pgfplots}
\usepackage{multirow}
\usepackage{empheq}
\usepackage{float}
\usepackage{afterpage}
\usepackage[export]{adjustbox}

\newcommand{\subf}[1]{%
  {\small\begin{tabular}[t]{@{}c@{}}
   \mbox{}\\[-\ht\strutbox]
   #1
   \end{tabular}}%
}

\allowdisplaybreaks

\begin{document}

\preprint{APS/123-QED}

\title{Silence is bliss in a Platonic relationship}

\author{Cheuk-Him Yeung}
\author{Tom Shearer}
\author{William J.\ Parnell}
\affiliation{%
Department of Mathematics, University of Manchester, Manchester M13 9PL, UK
}%

%
%

\date{\today}

\begin{abstract}
We describe an effective active cloaking strategy for the scalar Helmholtz equation in three dimensions where multipole active sources are located at the vertices of the Platonic solids. A ``silent zone'' is created interior to the imaginary Platonic solid and only the incident field remains in a defined region exterior to the silent zone and active source configuration.  This distribution of the sources ensures that the implementation of the cloaking strategy is extremely efficient. In particular, once the multipole source amplitudes required at a single source location are determined, the other source amplitudes can be calculated by simple post-processing involving multiplication of the multipole source vector by a rotation matrix. The general nature of the problem means that the technique is relevant to any scalar wave field, including both acoustics and electromagnetism.  
\end{abstract}

\maketitle

Over the last two decades significant excitement has been generated around the idea of cloaking objects, i.e.\ rendering them invisible to incident wave fields \cite{schurig2006metamaterial, miller2006perfect, chen2007acoustic, norris2008acoustic, fleury2014cloaking}. Cloaking strategies are either passive or active, the former requiring the fabrication of metamaterials that are able to manipulate the wave, steering it around the object regardless of the form of the incident field \cite{cai2007optical, silveirinha2007parallel, zhang2011broadband, craster2012acoustic}. In the context of active cloaking, sources are employed to suppress a field within a certain domain \cite{miller2006perfect,vasquez2009active, zheng2010exterior}. Indeed, using active sources to modify fields has been of interest in science, technology and engineering for many decades: Paul Lueg's patent in 1936 on the topic of anti-sound illustrates the use of active sources to suppress noise in specific regions of space \cite{guicking1990invention}.  Anti-sound and anti-vibration are huge areas of research and a plethora of techniques have been developed in order to address numerous problems in acoustic engineering \cite{nelson1991active, fuller1996active, guicking2007active}. An illustration of the relation between various methods in acoustics was provided by Cheer \cite{cheer2016active}. Specific choices of sources can ensure quiet zones and illusions \cite{zheng2010exterior,ma2013experiments} and recent work has optimised control sources to reduce the scattered field in passive scenarios and cases with flow for the cloaking of specific objects\cite{o2015active, eggler2019activeA, eggler2019activeB, house2020experimental}.

Over the last decade or so, interest has centred on active cloaking methods that are independent of the object to be cloaked. Miller's method of cloaking measured particle motion close to the surface of the cloaking region and  simultaneously excited necessary surface sources where each source amplitude depends on the measurements at all sensing points \cite{miller2006perfect}. This approach is rather limited as an active cloaking method because it cannot provide a relationship between the incident field and source amplitudes. Guevara-Vasquez et al.\ addressed this problem \cite{vasquez2009active}, describing an active source method that could use multipoles to create silent zones and simultaneously ensure that the active field is zero outside the silent zone. The integral equation formulation of the problem was employed, and subsequently converted to a linear system of equations for the source amplitudes. These amplitudes are linear functions of the incident field and it was shown, by construction, that active cloaking could be realised in two dimensions. An explicit form of the relations between the sources and incident field was provided in \cite{vasquez2011exterior}. In particular multipole sources were used, Miller's cloak was reproduced and numerical results were compared with SVD solutions of the linearised system \cite{vasquez2009active, vasquez2009broadband}.

Further progress was made on the form of the source coefficients in \cite{norris2012source}, where it was demonstrated that the integral representation of Vasquez et al.\ \cite{vasquez2009active} could be reduced to closed-form explicit formulas, which bypassed the requirement for a numerical solution. Analytical expressions were provided for general incidence and specifically for plane-wave incidence. This approach was then extended to two-dimensional (vector) elastodynamics \cite{norris2014active} and to the thin-plate equation \cite{futhazar2015active}. The attractive nature of this approach to active cloaking is that the source coefficients are independent of the object to be cloaked inside the silent zone. This is a benefit over approaches that are object-dependent where alternative approaches have to be taken in resonant regimes \cite{o2016active}. The latter approach however, does not suffer from large amplitudes in the vicinity of the active source regions. 

Although there has been extensive work regarding the active manipulation of sound in three-dimensions in general settings, see e.g.\  \cite{elliott2012robustness, ahrens2012analytic, onofrei2018synthesis,egarguin2020active}, the active cloaking approach that is independent of the object to be cloaked has thus far predominantly been conducted in the two-dimensional setting, with the sole exception of the work of Guevara-Vasquez et al \cite{vasquez2013transformation} who described the extension of the methods in \cite{vasquez2009active, vasquez2009broadband} for the scalar Helmholtz equation in three dimensions. Expressions were derived for the active field in terms of an integral of the incident field and subsequently illustrated with four active sources. 

Here we deduce integral expressions for active cloaking in three dimensions. We provide expressions for the active source coefficients and introduce a new methodology associated with distributions of the active sources on the vertices of the Platonic solids. This approach is fast and efficient to implement due to the sources residing on the circumsphere of the Platonic solid. Regardless of the form of the incident field, once the form of multipole source coefficients has been deduced for just \textit{one} of the active sources, the remaining source amplitudes can be calculated merely by solving a matrix equation in terms of the determined multipole source vector, thus providing a clear mechanism and complete exposition of three-dimensional active cloaking for the scalar Helmholtz equation. 

\begin{figure}
	\centering 
	\includegraphics[trim={2cm 2cm 1.5cm 2.5cm},clip,width=0.3\textwidth]{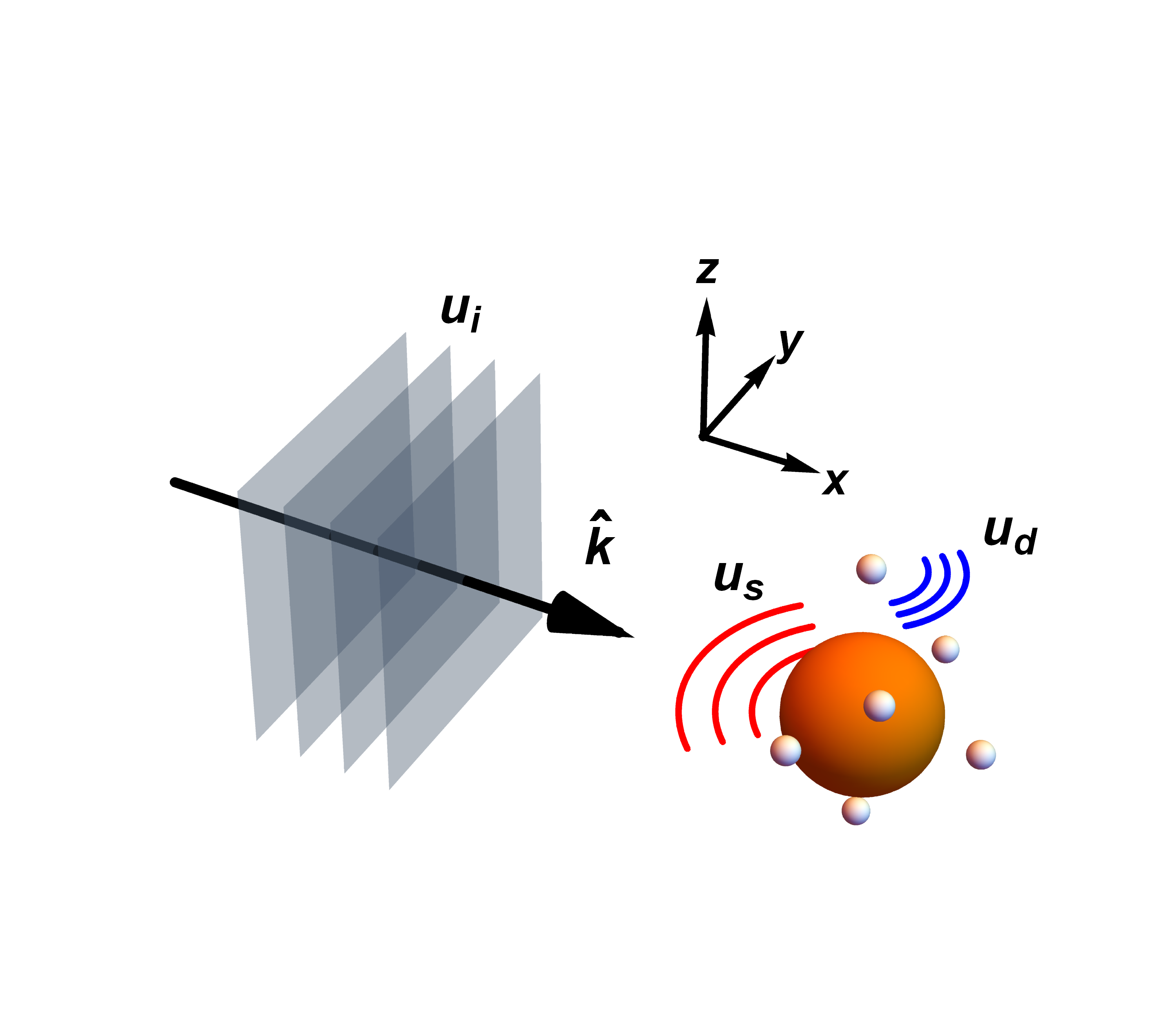}
	\caption{Schematic of the three-dimensional active exterior cloaking problem. The incident plane wave $u_i$ propagating in the direction $\widehat{\mathbf{k}}$ impinges upon an arbitrary object (the orange sphere) and is scattered in the form of the field $u_s$. The amplitudes of the $L$ active sources (the smaller spheres) are determined such that they produce an active field $u_d$ which cancels out the incident field $u_i$ in a volume containing the object and thus nullifies the scattered field $u_s$.}
	\label{schematic}
\end{figure}


We consider active exterior cloaking for time-harmonic waves (with dependence $e^{-i\omega t}$ where $\omega$ is the angular frequency and $t$ is time) governed by the three-dimensional homogeneous Helmholtz equation $(\nabla^2 + k^2) u(\mathbf{x}) = 0$ where $k=\omega/c$ is the wavenumber with $c$ the speed of wave. The scalar wave field of interest is $u(\mathbf{x})$, which in the context of acoustics is the velocity potential at the position $\mathbf{x}=(x,y,z)$.  

As illustrated in Fig.\ \ref{schematic} for the case of an incident field $u_i(\mathbf{x})$ that is planar, we introduce the active field $u_d(\mathbf{x})$ generated by $L$ multipole sources. The field subsequently scattered from an object interior to the active field is denoted by $u_s(\mathbf{x})$. These fields are conveniently described in terms of spherical waves:
\begin{align}
u_i(\mathbf{x}) &= e^{i\mathbf{k}\cdot\mathbf{x}} = \sum_{n=0}^\infty \sum_{m=-n}^n Q_{nm} U^m_n(\mathbf{x}), \label{u_i} \\
u_d(\mathbf{x}) &= \sum_{\ell=1}^{L} \sum_{n=0}^\infty \sum_{m=-n}^n q_{\ell,nm} V^m_n(\mathbf{x}-\mathbf{x}_\ell), \label{u_d} \\
u_s (\mathbf{x}) &= \sum_{n=0}^{\infty} \sum_{m=-n}^n a_{nm} V^m_n(\mathbf{x}), \label{u_s} 
\end{align}
where $U^m_n(\mathbf{x}) = j_n(k|\mathbf{x}|) Y^m_n(\widehat{\mathbf{x}})$ and $V^m_n(\mathbf{x}) = h_n^{(1)}(k|\mathbf{x}|) Y^m_n(\widehat{\mathbf{x}})$ are incoming and outgoing spherical waves respectively \cite{martin2006multiple}, noting that $j_n(k|\mathbf{x}|)$ is the spherical Bessel function of the first kind and $h_n^{(1)} (k|\mathbf{x}|)$ is the spherical Hankel function of the first kind. Further,  $Y^m_n(\widehat{\mathbf{x}})= A^m_n P^m_n(\cos \theta) e^{im\varphi}$ is the normalized spherical harmonic function and
\begin{align}
A^m_n &= \sqrt{\frac{2n+1}{4\pi} \frac{(n-m)!}{(n+m)!}}, \label{A}
\end{align}
where hats denote unit vectors and $P^m_n(\cos \theta)$ is the associated Legendre function with order $m$ and degree $n$ 
in terms of the polar angle $\theta$ and the azimuthal angle $\varphi$.

Referring to Fig.\ \ref{schematic}, we write $\mathbf{k}=k\widehat{\mathbf{k}}$ where $\widehat{\mathbf{k}}=(\sin\theta_i \cos\varphi_i, \sin\theta_i \sin\varphi_i,\cos\theta_i)$ is the unit propagating vector. For plane wave incidence, the coefficients $Q_{nm}=4\pi i^n \overline{Y^m_n(\widehat{\mathbf{k}})}$, thus depending only on the incident wave angles $\theta_i,\varphi_i$ with the overline denoting complex conjugate. In order to achieve active cloaking, source coefficients $q_{\ell, nm}$ are sought such that for some closed domain $C$ 
surrounded by the active sources at $\mathbf{x}_\ell \notin C$ where $\ell=1,2,\cdots,L$, we have $u_i(\mathbf{x}) + u_d(\mathbf{x}) = 0$ for $ \mathbf{x} \in C$ and 
$u_d(\mathbf{x}) \to 0$ as $|\mathbf{x}| \to \infty$. We require by the first condition that the active field $u_d$ interferes destructively with the incoming wave $u_i$ such that the total field vanishes in the region $C$. While it leads to the nullification of wave scattering from any object in $C$, we also stipulate by the second condition that the radiation of $u_d$ itself to the far field is minimized. This property will leave minimal evidence of the cloak to be detected. 



\begin{figure}[h!]  
	\centering
	\begin{tabular}{cc}
		(a) \subf{\includegraphics[trim={0cm 2cm 2.5cm 0cm},clip,width=3.6cm]{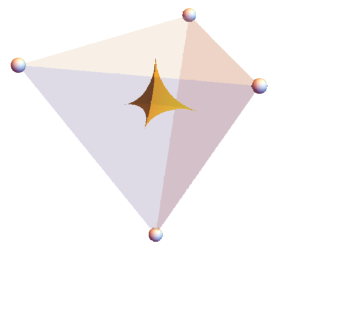}}
		&
		(b)\subf{\includegraphics[trim={1.2cm 1.2cm 1.8cm 1.2cm},clip,width=3.6cm]{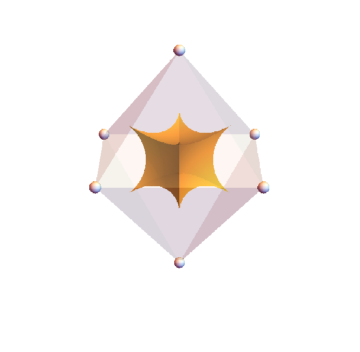}}
		\\
		(c)\subf{\includegraphics[trim={1.2cm 1.2cm 1.2cm 1.2cm},clip,width=3.6cm]{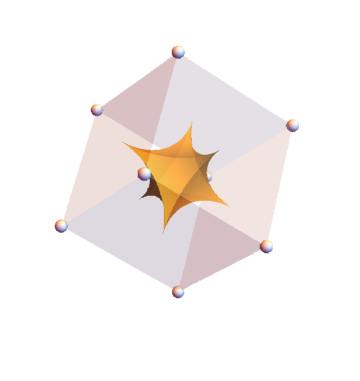}}
		&
		(d)\subf{\includegraphics[trim={1.2cm 1.2cm 1.2cm 1.2cm},clip,width=3.6cm]{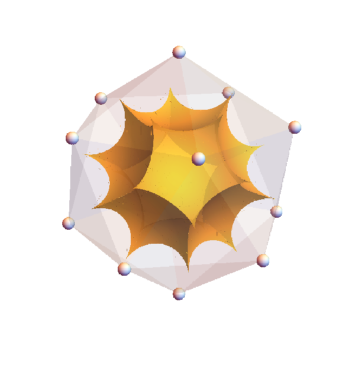}}
		\\
		(e)\subf{\includegraphics[trim={1.2cm 1.2cm 1.5cm 1.2cm},clip,width=3.6cm]{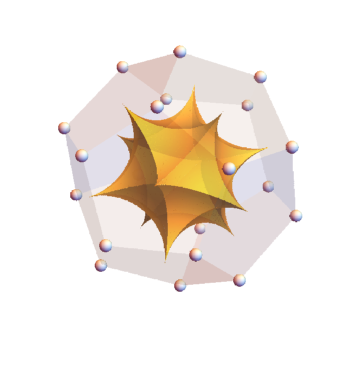}}
		&
		(f)\subf{\includegraphics[trim={0cm 0cm 0cm 0cm},clip,width=3.5cm]{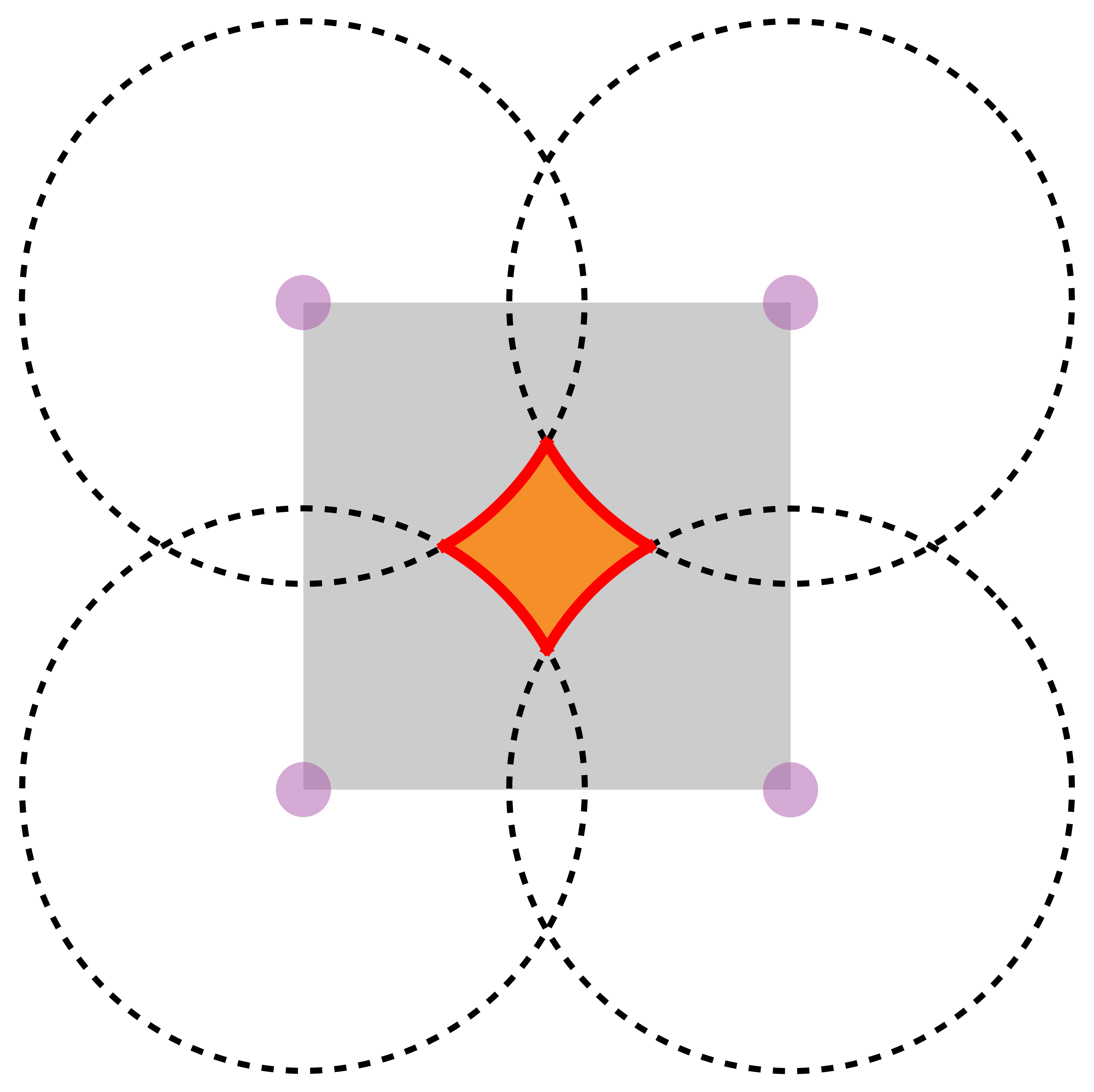}}
	\end{tabular}
	\caption{
		In (a) -- (e) the 
		cloaked regions are illustrated for the cases of the (a) tetrahedron (4 vertices), (b) octahedron (6 vertices), (c) cube (8 vertices) (d) icosahedron (12 vertices) and (e) dodecahedron (20 vertices) Platonic solids. In (f) a cross section through the plane $z=0$ in (b) is shown.} \label{fig1}
\end{figure}

Consider now that the active sources are located at the vertices of an imaginary Platonic solid. 
This arrangement will limit the total number of sources $L$ to five values \cite{euclid2012thirteen}
as illustrated in Fig.\ \ref{fig1}(a) -- (e). 
The geometry ensures that the sources reside on the circumsphere of the Platonic solids such that $|\mathbf{x}_\ell| = x_0$ for all $\ell$ in each case, with $x_0$ an arbitrary constant.
We further set $\mathbf{x}_1=(0,0,-x_0)$ in every case such that the active source with $\ell=1$ is always at the lowest position of the distribution in terms of the $z$ coordinate.  
To locate the remaining sources, we note that a Platonic solid consisting of $q$ $p$-sided regular polygonal faces around each vertex can be characterized by a set of two indices $(p,q)$. (For example, a regular tetrahedron with three equilateral triangles around each vertex has the indices $(3,3)$.) It is useful to define the length of each side of a Platonic solid in terms of $x_0$ as $sx_0$. We will show in Part 2 of Supplementary Material that the position of each active source can be fully determined using knowledge of $(p,q)$ and $s$, whose values for all the five cases are listed in Table \ref{geo}.  
In particular, since the source distribution is $q$-fold rotationally symmetric about the $z$ axis, if we assign the indices $\ell=2,\cdots,q+1$ to the $q$ vertices located immediately above the source $\ell=1$ in the counterclockwise direction, then their position vectors $\mathbf{x}_\ell$ are given by 
\begin{align}
\mathbf{x}_\ell = x_0\left(r_2 \cos \varphi_{\ell} , r_2 \sin \varphi_{\ell}, z_2 \right), \label{x_2}
\end{align}
where $z_2=(s^2-2)/2$, $r_2 = \sqrt{1-z_2^2}$ 
and $\varphi_{\ell}=2(\ell-2)\pi/q$. 
Further details about the indexing of sources with $1< \ell \leq L$ can be found also in Part 2 of Supplementary Material. 

\begin{table}[h!]
	\centering
	\begin{tabular}{c|c|c|c}
		\hline
		$L$ & $(p,q)$ & $s$ & $\mathcal{V}$ (in units of $x_0^3$)   \\ \hline
		4 & $(3,3)$ & $2\sqrt{6}/3$ & $0.0033$   \\ \hline
		6 & $(3,4)$ & $\sqrt{2}$ &$0.0675$  \\ \hline
		8 & $(4,3)$ & $2\sqrt{3}/3$ & $0.0538$  \\ \hline
		12 & $(3,5)$ & $2\sqrt{5(3-\phi)}/5$ & $0.4562$  \\ \hline
		20 & $(5,3)$ & $2\sqrt{3}(\phi-1)/3$ & $0.3692$ \\ \hline
	\end{tabular}
	\caption{The geometric properties of each Platonic distribution of active sources and the volumes of the respective cloaked regions $\mathcal{V}=|C|$ (corrected to four decimal places) when the sphere radius $a$ takes the lower limit in \eqref{Range}. Here $\phi=(1+\sqrt{5})/2$. }
	\label{geo}
\end{table}

To determine the source amplitudes $q_{\ell,nm}$ and the precise geometric shape of the cloaked region $C$, we employ the procedure in \cite{vasquez2013transformation} as our starting point and then use the approach in \cite{norris2012source} to obtain the convenient explicit expressions (see Part 1 of Supplementary Material) in the form
\begin{align}
q_{\ell, nm} = -ik^2\sum_{t=0}^\infty \sum_{s=-t}^t Q_{ts} q_{\ell, nm, ts}, \label{q_l} \end{align}
where we recall that $Q_{ts}$ are the coefficients associated with the expansion of the incident field \eqref{u_i} and 
\begin{align}
q_{\ell, nm, ts} &= \sum_{\nu=0}^\infty \sum_{\mu=-\nu}^\nu \widehat{S}^{s\mu}_{t\nu} (\mathbf{x}_\ell) D_{\nu n} \mathcal{I}_{n\nu}^{m\mu}(\mathbf{x}_\ell, \partial C_\ell). \label{q_2}
\end{align}
In \eqref{q_2}, $\widehat{S}^{s\mu}_{t\nu} (\mathbf{x}_\ell)$ (defined in \eqref{S} -- \eqref{cal_S} of Supplementary Material) is a coefficient depending on the position vector $\mathbf{x}_\ell$ and the quantities $D_{\nu n}, \mathcal{I}_{n\nu}^{m\mu}(\mathbf{x}_\ell, \partial C_\ell)$ take the respective forms
\begin{align}
D_{\nu n} &= j_\nu(ka x_0)j_n'(ka x_0) -j_\nu'(ka x_0)j_n(ka x_0), \label{D_vn} \\
\mathcal{I}_{n\nu}^{m\mu}(\mathbf{x}_\ell, \partial C_\ell) &= \int_{\partial C_\ell} \overline{Y^m_n(\widehat{\mathbf{y}-\mathbf{x}_\ell})} Y^\mu_\nu(\widehat{\mathbf{y}-\mathbf{x}_\ell}) \ dS(\mathbf{y}-\mathbf{x}_\ell),\label{I}
\end{align}
where prime denotes the derivative with respect to the argument and $a$ is any value within the range 
\begin{align}
	\frac{s}{2\sin(\pi/p)} \leq a < 1, \label{Range}
\end{align}
which is derived in Part 2 of Supplementary Material. 
In \eqref{I}, the surface integral $\mathcal{I}_{n\nu}^{m\mu}$ is performed over the face $\partial C_\ell$ parameterized by the vector $\mathbf{y}-\mathbf{x}_\ell$ for $\mathbf{y} \in \partial C_\ell$. (To the authors' knowledge, an analytic form of $\mathcal{I}_{n\nu}^{m\mu}$ is not available at this point. However, the integral can be simplified such that its numerical evaluation becomes significantly less expensive,
which will be discussed in Part 4 of Supplementary Material.) The formulations \eqref{q_l} -- \eqref{Range} hold under two conditions: 1) the domain $C$ is completely bounded by the union of faces $\partial C_\ell$ where $\ell=1,2,\cdots,L$; 2) 
$|\mathbf{y}-\mathbf{x}_\ell| = a x_0$ such that
$\partial C_\ell$ is 
a surface belonging to a sphere centred at $\mathbf{x}_\ell$ with radius $a x_0$. 
The cloaked region $C$ therefore consists of the domain interior to the union of surfaces formed by identical imaginary spheres of radius $a x_0$ 
located at the vertices. 
Although active cloaking for a plane wave $u_i(\mathbf{x}) = e^{i\mathbf{k} \cdot \mathbf{x}} $ is considered in particular, the source amplitude in the form \eqref{q_l} -- \eqref{Range} applies to a general incident wave since the expansion coefficient $Q_{nm}$ suffices to fully describe the nature of $u_i$ by \eqref{u_i}. 
Nonetheless, the case of plane wave incidence where $Q_{nm} = 4\pi i^n \overline{Y^m_n(\widehat{\mathbf{k}})}$ admits a more compact form of $q_{\ell,nm}$ as \eqref{q_l} is reducible to
\begin{align}
	q_{\ell,nm} = -ik^2 e^{i \mathbf{k} \cdot \mathbf{x}_\ell} \sum_{\nu=0}^\infty \sum_{\mu=-\nu}^{\nu}  Q_{\nu \mu} D_{\nu n}  \mathcal{I}_{n\nu}^{m\mu}(\mathbf{x}_\ell, \partial C_\ell). \label{q_plane_wave}
\end{align}
Note that for computational purpose, the infinite series in \eqref{q_plane_wave} needs to be truncated to a finite order. We discuss further in Part 3 of Supplementary Material how to choose this truncation parameter such that the source amplitude converges with a prescribed level of accuracy.
 
We show in Fig.\ \ref{fig1}(a) -- (e) the respective cloaked region $C$ inside the Platonic solid for each source distribution, with $a$ taken 
as the lower bound in \eqref{Range}.
(In this case, the volume $\mathcal{V}=|C|$ of the domain becomes the maximum.) 
In Fig.\ \ref{fig1}(f), a cross section through the plane $z=0$ for Fig.\ \ref{fig1}(b) is also illustrated.
Note that in every case each surface of integration $\partial C_\ell$ is delimited by a total of $q$ identical circular arcs, where $q$ depends on $L$ as indicated in Table \ref{geo}. For non-Platonic distributions of sources these arcs would \textit{not} be identical. 

We will first evaluate the surface integral $\mathcal{I}_{n\nu}^{m\mu}(\mathbf{x}_1, \partial C_1)$ for the active source $\ell=1$. 
A parametric form for the corresponding spherical face $\partial C_1$ can be conveniently derived using a local spherical coordinate system 
centred at the position $\mathbf{x}_1$ (see Part 2 of Supplementary Material). Since $\partial C_1$ possesses a $q$-fold rotational symmetry about the $z$ axis, it can be subdivided into $q$ congruent segments,
each being the region bounded by
\begin{gather}
	g_t (\varphi) \leq \cos \theta \leq 1, \label{cos}\\
	\varphi_t - \pi/q \leq \varphi \leq \varphi_t + \pi/q, \label{varphi}
\end{gather}
where
\begin{align}
	g_t (\varphi) &= \Big \{1+ \big[a^2h^2\cos^4 (\varphi-\varphi_t) - h(1-a^2) \nonumber \\
	& \ \ \ \times \cos^2 (\varphi-\varphi_t) \big]^{1/2} \Big \}  / \left\{ a \left[1+h\cos^2(\varphi-\varphi_t) \right] \right \}, \label{g} \\
	h &= (1-z_2)/(1+z_2), \label{h}
\end{align}
and $\varphi_{t}, z_2$ are defined as in \eqref{x_2}. The entire surface $\partial C_1$ is given by the union of segments parameterized by \eqref{cos} -- \eqref{h} over $t=2,\cdots,q+1$. 
Since all surfaces of integration $\partial C_\ell$ have an identical geometric shape  within each Platonic distribution of sources, this parametric form describes not only the face $\partial C_1$ but also the rest as long as a suitable coordinate transformation is applied to account for their different orientations. 
We may also compute the volumes of the cloaked region $C$ using the divergence theorem on \eqref{cos} -- \eqref{h}. 
In Table \ref{geo} we list the volume $\mathcal{V}$ for all the five cases illustrated in Fig.\ \ref{fig1}. 
It is observed that amongst the Platonic source distributions, the 
largest cloaked region is attained when $L=12$.   

\begin{figure*}[!t] 
	\centering
	\begin{tabular}{cccc}
		(a) &\subf{\includegraphics[trim={0cm 2.2cm 0cm 1cm},clip,width=7cm]{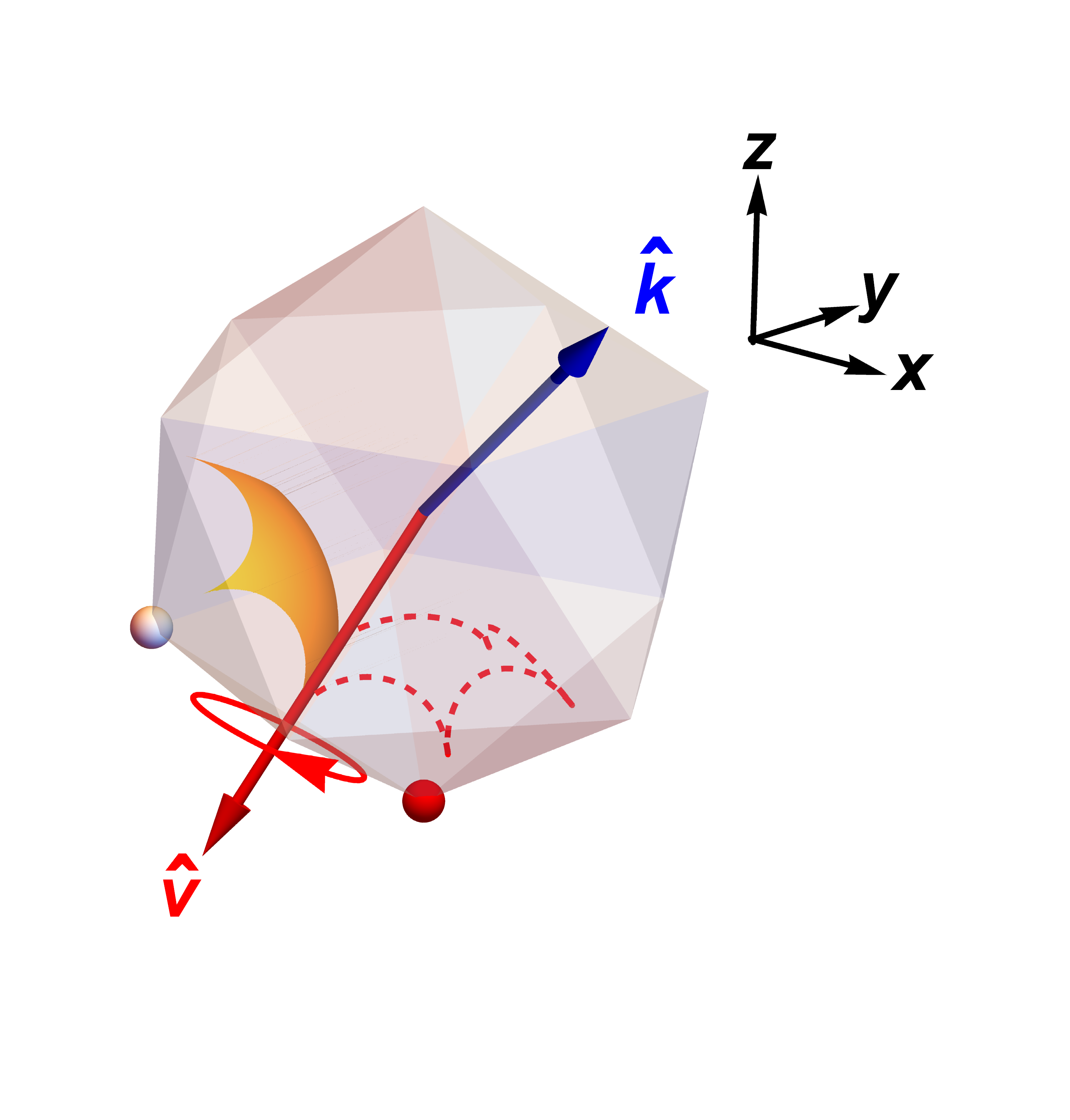}}&
		(b) &\subf{\includegraphics[trim={0cm 2.2cm 0cm 1cm},clip,width=7cm]{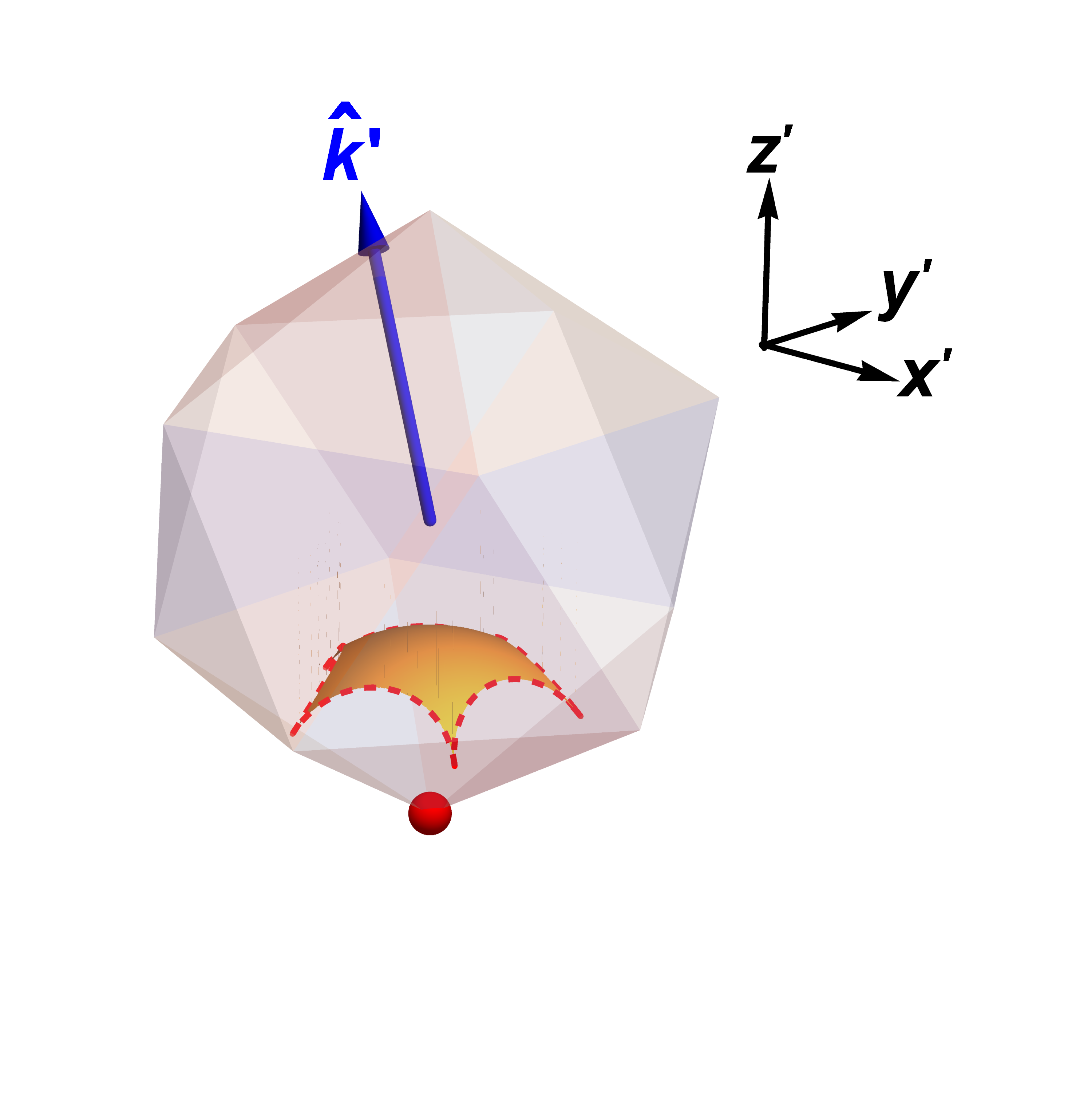}}
	\end{tabular}
	\caption{
	Visualization of the transformation between (a) the original space $\mathbf{x}=(x,y,z)$ and (b) the rotated space $\mathbf{x}'=(x',y',z')$. Under the rotation $\mathbf{R}(\widehat{\mathbf{v}},\Theta)$, the $\ell$th active source is mapped to the bottommost position in the source distribution replacing the source $\ell=1$ (the red small sphere) with the spherical bounding face $\partial C_\ell$ oriented in the same way as $\partial C_1$ (delimited by the red dotted curve) is in the frame $\mathbf{x}$. Meanwhile the propagating vector of the incident wave $\widehat{\mathbf{k}}$ is mapped to the new direction $\widehat{\mathbf{k}}'$.}
	\label{fig2}
\end{figure*}

Given the regularity and rotational symmetry of the Platonic source distribution, we can work out the amplitudes of all sources using only the knowledge of the explicit formula \eqref{q_plane_wave} and the surface integral $\mathcal{I}_{n\nu}^{m\mu}(\mathbf{x}_1, \partial C_1)$ instead of performing the integration for all the $L$ active sources. 
To elucidate the approach, it is useful to write the coefficient $q_{\ell,nm}$ as $q_{\ell,nm}(\widehat{\mathbf{k}},\mathbf{x}_\ell,\partial C_\ell)$ to indicate its dependence on the propagating vector $\widehat{\mathbf{k}}$ and the two geometric parameters.
For notational brevity, we also denote the surface integral as $\mathcal{I}_\ell$ by suppressing the indices and arguments in \eqref{I}.
Define a new coordinate system $\mathbf{x}'=(x',y',z')= \mathbf{R}(\widehat{\mathbf{v}},\Theta) \mathbf{x}$ where $\mathbf{R}(\widehat{\mathbf{v}},\Theta)$ is a rotation matrix with $\widehat{\mathbf{v}}$ the unit vector representing the rotation axis and $\Theta$ the rotation angle. As illustrated in Fig.\ \ref{fig2}, the idea is to rotate the original frame $\mathbf{x}$ such that in the rotated frame $\mathbf{x}'$, the $\ell$th active source takes up the bottommost location replacing the source $\ell=1$ and the corresponding bounding surface $\partial C_\ell$ 
has the same orientation as $\partial C_1$ does in the original frame $\mathbf{x}$. 
Under such transformation, the source amplitude in \eqref{q_plane_wave} becomes
\begin{align}
	q_{\ell,nm}'(\widehat{\mathbf{k}}',\mathbf{x}_\ell', \partial C_\ell') = -ik^2 e^{ik \widehat{\mathbf{k}}' \cdot \mathbf{x}_\ell'} \sum_{\nu=0}^\infty \sum_{\mu=-\nu}^{\nu}  Q_{\nu \mu}' D_{\nu n}'  \mathcal{I}_\ell', \label{q_plane_wave_prime}
\end{align}
where $\widehat{\mathbf{k}}'$, $\mathbf{x}_\ell'$ and $\partial C_\ell'$ are the new forms of the propagating vector, position vector and integration surface in the space $\mathbf{x}'$ respectively and $Q_{\nu\mu}'=4\pi i^\nu \overline{Y^\mu_\nu(\widehat{\mathbf{k}}')}$.
Note that by construction, $\mathbf{x}_\ell'=\mathbf{x}_1$ and $\partial C_\ell'=\partial C_1$.
As the forms of $D_{\nu n}$ and $\mathcal{I}_\ell$ suggest in \eqref{D_vn} and \eqref{I}, they are independent of $\widehat{\mathbf{k}}$.
The rotational symmetry of the Platonic solids and our choice of $\mathbf{R}$ thus ensure that 
they remain rotationally invariant across all $\ell$. 
In particular, we have $D_{\nu n}' = D_{\nu n}$ and $\mathcal{I}'_\ell = \mathcal{I}_1$.
The source amplitude $q_{\ell,nm}(\widehat{\mathbf{k}},\mathbf{x}_\ell,\partial C_\ell)$, is thus transformed to $q_{1,nm}(\widehat{\mathbf{k}}',\mathbf{x}_1,\partial C_1)$ in the rotated frame $\mathbf{x}'$. While the former has to be calculated by integrating over the faces $\partial C_\ell$ for all $\ell$, the latter can be evaluated by simply integrating over $\partial C_1$  
parameterized by \eqref{cos} -- \eqref{h} and replacing $\widehat{\mathbf{k}}$ by $\widehat{\mathbf{k}}'$.   
If we express the source coefficients as a vector $\mathbf{q}_{\ell,n}=\{ q_{\ell,nm} \}_{m=-n}^n$ , then it can be shown that the system of linear equations
\begin{align}
\mathbf{D}^{n}(\gamma, \beta, \alpha) \mathbf{q}_{\ell, n}(\widehat{\mathbf{k}},\mathbf{x}_\ell,\partial C_\ell) = \mathbf{q}_{1,n}(\widehat{\mathbf{k}}',\mathbf{x}_1,\partial C_1) \label{Dq=q}
\end{align}
holds where $\mathbf{D}^{n}(\gamma, \beta, \alpha)$ (defined in \eqref{3d_o_D} -- \eqref{C} of Supplementary Material) is the Wigner D-matrix \cite{wigner2012group} with dimensions $(2n+1) \times (2n+1)$ and $\gamma, \beta, \alpha$ are the Euler angles \cite{varshalovich1988description} of the matrix $\mathbf{R}(\widehat{\mathbf{v}},\Theta)$ such that
\begin{align}
\mathbf{R}(\widehat{\mathbf{v}},\Theta)= \mathbf{R} (\widehat{\mathbf{e}}_z,\gamma) \mathbf{R} (\widehat{\mathbf{e}}_y,\beta) \mathbf{R} (\widehat{\mathbf{e}}_z,\alpha) \label{R=RRR}
\end{align}
with $\widehat{\mathbf{e}}_y, \widehat{\mathbf{e}}_z$ the unit vectors along the $y,z$ axes. By solving \eqref{Dq=q} for $n=0,1,\cdots,N$ with $N$ a positive integer 
, we can retrieve the full set of amplitudes $q_{\ell,nm}(\widehat{\mathbf{k}},\mathbf{x}_\ell,\partial C_\ell)$ for the $\ell$th active source up to the $N$th order multipole. The configuration of the cloaking device can be completed by repeating this procedure with the corresponding $\widehat{\mathbf{k}}'$ and Euler angles for different sources. The benefit of employing active sources distributed on the vertices of the Platonic solids is therefore that once one set of source coefficients is determined for a given location, those for others follow by simple post-processing operations. More technical details about this approach including the exact form of the rotation matrix $\mathbf{R}$ can be found in Part 5 of Supplementary Material.





\begin{figure}[h!] 
	\centering
	\begin{tabular}{c}
		(a)\subf{\includegraphics[trim={0cm 0cm 0cm 0cm},clip,width=6cm]{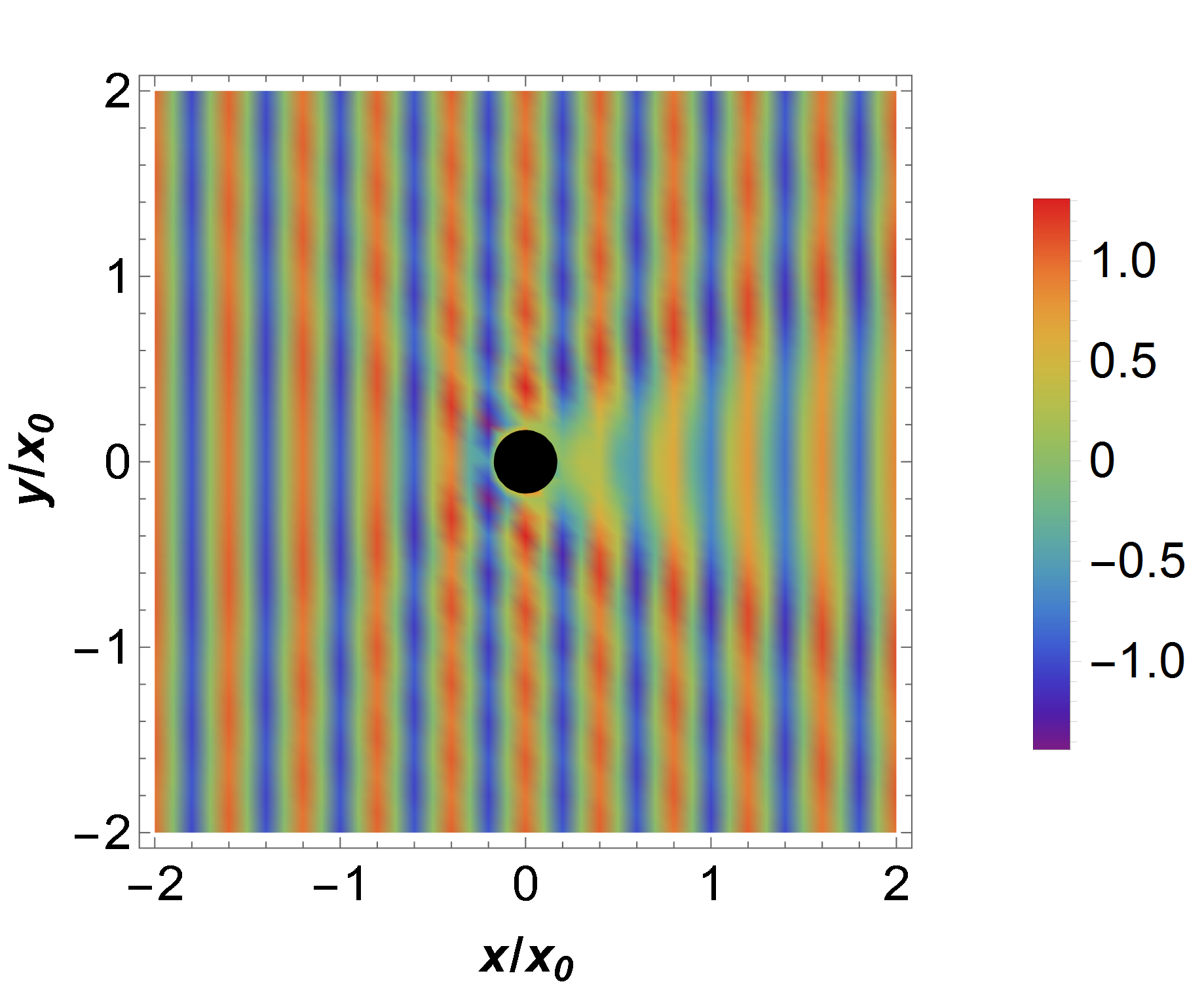}}
		\\
		(b)\subf{\includegraphics[trim={0cm 0cm 0cm 0cm},clip,width=6cm]{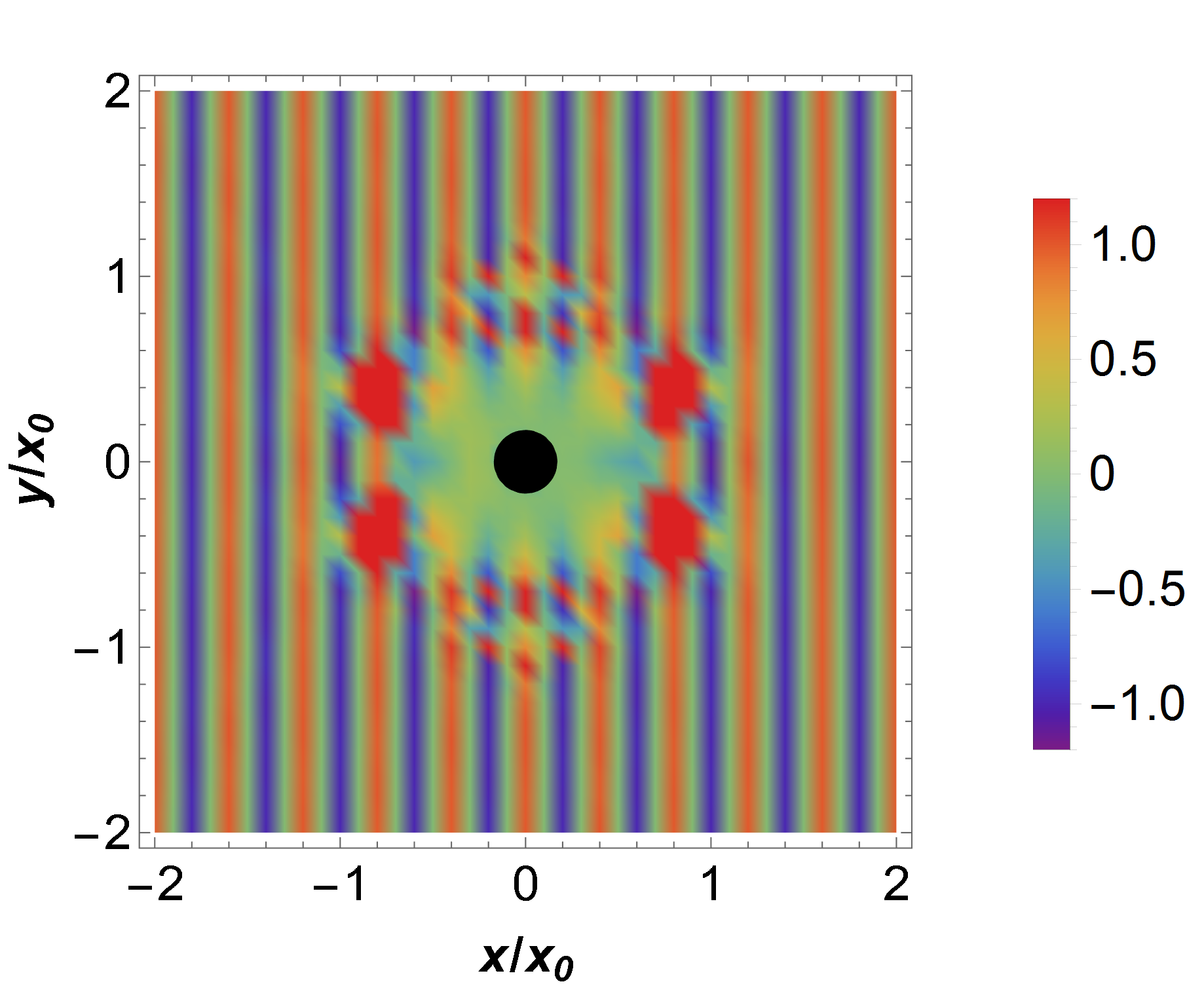}}
	\end{tabular}
	\caption{
		The real part of the total wave field $u$ on the cross section $z=0$ with a sound-soft sphere 
		subject to an incident plane wave with angles of incidence $\theta_i=\pi/2, \varphi_i=0$ 
		when the cloaking devices are switched off (a) and on (b). 
		Here the source distance is $kx_0=5\pi$ and the sphere radius is $kA \approx 0.8579\pi$.
		A total of $L=20$ active sources are used, each consisting of multipoles up to 
		$N=10$.  
	} \label{fig3}
\end{figure}

The reduction of the scattered field from an obstacle by the active exterior cloaking method discussed
is illustrated in Fig.\ \ref{fig3} with $L=20$ active sources distributed at the vertices of an imaginary regular dodecahedron with source distance $kx_0= 5\pi$. Here we consider the problem in the context of acoustics but the principles are similar for other scalar waves. The incident plane wave is propagating in the positive $x$ direction with $\theta_i=\pi/2, \varphi_i=0$. The scattering object inside the cloaked region $C$ is a sound-soft sphere with radius $kA =3(1-a)kx_0 \approx 0.8579\pi$ where $a=(2\sqrt{6}/3)/[2\sin(\pi/3)]$. Note that here $a$ is taken as the minimum radius permissible of the imaginary spheres bounding $C$ for $L=4$, which means that the scattering sphere has a radius three times that of the inscribed sphere of $C$ when $L=4$. 
In both subplots the real part of the total wave field $u$ on the cross section $z=0$ is shown. In Fig.\ \ref{fig3}(a), the cloaking devices are inactive and a prominent scattering pattern including distorted wavefronts and a shadow region behind the sphere is observed. In Fig.\ \ref{fig3}(b), 
the cloaking devices are activated and the multipole order of each active source is taken as $N=10$. The series expansions for the source coefficients in \eqref{q_plane_wave} are truncated such that the active field produced by each source is within $1\%$ relative error. The sphere now resides in a region with literally zero wave amplitude. The straighter wavefronts and the absence of the shadow region in the figure indicate that the incident wave is scattered only slightly by the sphere inside the quiet zone and the sources radiate little into the exterior of the silent region, demonstrating the effectiveness of the cloak. Note that the wave field diverges within the small neighbourhoods centred at the active point sources. In practice these large fields are confined within the finite-sized sources. For visualization purpose, we have cropped the excessively large parts to a sufficiently small value.   

While it is seen that our cloaking method can effectively suppress scattering on a local cross section, the cloaking effect can be assessed 
globally by examining the total power $W$ radiated by the entire system into the far field. 
We may look at the quantity $\sigma$ defined by
\begin{align}
	\sigma &= \frac{W(u_d\neq 0)}{W(u_d=0)} 
	, \label{sigma}
\end{align}
which, in the context of acoustic wave, is 
the ratio between the sound power detected in the far field after (when $u_d\neq 0$) and before (when $u_d= 0$) the cloak is activated. 
Detailed expressions for $W(u_d\neq0)$ and $W(u_d=0)$ are derived in Part 6 of Supplementary Material. 
In Fig. \ref{sc_plot}, we plot the sound power level (SWL) defined by $\text{SWL}=10\log \sigma$ against the nondimensionalized radius $kA$ of the sound-soft sphere inside the silent zone for the five source distributions depicted in Fig. \ref{fig1}. The modal order $N$, the sphere radius $A$ and the direction of incident wave remain the same as those in Fig. \ref{fig3}. The SWL is simulated for wavenumbers in the range $\pi/2 \leq kx_0 \leq 6\pi$ at an interval of $\pi/2$ (or $0.0858\pi \leq kA \leq 1.0294\pi$ at an interval of around $0.0858\pi$). Note that at each observation point, the truncation parameter of the corresponding source amplitude is chosen to achieve a relative error less than $1\%$ in the active field generated by each source. For our cloaking approach to work we require $\sigma<1$ and thus $10 \log \sigma <0$. It is observed that 
in all cases of $L$ the SWL remains negative for the range of $kA$ studied here with a more significant reduction attained for lower frequencies. 
As the number of sources $L$ increases the cloaking effect improves in general with a maximum reduction of about $70$ decibels achieved when $L=20$ at $kA\approx 0.1716\pi$. Further study about how changes in different parameters affect the performance of the cloaking system can be found also in Part 6 of Supplementary Material.


\begin{figure}[h!] 
	\centering
	\includegraphics[trim={0cm 0cm 0cm 0cm},clip,width=8.5cm]{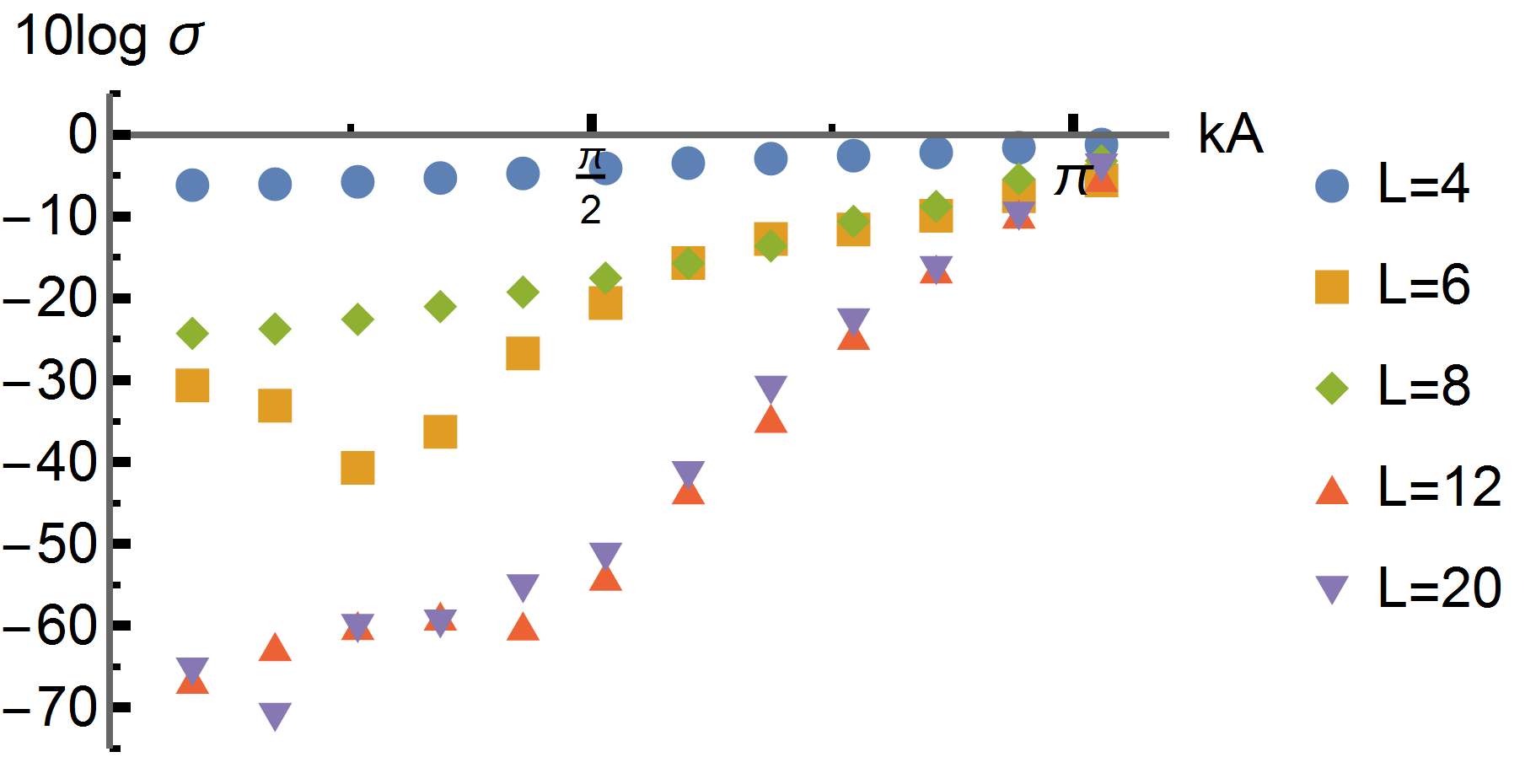}
	\caption{The sound power level $\text{SWL}=10\log \sigma$ versus the nondimensionalized radius $kA$ of the sound-soft sphere inside the cloaked region for five different number of sources $L$ and $\pi/2 \leq kx_0 \leq 6\pi$ at an interval of $\pi/2$ (or $0.0858\pi \leq kA \leq 1.0294\pi$ at an interval of around $0.0858\pi$). 
	The other parameters are the same as those in Fig. \ref{fig3}.} 
		\label{sc_plot}
\end{figure}

In summary, we have formulated an efficient three-dimensional active exterior cloaking strategy for the scalar Helmholtz equation, which employs multipolar sources distributed at the vertices of the Platonic solids to create a domain with zero total wave amplitude. This approach minimizes the incident wave impinging upon the object inside the region to suppress the scattered wave whilst simultaneously ensuring that the active source radiation is minimized. The Platonic distribution of the active sources means that we only need to determine the source amplitudes 
as an integral expression at one source location in terms of the incident field and those at other locations will follow from simple post-processing, by exploiting the symmetry and regularity of the Platonic solids.  

This work was supported by a University of Manchester President's scholarship for Yeung (2017-21) and by the Engineering and Physical Sciences Research Council (grant EP/L018039/1) for Parnell. 

\nocite{*}


\bibliographystyle{apsrev4-2} 
\bibliography{bibtexrefsCHtrial}

\begin{thebibliography}{50}%
\makeatletter
\providecommand \@ifxundefined [1]{%
 \@ifx{#1\undefined}
}%
\providecommand \@ifnum [1]{%
 \ifnum #1\expandafter \@firstoftwo
 \else \expandafter \@secondoftwo
 \fi
}%
\providecommand \@ifx [1]{%
 \ifx #1\expandafter \@firstoftwo
 \else \expandafter \@secondoftwo
 \fi
}%
\providecommand \natexlab [1]{#1}%
\providecommand \enquote  [1]{``#1''}%
\providecommand \bibnamefont  [1]{#1}%
\providecommand \bibfnamefont [1]{#1}%
\providecommand \citenamefont [1]{#1}%
\providecommand \href@noop [0]{\@secondoftwo}%
\providecommand \href [0]{\begingroup \@sanitize@url \@href}%
\providecommand \@href[1]{\@@startlink{#1}\@@href}%
\providecommand \@@href[1]{\endgroup#1\@@endlink}%
\providecommand \@sanitize@url [0]{\catcode `\\12\catcode `\$12\catcode
  `\&12\catcode `\#12\catcode `\^12\catcode `\_12\catcode `\%12\relax}%
\providecommand \@@startlink[1]{}%
\providecommand \@@endlink[0]{}%
\providecommand \url  [0]{\begingroup\@sanitize@url \@url }%
\providecommand \@url [1]{\endgroup\@href {#1}{\urlprefix }}%
\providecommand \urlprefix  [0]{URL }%
\providecommand \Eprint [0]{\href }%
\providecommand \doibase [0]{https://doi.org/}%
\providecommand \selectlanguage [0]{\@gobble}%
\providecommand \bibinfo  [0]{\@secondoftwo}%
\providecommand \bibfield  [0]{\@secondoftwo}%
\providecommand \translation [1]{[#1]}%
\providecommand \BibitemOpen [0]{}%
\providecommand \bibitemStop [0]{}%
\providecommand \bibitemNoStop [0]{.\EOS\space}%
\providecommand \EOS [0]{\spacefactor3000\relax}%
\providecommand \BibitemShut  [1]{\csname bibitem#1\endcsname}%
\let\auto@bib@innerbib\@empty
\bibitem [{\citenamefont {Schurig}\ \emph {et~al.}(2006)\citenamefont
  {Schurig}, \citenamefont {Mock}, \citenamefont {Justice}, \citenamefont
  {Cummer}, \citenamefont {Pendry}, \citenamefont {Starr},\ and\ \citenamefont
  {Smith}}]{schurig2006metamaterial}%
  \BibitemOpen
  \bibfield  {author} {\bibinfo {author} {\bibfnamefont {D.}~\bibnamefont
  {Schurig}}, \bibinfo {author} {\bibfnamefont {J.~J.}\ \bibnamefont {Mock}},
  \bibinfo {author} {\bibfnamefont {B.}~\bibnamefont {Justice}}, \bibinfo
  {author} {\bibfnamefont {S.~A.}\ \bibnamefont {Cummer}}, \bibinfo {author}
  {\bibfnamefont {J.~B.}\ \bibnamefont {Pendry}}, \bibinfo {author}
  {\bibfnamefont {A.~F.}\ \bibnamefont {Starr}},\ and\ \bibinfo {author}
  {\bibfnamefont {D.~R.}\ \bibnamefont {Smith}},\ }\href@noop {} {\bibfield
  {journal} {\bibinfo  {journal} {Science}\ }\textbf {\bibinfo {volume}
  {314}},\ \bibinfo {pages} {977} (\bibinfo {year} {2006})}\BibitemShut
  {NoStop}%
\bibitem [{\citenamefont {Miller}(2006)}]{miller2006perfect}%
  \BibitemOpen
  \bibfield  {author} {\bibinfo {author} {\bibfnamefont {D.~A.}\ \bibnamefont
  {Miller}},\ }\href@noop {} {\bibfield  {journal} {\bibinfo  {journal} {Optics
  Express}\ }\textbf {\bibinfo {volume} {14}},\ \bibinfo {pages} {12457}
  (\bibinfo {year} {2006})}\BibitemShut {NoStop}%
\bibitem [{\citenamefont {Chen}\ and\ \citenamefont
  {Chan}(2007)}]{chen2007acoustic}%
  \BibitemOpen
  \bibfield  {author} {\bibinfo {author} {\bibfnamefont {H.}~\bibnamefont
  {Chen}}\ and\ \bibinfo {author} {\bibfnamefont {C.}~\bibnamefont {Chan}},\
  }\href@noop {} {\bibfield  {journal} {\bibinfo  {journal} {Applied physics
  letters}\ }\textbf {\bibinfo {volume} {91}},\ \bibinfo {pages} {183518}
  (\bibinfo {year} {2007})}\BibitemShut {NoStop}%
\bibitem [{\citenamefont {Norris}(2008)}]{norris2008acoustic}%
  \BibitemOpen
  \bibfield  {author} {\bibinfo {author} {\bibfnamefont {A.~N.}\ \bibnamefont
  {Norris}},\ }\href@noop {} {\bibfield  {journal} {\bibinfo  {journal}
  {Proceedings of the Royal Society A: Mathematical, Physical and Engineering
  Sciences}\ }\textbf {\bibinfo {volume} {464}},\ \bibinfo {pages} {2411}
  (\bibinfo {year} {2008})}\BibitemShut {NoStop}%
\bibitem [{\citenamefont {Fleury}\ and\ \citenamefont
  {Al{\`u}}(2014)}]{fleury2014cloaking}%
  \BibitemOpen
  \bibfield  {author} {\bibinfo {author} {\bibfnamefont {R.}~\bibnamefont
  {Fleury}}\ and\ \bibinfo {author} {\bibfnamefont {A.}~\bibnamefont
  {Al{\`u}}},\ }in\ \href@noop {} {\emph {\bibinfo {booktitle} {Forum for
  Electromagnetic Research Methods and Application Technologies (FERMAT)}}},\
  Vol.~\bibinfo {volume} {1}\ (\bibinfo {year} {2014})\BibitemShut {NoStop}%
\bibitem [{\citenamefont {Cai}\ \emph {et~al.}(2007)\citenamefont {Cai},
  \citenamefont {Chettiar}, \citenamefont {Kildishev},\ and\ \citenamefont
  {Shalaev}}]{cai2007optical}%
  \BibitemOpen
  \bibfield  {author} {\bibinfo {author} {\bibfnamefont {W.}~\bibnamefont
  {Cai}}, \bibinfo {author} {\bibfnamefont {U.~K.}\ \bibnamefont {Chettiar}},
  \bibinfo {author} {\bibfnamefont {A.~V.}\ \bibnamefont {Kildishev}},\ and\
  \bibinfo {author} {\bibfnamefont {V.~M.}\ \bibnamefont {Shalaev}},\
  }\href@noop {} {\bibfield  {journal} {\bibinfo  {journal} {Nature photonics}\
  }\textbf {\bibinfo {volume} {1}},\ \bibinfo {pages} {224} (\bibinfo {year}
  {2007})}\BibitemShut {NoStop}%
\bibitem [{\citenamefont {Silveirinha}\ \emph {et~al.}(2007)\citenamefont
  {Silveirinha}, \citenamefont {Al{\`u}},\ and\ \citenamefont
  {Engheta}}]{silveirinha2007parallel}%
  \BibitemOpen
  \bibfield  {author} {\bibinfo {author} {\bibfnamefont {M.~G.}\ \bibnamefont
  {Silveirinha}}, \bibinfo {author} {\bibfnamefont {A.}~\bibnamefont
  {Al{\`u}}},\ and\ \bibinfo {author} {\bibfnamefont {N.}~\bibnamefont
  {Engheta}},\ }\href@noop {} {\bibfield  {journal} {\bibinfo  {journal}
  {Physical Review E}\ }\textbf {\bibinfo {volume} {75}},\ \bibinfo {pages}
  {036603} (\bibinfo {year} {2007})}\BibitemShut {NoStop}%
\bibitem [{\citenamefont {Zhang}\ \emph {et~al.}(2011)\citenamefont {Zhang},
  \citenamefont {Xia},\ and\ \citenamefont {Fang}}]{zhang2011broadband}%
  \BibitemOpen
  \bibfield  {author} {\bibinfo {author} {\bibfnamefont {S.}~\bibnamefont
  {Zhang}}, \bibinfo {author} {\bibfnamefont {C.}~\bibnamefont {Xia}},\ and\
  \bibinfo {author} {\bibfnamefont {N.}~\bibnamefont {Fang}},\ }\href@noop {}
  {\bibfield  {journal} {\bibinfo  {journal} {Physical review letters}\
  }\textbf {\bibinfo {volume} {106}},\ \bibinfo {pages} {024301} (\bibinfo
  {year} {2011})}\BibitemShut {NoStop}%
\bibitem [{\citenamefont {Craster}\ and\ \citenamefont
  {Guenneau}(2012)}]{craster2012acoustic}%
  \BibitemOpen
  \bibfield  {author} {\bibinfo {author} {\bibfnamefont {R.~V.}\ \bibnamefont
  {Craster}}\ and\ \bibinfo {author} {\bibfnamefont {S.}~\bibnamefont
  {Guenneau}},\ }\href@noop {} {\emph {\bibinfo {title} {Acoustic
  metamaterials: Negative refraction, imaging, lensing and cloaking}}},\ Vol.\
  \bibinfo {volume} {166}\ (\bibinfo  {publisher} {Springer Science \& Business
  Media},\ \bibinfo {year} {2012})\BibitemShut {NoStop}%
\bibitem [{\citenamefont {Vasquez}\ \emph
  {et~al.}(2009{\natexlab{a}})\citenamefont {Vasquez}, \citenamefont {Milton},\
  and\ \citenamefont {Onofrei}}]{vasquez2009active}%
  \BibitemOpen
  \bibfield  {author} {\bibinfo {author} {\bibfnamefont {F.~G.}\ \bibnamefont
  {Vasquez}}, \bibinfo {author} {\bibfnamefont {G.~W.}\ \bibnamefont
  {Milton}},\ and\ \bibinfo {author} {\bibfnamefont {D.}~\bibnamefont
  {Onofrei}},\ }\href@noop {} {\bibfield  {journal} {\bibinfo  {journal}
  {Physical review letters}\ }\textbf {\bibinfo {volume} {103}},\ \bibinfo
  {pages} {073901} (\bibinfo {year} {2009}{\natexlab{a}})}\BibitemShut
  {NoStop}%
\bibitem [{\citenamefont {Zheng}\ \emph {et~al.}(2010)\citenamefont {Zheng},
  \citenamefont {Xiao}, \citenamefont {Lai},\ and\ \citenamefont
  {Chan}}]{zheng2010exterior}%
  \BibitemOpen
  \bibfield  {author} {\bibinfo {author} {\bibfnamefont {H.}~\bibnamefont
  {Zheng}}, \bibinfo {author} {\bibfnamefont {J.}~\bibnamefont {Xiao}},
  \bibinfo {author} {\bibfnamefont {Y.}~\bibnamefont {Lai}},\ and\ \bibinfo
  {author} {\bibfnamefont {C.}~\bibnamefont {Chan}},\ }\href@noop {} {\bibfield
   {journal} {\bibinfo  {journal} {Physical Review B}\ }\textbf {\bibinfo
  {volume} {81}},\ \bibinfo {pages} {195116} (\bibinfo {year}
  {2010})}\BibitemShut {NoStop}%
\bibitem [{\citenamefont {Guicking}(1990)}]{guicking1990invention}%
  \BibitemOpen
  \bibfield  {author} {\bibinfo {author} {\bibfnamefont {D.}~\bibnamefont
  {Guicking}},\ }\href@noop {} {\bibfield  {journal} {\bibinfo  {journal} {The
  Journal of the Acoustical Society of America}\ }\textbf {\bibinfo {volume}
  {87}},\ \bibinfo {pages} {2251} (\bibinfo {year} {1990})}\BibitemShut
  {NoStop}%
\bibitem [{\citenamefont {Nelson}\ and\ \citenamefont
  {Elliott}(1991)}]{nelson1991active}%
  \BibitemOpen
  \bibfield  {author} {\bibinfo {author} {\bibfnamefont {P.~A.}\ \bibnamefont
  {Nelson}}\ and\ \bibinfo {author} {\bibfnamefont {S.~J.}\ \bibnamefont
  {Elliott}},\ }\href@noop {} {\emph {\bibinfo {title} {Active control of
  sound}}}\ (\bibinfo  {publisher} {Academic press},\ \bibinfo {year}
  {1991})\BibitemShut {NoStop}%
\bibitem [{\citenamefont {Fuller}\ \emph {et~al.}(1996)\citenamefont {Fuller},
  \citenamefont {Elliott},\ and\ \citenamefont {Nelson}}]{fuller1996active}%
  \BibitemOpen
  \bibfield  {author} {\bibinfo {author} {\bibfnamefont {C.~C.}\ \bibnamefont
  {Fuller}}, \bibinfo {author} {\bibfnamefont {S.}~\bibnamefont {Elliott}},\
  and\ \bibinfo {author} {\bibfnamefont {P.~A.}\ \bibnamefont {Nelson}},\
  }\href@noop {} {\emph {\bibinfo {title} {Active control of vibration}}}\
  (\bibinfo  {publisher} {Academic Press},\ \bibinfo {year} {1996})\BibitemShut
  {NoStop}%
\bibitem [{\citenamefont {Guicking}(2007)}]{guicking2007active}%
  \BibitemOpen
  \bibfield  {author} {\bibinfo {author} {\bibfnamefont {D.}~\bibnamefont
  {Guicking}},\ }\href@noop {} {\bibfield  {journal} {\bibinfo  {journal}
  {Oscillations, Waves and Interactions--60 Years Drittes Physikalisches
  Institute}\ ,\ \bibinfo {pages} {107}} (\bibinfo {year} {2007})}\BibitemShut
  {NoStop}%
\bibitem [{\citenamefont {Cheer}(2016)}]{cheer2016active}%
  \BibitemOpen
  \bibfield  {author} {\bibinfo {author} {\bibfnamefont {J.}~\bibnamefont
  {Cheer}},\ }\href@noop {} {\bibfield  {journal} {\bibinfo  {journal} {The
  Journal of the Acoustical Society of America}\ }\textbf {\bibinfo {volume}
  {140}},\ \bibinfo {pages} {1502} (\bibinfo {year} {2016})}\BibitemShut
  {NoStop}%
\bibitem [{\citenamefont {Ma}\ \emph {et~al.}(2013)\citenamefont {Ma},
  \citenamefont {Mei}, \citenamefont {Zhu}, \citenamefont {Jin},\ and\
  \citenamefont {Cui}}]{ma2013experiments}%
  \BibitemOpen
  \bibfield  {author} {\bibinfo {author} {\bibfnamefont {Q.}~\bibnamefont
  {Ma}}, \bibinfo {author} {\bibfnamefont {Z.~L.}\ \bibnamefont {Mei}},
  \bibinfo {author} {\bibfnamefont {S.~K.}\ \bibnamefont {Zhu}}, \bibinfo
  {author} {\bibfnamefont {T.~Y.}\ \bibnamefont {Jin}},\ and\ \bibinfo {author}
  {\bibfnamefont {T.~J.}\ \bibnamefont {Cui}},\ }\href@noop {} {\bibfield
  {journal} {\bibinfo  {journal} {Physical review letters}\ }\textbf {\bibinfo
  {volume} {111}},\ \bibinfo {pages} {173901} (\bibinfo {year}
  {2013})}\BibitemShut {NoStop}%
\bibitem [{\citenamefont {O'Neill}\ \emph {et~al.}(2015)\citenamefont
  {O'Neill}, \citenamefont {Selsil}, \citenamefont {McPhedran}, \citenamefont
  {Movchan},\ and\ \citenamefont {Movchan}}]{o2015active}%
  \BibitemOpen
  \bibfield  {author} {\bibinfo {author} {\bibfnamefont {J.}~\bibnamefont
  {O'Neill}}, \bibinfo {author} {\bibfnamefont {{\"O}.}~\bibnamefont {Selsil}},
  \bibinfo {author} {\bibfnamefont {R.}~\bibnamefont {McPhedran}}, \bibinfo
  {author} {\bibfnamefont {A.}~\bibnamefont {Movchan}},\ and\ \bibinfo {author}
  {\bibfnamefont {N.}~\bibnamefont {Movchan}},\ }\href@noop {} {\bibfield
  {journal} {\bibinfo  {journal} {The Quarterly Journal of Mechanics and
  Applied Mathematics}\ }\textbf {\bibinfo {volume} {68}},\ \bibinfo {pages}
  {263} (\bibinfo {year} {2015})}\BibitemShut {NoStop}%
\bibitem [{\citenamefont {Eggler}\ \emph
  {et~al.}(2019{\natexlab{a}})\citenamefont {Eggler}, \citenamefont {Chung},
  \citenamefont {Montiel}, \citenamefont {Pan},\ and\ \citenamefont
  {Kessissoglou}}]{eggler2019activeA}%
  \BibitemOpen
  \bibfield  {author} {\bibinfo {author} {\bibfnamefont {D.}~\bibnamefont
  {Eggler}}, \bibinfo {author} {\bibfnamefont {H.}~\bibnamefont {Chung}},
  \bibinfo {author} {\bibfnamefont {F.}~\bibnamefont {Montiel}}, \bibinfo
  {author} {\bibfnamefont {J.}~\bibnamefont {Pan}},\ and\ \bibinfo {author}
  {\bibfnamefont {N.}~\bibnamefont {Kessissoglou}},\ }\href@noop {} {\bibfield
  {journal} {\bibinfo  {journal} {Wave Motion}\ }\textbf {\bibinfo {volume}
  {87}},\ \bibinfo {pages} {106} (\bibinfo {year}
  {2019}{\natexlab{a}})}\BibitemShut {NoStop}%
\bibitem [{\citenamefont {Eggler}\ \emph
  {et~al.}(2019{\natexlab{b}})\citenamefont {Eggler}, \citenamefont {Karimi},\
  and\ \citenamefont {Kessissoglou}}]{eggler2019activeB}%
  \BibitemOpen
  \bibfield  {author} {\bibinfo {author} {\bibfnamefont {D.}~\bibnamefont
  {Eggler}}, \bibinfo {author} {\bibfnamefont {M.}~\bibnamefont {Karimi}},\
  and\ \bibinfo {author} {\bibfnamefont {N.}~\bibnamefont {Kessissoglou}},\
  }\href@noop {} {\bibfield  {journal} {\bibinfo  {journal} {The Journal of the
  Acoustical Society of America}\ }\textbf {\bibinfo {volume} {146}},\ \bibinfo
  {pages} {586} (\bibinfo {year} {2019}{\natexlab{b}})}\BibitemShut {NoStop}%
\bibitem [{\citenamefont {House}\ \emph {et~al.}(2020)\citenamefont {House},
  \citenamefont {Cheer},\ and\ \citenamefont {Daley}}]{house2020experimental}%
  \BibitemOpen
  \bibfield  {author} {\bibinfo {author} {\bibfnamefont {C.}~\bibnamefont
  {House}}, \bibinfo {author} {\bibfnamefont {J.}~\bibnamefont {Cheer}},\ and\
  \bibinfo {author} {\bibfnamefont {S.}~\bibnamefont {Daley}},\ }\href@noop {}
  {\bibfield  {journal} {\bibinfo  {journal} {Applied Acoustics}\ }\textbf
  {\bibinfo {volume} {170}},\ \bibinfo {pages} {107436} (\bibinfo {year}
  {2020})}\BibitemShut {NoStop}%
\bibitem [{\citenamefont {Vasquez}\ \emph {et~al.}(2011)\citenamefont
  {Vasquez}, \citenamefont {Milton},\ and\ \citenamefont
  {Onofrei}}]{vasquez2011exterior}%
  \BibitemOpen
  \bibfield  {author} {\bibinfo {author} {\bibfnamefont {F.~G.}\ \bibnamefont
  {Vasquez}}, \bibinfo {author} {\bibfnamefont {G.~W.}\ \bibnamefont
  {Milton}},\ and\ \bibinfo {author} {\bibfnamefont {D.}~\bibnamefont
  {Onofrei}},\ }\href@noop {} {\bibfield  {journal} {\bibinfo  {journal} {Wave
  Motion}\ }\textbf {\bibinfo {volume} {48}},\ \bibinfo {pages} {515} (\bibinfo
  {year} {2011})}\BibitemShut {NoStop}%
\bibitem [{\citenamefont {Vasquez}\ \emph
  {et~al.}(2009{\natexlab{b}})\citenamefont {Vasquez}, \citenamefont {Milton},\
  and\ \citenamefont {Onofrei}}]{vasquez2009broadband}%
  \BibitemOpen
  \bibfield  {author} {\bibinfo {author} {\bibfnamefont {F.~G.}\ \bibnamefont
  {Vasquez}}, \bibinfo {author} {\bibfnamefont {G.~W.}\ \bibnamefont
  {Milton}},\ and\ \bibinfo {author} {\bibfnamefont {D.}~\bibnamefont
  {Onofrei}},\ }\href@noop {} {\bibfield  {journal} {\bibinfo  {journal}
  {Optics Express}\ }\textbf {\bibinfo {volume} {17}},\ \bibinfo {pages}
  {14800} (\bibinfo {year} {2009}{\natexlab{b}})}\BibitemShut {NoStop}%
\bibitem [{\citenamefont {Norris}\ \emph {et~al.}(2012)\citenamefont {Norris},
  \citenamefont {Amirkulova},\ and\ \citenamefont
  {Parnell}}]{norris2012source}%
  \BibitemOpen
  \bibfield  {author} {\bibinfo {author} {\bibfnamefont {A.~N.}\ \bibnamefont
  {Norris}}, \bibinfo {author} {\bibfnamefont {F.~A.}\ \bibnamefont
  {Amirkulova}},\ and\ \bibinfo {author} {\bibfnamefont {W.~J.}\ \bibnamefont
  {Parnell}},\ }\href@noop {} {\bibfield  {journal} {\bibinfo  {journal}
  {Inverse Problems}\ }\textbf {\bibinfo {volume} {28}},\ \bibinfo {pages}
  {105002} (\bibinfo {year} {2012})}\BibitemShut {NoStop}%
\bibitem [{\citenamefont {Norris}\ \emph {et~al.}(2014)\citenamefont {Norris},
  \citenamefont {Amirkulova},\ and\ \citenamefont
  {Parnell}}]{norris2014active}%
  \BibitemOpen
  \bibfield  {author} {\bibinfo {author} {\bibfnamefont {A.~N.}\ \bibnamefont
  {Norris}}, \bibinfo {author} {\bibfnamefont {F.~A.}\ \bibnamefont
  {Amirkulova}},\ and\ \bibinfo {author} {\bibfnamefont {W.~J.}\ \bibnamefont
  {Parnell}},\ }\href@noop {} {\bibfield  {journal} {\bibinfo  {journal}
  {Mathematics and Mechanics of Solids}\ }\textbf {\bibinfo {volume} {19}},\
  \bibinfo {pages} {603} (\bibinfo {year} {2014})}\BibitemShut {NoStop}%
\bibitem [{\citenamefont {Futhazar}\ \emph {et~al.}(2015)\citenamefont
  {Futhazar}, \citenamefont {Parnell},\ and\ \citenamefont
  {Norris}}]{futhazar2015active}%
  \BibitemOpen
  \bibfield  {author} {\bibinfo {author} {\bibfnamefont {G.}~\bibnamefont
  {Futhazar}}, \bibinfo {author} {\bibfnamefont {W.~J.}\ \bibnamefont
  {Parnell}},\ and\ \bibinfo {author} {\bibfnamefont {A.~N.}\ \bibnamefont
  {Norris}},\ }\href@noop {} {\bibfield  {journal} {\bibinfo  {journal}
  {Journal of Sound and Vibration}\ }\textbf {\bibinfo {volume} {356}},\
  \bibinfo {pages} {1} (\bibinfo {year} {2015})}\BibitemShut {NoStop}%
\bibitem [{\citenamefont {O'Neill}\ \emph {et~al.}(2016)\citenamefont
  {O'Neill}, \citenamefont {Selsil}, \citenamefont {McPhedran}, \citenamefont
  {Movchan}, \citenamefont {Movchan},\ and\ \citenamefont
  {Henderson~Moggach}}]{o2016active}%
  \BibitemOpen
  \bibfield  {author} {\bibinfo {author} {\bibfnamefont {J.}~\bibnamefont
  {O'Neill}}, \bibinfo {author} {\bibfnamefont {{\"O}.}~\bibnamefont {Selsil}},
  \bibinfo {author} {\bibfnamefont {R.}~\bibnamefont {McPhedran}}, \bibinfo
  {author} {\bibfnamefont {A.}~\bibnamefont {Movchan}}, \bibinfo {author}
  {\bibfnamefont {N.}~\bibnamefont {Movchan}},\ and\ \bibinfo {author}
  {\bibfnamefont {C.}~\bibnamefont {Henderson~Moggach}},\ }\href@noop {}
  {\bibfield  {journal} {\bibinfo  {journal} {The Quarterly Journal of
  Mechanics and Applied Mathematics}\ }\textbf {\bibinfo {volume} {69}},\
  \bibinfo {pages} {115} (\bibinfo {year} {2016})}\BibitemShut {NoStop}%
\bibitem [{\citenamefont {Elliott}\ \emph {et~al.}(2012)\citenamefont
  {Elliott}, \citenamefont {Cheer}, \citenamefont {Choi},\ and\ \citenamefont
  {Kim}}]{elliott2012robustness}%
  \BibitemOpen
  \bibfield  {author} {\bibinfo {author} {\bibfnamefont {S.~J.}\ \bibnamefont
  {Elliott}}, \bibinfo {author} {\bibfnamefont {J.}~\bibnamefont {Cheer}},
  \bibinfo {author} {\bibfnamefont {J.-W.}\ \bibnamefont {Choi}},\ and\
  \bibinfo {author} {\bibfnamefont {Y.}~\bibnamefont {Kim}},\ }\href@noop {}
  {\bibfield  {journal} {\bibinfo  {journal} {IEEE Transactions on Audio,
  Speech, and Language Processing}\ }\textbf {\bibinfo {volume} {20}},\
  \bibinfo {pages} {2123} (\bibinfo {year} {2012})}\BibitemShut {NoStop}%
\bibitem [{\citenamefont {Ahrens}(2012)}]{ahrens2012analytic}%
  \BibitemOpen
  \bibfield  {author} {\bibinfo {author} {\bibfnamefont {J.}~\bibnamefont
  {Ahrens}},\ }\href@noop {} {\emph {\bibinfo {title} {Analytic methods of
  sound field synthesis}}}\ (\bibinfo  {publisher} {Springer Science \&
  Business Media},\ \bibinfo {year} {2012})\BibitemShut {NoStop}%
\bibitem [{\citenamefont {Onofrei}\ and\ \citenamefont
  {Platt}(2018)}]{onofrei2018synthesis}%
  \BibitemOpen
  \bibfield  {author} {\bibinfo {author} {\bibfnamefont {D.}~\bibnamefont
  {Onofrei}}\ and\ \bibinfo {author} {\bibfnamefont {E.}~\bibnamefont
  {Platt}},\ }\href@noop {} {\bibfield  {journal} {\bibinfo  {journal} {Wave
  Motion}\ }\textbf {\bibinfo {volume} {77}},\ \bibinfo {pages} {12} (\bibinfo
  {year} {2018})}\BibitemShut {NoStop}%
\bibitem [{\citenamefont {Egarguin}\ \emph {et~al.}(2020)\citenamefont
  {Egarguin}, \citenamefont {Zeng}, \citenamefont {Onofrei},\ and\
  \citenamefont {Chen}}]{egarguin2020active}%
  \BibitemOpen
  \bibfield  {author} {\bibinfo {author} {\bibfnamefont {N.~J.~A.}\
  \bibnamefont {Egarguin}}, \bibinfo {author} {\bibfnamefont {S.}~\bibnamefont
  {Zeng}}, \bibinfo {author} {\bibfnamefont {D.}~\bibnamefont {Onofrei}},\ and\
  \bibinfo {author} {\bibfnamefont {J.}~\bibnamefont {Chen}},\ }\href@noop {}
  {\bibfield  {journal} {\bibinfo  {journal} {Wave Motion}\ }\textbf {\bibinfo
  {volume} {94}},\ \bibinfo {pages} {102523} (\bibinfo {year}
  {2020})}\BibitemShut {NoStop}%
\bibitem [{\citenamefont {Vasquez}\ \emph {et~al.}(2013)\citenamefont
  {Vasquez}, \citenamefont {Milton}, \citenamefont {Onofrei},\ and\
  \citenamefont {Seppecher}}]{vasquez2013transformation}%
  \BibitemOpen
  \bibfield  {author} {\bibinfo {author} {\bibfnamefont {F.~G.}\ \bibnamefont
  {Vasquez}}, \bibinfo {author} {\bibfnamefont {G.~W.}\ \bibnamefont {Milton}},
  \bibinfo {author} {\bibfnamefont {D.}~\bibnamefont {Onofrei}},\ and\ \bibinfo
  {author} {\bibfnamefont {P.}~\bibnamefont {Seppecher}},\ }in\ \href@noop {}
  {\emph {\bibinfo {booktitle} {Acoustic Metamaterials}}}\ (\bibinfo
  {publisher} {Springer},\ \bibinfo {year} {2013})\ pp.\ \bibinfo {pages}
  {289--318}\BibitemShut {NoStop}%
\bibitem [{\citenamefont {Martin}(2006)}]{martin2006multiple}%
  \BibitemOpen
  \bibfield  {author} {\bibinfo {author} {\bibfnamefont {P.~A.}\ \bibnamefont
  {Martin}},\ }\href@noop {} {\emph {\bibinfo {title} {Multiple scattering:
  interaction of time-harmonic waves with N obstacles}}},\ \bibinfo {number}
  {107}\ (\bibinfo  {publisher} {Cambridge University Press},\ \bibinfo {year}
  {2006})\BibitemShut {NoStop}%
\bibitem [{\citenamefont {Euclid}(2012)}]{euclid2012thirteen}%
  \BibitemOpen
  \bibfield  {author} {\bibinfo {author} {\bibfnamefont {E.}~\bibnamefont
  {Euclid}},\ }\href@noop {} {\emph {\bibinfo {title} {The Thirteen Books of
  the Elements, Vol. 1}}}\ (\bibinfo  {publisher} {Dover Publications},\
  \bibinfo {year} {2012})\BibitemShut {NoStop}%
\bibitem [{\citenamefont {Wigner}(2012)}]{wigner2012group}%
  \BibitemOpen
  \bibfield  {author} {\bibinfo {author} {\bibfnamefont {E.}~\bibnamefont
  {Wigner}},\ }\href@noop {} {\emph {\bibinfo {title} {Group theory: and its
  application to the quantum mechanics of atomic spectra}}},\ Vol.~\bibinfo
  {volume} {5}\ (\bibinfo  {publisher} {Elsevier},\ \bibinfo {year}
  {2012})\BibitemShut {NoStop}%
\bibitem [{\citenamefont {Varshalovich}\ \emph {et~al.}(1988)\citenamefont
  {Varshalovich}, \citenamefont {Moskalev},\ and\ \citenamefont
  {Khersonskii}}]{varshalovich1988description}%
  \BibitemOpen
  \bibfield  {author} {\bibinfo {author} {\bibfnamefont {D.}~\bibnamefont
  {Varshalovich}}, \bibinfo {author} {\bibfnamefont {A.}~\bibnamefont
  {Moskalev}},\ and\ \bibinfo {author} {\bibfnamefont {V.}~\bibnamefont
  {Khersonskii}},\ }in\ \href@noop {} {\emph {\bibinfo {booktitle} {Quantum
  Theory of Angular Momentum}}}\ (\bibinfo  {publisher} {World Scientific},\
  \bibinfo {year} {1988})\ pp.\ \bibinfo {pages} {21--23}\BibitemShut {NoStop}%
\bibitem [{\citenamefont {Colton}\ and\ \citenamefont
  {Kress}(2013)}]{colton2013integral}%
  \BibitemOpen
  \bibfield  {author} {\bibinfo {author} {\bibfnamefont {D.}~\bibnamefont
  {Colton}}\ and\ \bibinfo {author} {\bibfnamefont {R.}~\bibnamefont {Kress}},\
  }\href@noop {} {\emph {\bibinfo {title} {Integral equation methods in
  scattering theory}}}\ (\bibinfo  {publisher} {SIAM},\ \bibinfo {year}
  {2013})\BibitemShut {NoStop}%
\bibitem [{\citenamefont {Kinsler}\ \emph {et~al.}(1999)\citenamefont
  {Kinsler}, \citenamefont {Frey}, \citenamefont {Coppens},\ and\ \citenamefont
  {Sanders}}]{kinsler1999fundamentals}%
  \BibitemOpen
  \bibfield  {author} {\bibinfo {author} {\bibfnamefont {L.~E.}\ \bibnamefont
  {Kinsler}}, \bibinfo {author} {\bibfnamefont {A.~R.}\ \bibnamefont {Frey}},
  \bibinfo {author} {\bibfnamefont {A.~B.}\ \bibnamefont {Coppens}},\ and\
  \bibinfo {author} {\bibfnamefont {J.~V.}\ \bibnamefont {Sanders}},\
  }\href@noop {} {\emph {\bibinfo {title} {Fundamentals of acoustics}}}\
  (\bibinfo {year} {1999})\BibitemShut {NoStop}%
\bibitem [{\citenamefont {Weber}\ and\ \citenamefont
  {Arfken}(2005)}]{weber2005mathematical}%
  \BibitemOpen
  \bibfield  {author} {\bibinfo {author} {\bibfnamefont {H.-J.}\ \bibnamefont
  {Weber}}\ and\ \bibinfo {author} {\bibfnamefont {G.~B.}\ \bibnamefont
  {Arfken}},\ }\href@noop {} {\emph {\bibinfo {title} {Mathematical methods for
  physicists}}}\ (\bibinfo  {publisher} {Elsevier Academic},\ \bibinfo {year}
  {2005})\BibitemShut {NoStop}%
\bibitem [{\citenamefont {Gradshteyn}\ and\ \citenamefont
  {Ryzhik}(2014)}]{gradshteyn2014table}%
  \BibitemOpen
  \bibfield  {author} {\bibinfo {author} {\bibfnamefont {I.~S.}\ \bibnamefont
  {Gradshteyn}}\ and\ \bibinfo {author} {\bibfnamefont {I.~M.}\ \bibnamefont
  {Ryzhik}},\ }\href@noop {} {\emph {\bibinfo {title} {Table of integrals,
  series, and products}}}\ (\bibinfo  {publisher} {Academic press},\ \bibinfo
  {year} {2014})\BibitemShut {NoStop}%
\bibitem [{\citenamefont {Kennedy}\ \emph {et~al.}(2007)\citenamefont
  {Kennedy}, \citenamefont {Sadeghi}, \citenamefont {Abhayapala},\ and\
  \citenamefont {Jones}}]{kennedy2007intrinsic}%
  \BibitemOpen
  \bibfield  {author} {\bibinfo {author} {\bibfnamefont {R.~A.}\ \bibnamefont
  {Kennedy}}, \bibinfo {author} {\bibfnamefont {P.}~\bibnamefont {Sadeghi}},
  \bibinfo {author} {\bibfnamefont {T.~D.}\ \bibnamefont {Abhayapala}},\ and\
  \bibinfo {author} {\bibfnamefont {H.~M.}\ \bibnamefont {Jones}},\ }\href@noop
  {} {\bibfield  {journal} {\bibinfo  {journal} {IEEE Transactions on Signal
  processing}\ }\textbf {\bibinfo {volume} {55}},\ \bibinfo {pages} {2542}
  (\bibinfo {year} {2007})}\BibitemShut {NoStop}%
\bibitem [{\citenamefont {Abhayapala}\ \emph {et~al.}(2003)\citenamefont
  {Abhayapala}, \citenamefont {Pollock},\ and\ \citenamefont
  {Kennedy}}]{abhayapala2003characterization}%
  \BibitemOpen
  \bibfield  {author} {\bibinfo {author} {\bibfnamefont {T.~D.}\ \bibnamefont
  {Abhayapala}}, \bibinfo {author} {\bibfnamefont {T.~S.}\ \bibnamefont
  {Pollock}},\ and\ \bibinfo {author} {\bibfnamefont {R.~A.}\ \bibnamefont
  {Kennedy}},\ }in\ \href@noop {} {\emph {\bibinfo {booktitle} {2003 IEEE 58th
  Vehicular Technology Conference. VTC 2003-Fall (IEEE Cat. No. 03CH37484)}}},\
  Vol.~\bibinfo {volume} {1}\ (\bibinfo {organization} {IEEE},\ \bibinfo {year}
  {2003})\ pp.\ \bibinfo {pages} {123--127}\BibitemShut {NoStop}%
\bibitem [{\citenamefont {Abramowitz}\ \emph {et~al.}(1988)\citenamefont
  {Abramowitz}, \citenamefont {Stegun},\ and\ \citenamefont
  {Romer}}]{abramowitz1988handbook}%
  \BibitemOpen
  \bibfield  {author} {\bibinfo {author} {\bibfnamefont {M.}~\bibnamefont
  {Abramowitz}}, \bibinfo {author} {\bibfnamefont {I.~A.}\ \bibnamefont
  {Stegun}},\ and\ \bibinfo {author} {\bibfnamefont {R.~H.}\ \bibnamefont
  {Romer}},\ }\href@noop {} {\bibinfo {title} {Handbook of mathematical
  functions with formulas, graphs, and mathematical tables}} (\bibinfo {year}
  {1988})\BibitemShut {NoStop}%
\bibitem [{\citenamefont {Loh{\"o}fer}(1998)}]{lohofer1998inequalities}%
  \BibitemOpen
  \bibfield  {author} {\bibinfo {author} {\bibfnamefont {G.}~\bibnamefont
  {Loh{\"o}fer}},\ }\href@noop {} {\bibfield  {journal} {\bibinfo  {journal}
  {Journal of Approximation Theory}\ }\textbf {\bibinfo {volume} {95}},\
  \bibinfo {pages} {178} (\bibinfo {year} {1998})}\BibitemShut {NoStop}%
\bibitem [{\citenamefont {Srivastava}\ and\ \citenamefont
  {Choi}(2012)}]{srivastava2012zeta}%
  \BibitemOpen
  \bibfield  {author} {\bibinfo {author} {\bibfnamefont {H.~M.}\ \bibnamefont
  {Srivastava}}\ and\ \bibinfo {author} {\bibfnamefont {J.}~\bibnamefont
  {Choi}},\ }\href@noop {} {\emph {\bibinfo {title} {Zeta and q-Zeta functions
  and associated series and integrals}}}\ (\bibinfo  {publisher} {Elsevier},\
  \bibinfo {year} {2012})\BibitemShut {NoStop}%
\bibitem [{\citenamefont {Rodriguez}(1840)}]{rodriguez1840lois}%
  \BibitemOpen
  \bibfield  {author} {\bibinfo {author} {\bibfnamefont {O.}~\bibnamefont
  {Rodriguez}},\ }\href@noop {} {\bibfield  {journal} {\bibinfo  {journal} {J
  Mathematiques Pures Appliquees}\ }\textbf {\bibinfo {volume} {5}},\ \bibinfo
  {pages} {380} (\bibinfo {year} {1840})}\BibitemShut {NoStop}%
\bibitem [{\citenamefont {Man}(2016)}]{man2016wigner}%
  \BibitemOpen
  \bibfield  {author} {\bibinfo {author} {\bibfnamefont {P.~P.}\ \bibnamefont
  {Man}},\ }\href@noop {} {\bibfield  {journal} {\bibinfo  {journal} {Concepts
  in Magnetic Resonance Part A}\ }\textbf {\bibinfo {volume} {45}},\ \bibinfo
  {pages} {e21385} (\bibinfo {year} {2016})}\BibitemShut {NoStop}%
\bibitem [{\citenamefont {Waterman}(1969)}]{waterman1969new}%
  \BibitemOpen
  \bibfield  {author} {\bibinfo {author} {\bibfnamefont {P.}~\bibnamefont
  {Waterman}},\ }\href@noop {} {\bibfield  {journal} {\bibinfo  {journal} {The
  journal of the acoustical society of America}\ }\textbf {\bibinfo {volume}
  {45}},\ \bibinfo {pages} {1417} (\bibinfo {year} {1969})}\BibitemShut
  {NoStop}%
\bibitem [{\citenamefont {Shearer}\ \emph {et~al.}(2015)\citenamefont
  {Shearer}, \citenamefont {Parnell},\ and\ \citenamefont
  {Abrahams}}]{shearer2015antiplane}%
  \BibitemOpen
  \bibfield  {author} {\bibinfo {author} {\bibfnamefont {T.}~\bibnamefont
  {Shearer}}, \bibinfo {author} {\bibfnamefont {W.~J.}\ \bibnamefont
  {Parnell}},\ and\ \bibinfo {author} {\bibfnamefont {I.~D.}\ \bibnamefont
  {Abrahams}},\ }\href@noop {} {\bibfield  {journal} {\bibinfo  {journal}
  {Proceedings of the Royal Society A: Mathematical, Physical and Engineering
  Sciences}\ }\textbf {\bibinfo {volume} {471}},\ \bibinfo {pages} {20150450}
  (\bibinfo {year} {2015})}\BibitemShut {NoStop}%
\bibitem [{\citenamefont {Olver}(2000)}]{olver2000applications}%
  \BibitemOpen
  \bibfield  {author} {\bibinfo {author} {\bibfnamefont {P.~J.}\ \bibnamefont
  {Olver}},\ }\href@noop {} {\emph {\bibinfo {title} {Applications of Lie
  groups to differential equations}}},\ Vol.\ \bibinfo {volume} {107}\
  (\bibinfo  {publisher} {Springer Science \& Business Media},\ \bibinfo {year}
  {2000})\BibitemShut {NoStop}%
\end{thebibliography}%

\newpage

\appendix
\section{Supplementary Material}
\subsection{1. Integral form of the source coefficients $q_{\ell,nm}$}
For a field $u$ satisfying the homogeneous Helmholtz equation $(\nabla^2+k^2)u=0$ in a closed arbitrary domain $\widetilde{C}$ bounded by the surface $\partial \widetilde{C}$ without 
any radiating source inside, the Kirchhoff-Helmholtz integral equation \cite{colton2013integral} states that
\begin{align}
& \ \ \ \int_{\partial \widetilde{C}} \left[u(\mathbf{y}) \nabla_{\mathbf{y}} G(\mathbf{x},\mathbf{y})\cdot \mathbf{n} - G(\mathbf{x}, \mathbf{y}) \nabla_\mathbf{y} u(\mathbf{y}) \cdot \mathbf{n} \right] \ dS(\mathbf{y}) \nonumber \\
&=
\begin{cases}
u(\mathbf{x}), \ \ \  & \mathbf{x}\in \widetilde{C} \\
0, \ \ \ & \mathbf{x}\in \mathbb{R}^3 \setminus \widetilde{C}
\end{cases}
, \label{kh_int}
\end{align}
where $G(\mathbf{x},\mathbf{y})$ is the Green's function in three-dimensional free space \cite{kinsler1999fundamentals} defined by
\begin{align}
G(\mathbf{x},\mathbf{y}) = - \frac{e^{ik(\mathbf{x}-\mathbf{y})}}{4\pi|\mathbf{x}-\mathbf{y}|} = -\frac{ik}{\sqrt{4\pi}} V^0_0(\mathbf{x}-\mathbf{y}) \label{G}
\end{align}
and $\nabla_\mathbf{y} f(\mathbf{y})\cdot \mathbf{n}$ is the normal derivative of $f$ at $\mathbf{y} \in \partial \widetilde{C}$ with the outward-pointing unit normal $\mathbf{n}$. As illustrated in Fig.\ \ref{C_tilde}, we designate a region $C \subseteq \widetilde{C}$ such that for $\mathbf{x} \in C, u_i(\mathbf{x}) + u_d(\mathbf{x}) = 0$. Applying this condition and substituting \eqref{G} into the first case of \eqref{kh_int}, we can write the active field $u_d$ in terms of the incident field $u_i$ as
\begin{align}
u_d(\mathbf{x}) &= -u_i(\mathbf{x}) \nonumber \\
&= \frac{ik}{\sqrt{4\pi}} \int_{\partial \widetilde{C}} [u_i(\mathbf{y}) \nabla_{\mathbf{y}} V_0^0(\mathbf{x}-\mathbf{y}) \cdot \mathbf{n} \nonumber \\
& \ \ \ - V_0^0(\mathbf{x}-\mathbf{y}) \nabla_\mathbf{y} u_i(\mathbf{y}) \cdot \mathbf{n} ] \ dS(\mathbf{y}). \label{di}
\end{align}

The expression \eqref{di} implies from a physical perspective that $u_d$ is equivalent to the superposition of a continuous source distribution on the bounding face $\partial \widetilde{C}$ with amplitudes depending only on information related to $u_i$ on $\partial \widetilde{C}$. The next step is to replace this distribution with discrete point sources located at $\mathbf{x}_\ell \notin \widetilde{C}$ where $\ell=1,2,\cdots,L$. We first partition $\partial \widetilde{C}$ into the segments $\partial \widetilde{C}_\ell$ such that 
\begin{align}
u_d(\mathbf{x}) &= \frac{ik}{\sqrt{4\pi}} \sum_{\ell=1}^{L} \int_{\partial \widetilde{C}_\ell} [u_i(\mathbf{y}) \nabla_{\mathbf{y}} V_0^0(\mathbf{x}-\mathbf{y}) \cdot \mathbf{n} \nonumber \\
& \ \ \ - V_0^0(\mathbf{x}-\mathbf{y}) \nabla_\mathbf{y} u_i(\mathbf{y}) \cdot \mathbf{n} ] \ dS(\mathbf{y}). \label{dC_l}
\end{align} 
Given two fixed spatial points $\mathbf{r}_0, \mathbf{r}_0' \in \mathbb{R}^3$, a general position vector $\mathbf{r}-\mathbf{r}_0$ can be decomposed as $\mathbf{r}-\mathbf{r}_0 = \mathbf{r}-\mathbf{r}_0' + \mathbf{r}_0'-\mathbf{r}_0$. Take $\mathbf{r}_>$ and $\mathbf{r}_<$ such that
\begin{align}
	|\mathbf{r}_>| &= \max(|\mathbf{r}-\mathbf{r}_0'|, |\mathbf{r}_0'-\mathbf{r}_0|), \label{r_>} \\
	|\mathbf{r}_<| &= \min(|\mathbf{r}-\mathbf{r}_0'|, |\mathbf{r}_0'-\mathbf{r}_0|). \label{r_<}
\end{align}
The addition theorem for spherical wavefunctions \cite{martin2006multiple} states that
\begin{align}
&\ \ \ z_n(k|\mathbf{r}-\mathbf{r}_0|) Y_n^m(\widehat{\mathbf{r}-\mathbf{r}_0}) \nonumber \\
&= \sum_{\nu=0}^\infty \sum_{\mu=-\nu}^{\nu} 
\begin{dcases}
	\widehat{S}^{m\mu}_{n\nu}(\mathbf{r}_<) z_\nu(k|\mathbf{r}_>|) Y_\nu^\mu(\widehat{\mathbf{r}_>}), & \mathbf{r}_> = \mathbf{r}-\mathbf{r}_0' \\
	S^{m\mu}_{n\nu}(\mathbf{r}_>) j_\nu(k|\mathbf{r}_<|) Y_\nu^\mu(\widehat{\mathbf{r}_<}), & \mathbf{r}_< = \mathbf{r}-\mathbf{r}_0'	
\end{dcases}
, \label{add}
\end{align}
where $z_n(.)$ is the spherical Bessel or Hankel function of any kind and $\widehat{S}^{m\mu}_{n\nu}(\mathbf{r}_<), S^{m\mu}_{n\nu}(\mathbf{r}_>)$ are constants dependent on the position vectors $\mathbf{r}_<, \mathbf{r}_>$ such that
\begin{align}
	&\ \ \ 
	\begin{pmatrix}
		\widehat{S}^{m\mu}_{n\nu} (\mathbf{r}_<)\\
		S^{m\mu}_{n\nu} (\mathbf{r}_>)
	\end{pmatrix}
	 \nonumber \\
	&= 4\pi (-1)^m \sum_{q=0}^{(n+\nu-q_0)/2} i^{\nu-n+q_0+2q}   
	\begin{pmatrix}
		j_{q_0+2q} (k|\mathbf{r}_<|) \overline{Y_{q_0+2q}^{\mu-m} (\widehat{\mathbf{r}_<})} \\
		z_{q_0+2q} (k|\mathbf{r}_>|) \overline{Y_{q_0+2q}^{\mu-m} (\widehat{\mathbf{r}_>})}	
	\end{pmatrix} 
	 \nonumber \\
	& \ \ \ \times \mathcal{G} (n,m,\nu,-\mu,q_0+2q). \label{S}
\end{align}
In \eqref{S}, $q_0$ and $\mathcal{G}(n,m,\nu,-\mu,q)$ are defined as 
\begin{align}
&q_0 =
\begin{cases}
|n-\nu|, \ \ & |n-\nu| \geq |m-\mu| \\
|m-\mu|, \ \ & |n-\nu| < |m-\mu| \ \textnormal{and} \\
& n+\nu + |m-\mu| \ \textnormal{is even} \\
|m-\mu|+1, \ \ & |n-\nu| < |m-\mu| \ \textnormal{and} \\
& n+\nu + |m-\mu| \ \textnormal{is odd}
\end{cases}, \label{q_0} \\
& \ \ \ \mathcal{G}(n,m,\nu,-\mu,q) \nonumber \\
&= \frac{(-1)^{m-\mu}}{2} \mathcal{S} \sqrt{\frac{(n-m)!(\nu+\mu)!(q+m-\mu)!}{(n+m)!(\nu-\mu)!(q-m+\mu)!}} \nonumber \\
& \ \ \ \times \int_{-1}^1 P^m_n(t) P^{-\mu}_\nu(t) P^{-m+\mu}_q(t) \ dt, \label{cal_G}  \\
&\mathcal{S} = \sqrt{\frac{(2n+1)(2\nu+1)(2q+1)}{4\pi}}. \label{cal_S}
\end{align} 
Set $z_n(k|\mathbf{r}-\mathbf{r}_0|) = h_n^{(1)}(k|\mathbf{r}-\mathbf{r}_0|), \mathbf{r}=\mathbf{x}, \mathbf{r}_0=\mathbf{y} \ \text{and} \ \mathbf{r}_0'=\mathbf{x}_\ell$ as indicated in Fig.\ \ref{C_tilde}. Since $|\mathbf{x}-\mathbf{x}_\ell| > |\mathbf{y}-\mathbf{x}_\ell|$ for $\mathbf{y} \in \partial \widetilde{C}_\ell$, with the first case of \eqref{add} we can show that ,
\begin{align}
	V^0_0(\mathbf{x}-\mathbf{y}) = \sqrt{4\pi} \sum_{n=0}^\infty \sum_{m=-n}^n \overline{U^{m}_n ( \mathbf{y}-\mathbf{x}_\ell)} V^m_n (\mathbf{x}-\mathbf{x}_\ell). \label{V_0^0}
\end{align}
Now \eqref{V_0^0} has the same basis function $V^m_n (\mathbf{x} - \mathbf{x}_\ell)$ as the ansatz in \eqref{u_d}. Similarly, with the second case of \eqref{add} it can be proved that
\begin{align}
u_i(\mathbf{y}) =\sum_{n=0}^\infty \sum_{m=-n}^n Q_{nm} \sum_{\nu=0}^\infty \sum_{\mu=-\nu}^\nu \widehat{S}^{m\mu}_{n\nu} (\mathbf{x}_\ell) U^\mu_\nu (\mathbf{y}-\mathbf{x}_\ell) \label{u_i_add}
\end{align}
if we take $z_n(k|\mathbf{r}-\mathbf{r}_0|) = j_n(k|\mathbf{r}-\mathbf{r}_0|), \mathbf{r}=\mathbf{y}, \mathbf{r}_0=\mathbf{0} \ \text{and} \ \mathbf{r}_0'=\mathbf{x}_\ell$ as $|\mathbf{y}-\mathbf{x}_\ell|<|\mathbf{x}_\ell|$
. Note that the expressions \eqref{V_0^0} and \eqref{u_i_add} hold for all $\mathbf{x} \in C$ where $|\mathbf{x}-\mathbf{x}_\ell|> |\mathbf{y}-\mathbf{x}_\ell|$. 
The arbitrary nature of the domain $\widetilde{C}$ means that we can deform $\partial \widetilde{C}$ onto the bounding face $\partial C$ of the cloaked region $C$ and
take 
$|\mathbf{y} - \mathbf{x}_\ell| = a_\ell x_0 \ \text{for} \ \mathbf{y}\in \partial C_\ell$ with $a_\ell x_0$ a positive constant (possibly different for each $\ell$). As a result, $\partial C_\ell$ is a continuous set of points located on a sphere with centre $\mathbf{x}_\ell$ and radius $a_\ell x_0$. If we repeat this procedure for $\ell=1,2,\cdots,L$, then the cloaked region $C$ will be the closed domain interior to a region $D$ consisting of a union of $L$ imaginary spheres $S_\ell$ such that
\begin{align}
D=\bigcup_{\ell=1}^L S_\ell = \bigcup_{\ell=1}^L \{\mathbf{x}:|\mathbf{x}-\mathbf{x}_\ell| \leq a_\ell x_0 \}. \label{D}
\end{align}

Substituting \eqref{V_0^0} and \eqref{u_i_add} into \eqref{dC_l} 
and comparing with the ansatz for the active field $u_d$ \eqref{u_d}, we can show that the source coefficient $q_{\ell, nm}$ can be written in terms of the expansion coefficient of the incident field $Q_{nm}$ in the form
\begin{align}
q_{\ell, nm} = ik\sum_{t=0}^\infty \sum_{s=-t}^t Q_{ts} q_{\ell, nm, ts}', \label{sc} 
\end{align}
where
\begin{align}
&\ \ \ q_{\ell, nm, ts}' \nonumber \\
&= \sum_{\nu=0}^\infty \sum_{\mu=-\nu}^\nu \widehat{S}^{s\mu}_{t\nu} (\mathbf{x}_\ell) 
\int_{\partial C_\ell} \big[U^\mu_\nu (\mathbf{y}-\mathbf{x}_\ell) \nabla_\mathbf{y} \overline{U^m_n (\mathbf{y}-\mathbf{x}_\ell)} \cdot \mathbf{n}  \nonumber \\
& \ \ \ - \overline{U^m_n (\mathbf{y}-\mathbf{x}_\ell)} \nabla_\mathbf{y} U^\mu_\nu (\mathbf{y}-\mathbf{x}_\ell) \cdot \mathbf{n} \big] \ dS(\mathbf{y}). \label{sc_2}
\end{align}
Using the property that the outward unit normal $\mathbf{n}$ points towards $\mathbf{x}_\ell$ given the geometry of $\partial C_\ell$, we can replace the parameterizing vector $\mathbf{y}$ by $\mathbf{y}-\mathbf{x}_\ell$ and evaluate the derivative terms as
\begin{align}
\nabla_{\mathbf{y}} U^m_n(\mathbf{y}-\mathbf{x}_\ell) \cdot \mathbf{n} = -kj_n'(ka_\ell x_0) Y^m_n(\widehat{\mathbf{y}-\mathbf{x}_\ell}). \label{der}
\end{align}
The expressions \eqref{q_l} -- \eqref{I} can be obtained by substituting \eqref{der} into \eqref{sc} -- \eqref{sc_2}.

Note that the form of the source amplitude \eqref{q_l} -- \eqref{I} can be simplified further when the incident wave $u_i$ is a plane wave with 
$u_i(\mathbf{y}) = e^{i\mathbf{k} \cdot \mathbf{y}}$, which is evident if we write
\begin{align}
e^{i\mathbf{k} \cdot \mathbf{y}} &= e^{i\mathbf{k} \cdot \mathbf{x}_\ell} e^{i\mathbf{k} \cdot \left(\mathbf{y}-\mathbf{x}_\ell \right)} \nonumber \\
&=  e^{i\mathbf{k} \cdot \mathbf{x}_\ell} \sum_{\nu=0}^\infty \sum_{\mu=-\nu}^\nu Q_{\nu \mu} U^\mu_\nu (\mathbf{y}-\mathbf{x}_\ell) \label{e_iky-x_l}  
\end{align}
with $Q_{\nu \mu} = 4\pi i^\nu \overline{Y^\mu_\nu(\widehat{\mathbf{k}})}$. Comparing \eqref{e_iky-x_l} with \eqref{u_i_add} results in
\begin{align}
\sum_{t=0}^\infty \sum_{s=-t}^t Q_{ts} \widehat{S}^{s\mu}_{t\nu}(\mathbf{x}_\ell) = e^{i\mathbf{k} \cdot \mathbf{x}_\ell} Q_{\nu \mu}. \label{Q=Q}
\end{align}
The simplified expression \eqref{q_plane_wave} will then follow upon the substitution of \eqref{Q=Q} into \eqref{q_l} -- \eqref{I}.

As the total wave amplitude is driven to zero within $C$ with $q_{\ell, nm}$ taking the forms \eqref{sc} -- \eqref{sc_2}, it remains to be shown that the 
active sources produce negligibly small radiation to the far field. We first notice
by the ansatz \eqref{u_d} that the integral form of $u_d(\mathbf{x})$ in \eqref{di} should be defined for all $\mathbf{x}\in \mathbb{R}^3$ (although it is equal to $-u_i(\mathbf{x})$ only for $\mathbf{x}\in C$). The derivation of the source coefficient in the form \eqref{sc} involves expanding the Green's function in the integrand of \eqref{di} using the first case of the addition formula \eqref{add}, which holds only for $\mathbf{x} \in \mathbb{R}^3 \setminus D$ with $D$ defined in \eqref{D}.
From \eqref{kh_int}, the Kirchhoff-Helmholtz integral equation is identically zero for $\mathbf{x} \in \mathbb{R}^3 \setminus C$.
For multipolar sources of sufficiently high order, 
one can thus deduce that  $u_d(\mathbf{x})$, in the form \eqref{u_d} and \eqref{sc_2}, vanishes identically for $\mathbf{x}\in \mathbb{R}^3 \setminus \left(C \cup D \right)$, which is the region exterior to 
the imaginary spheres centred at the active sources.
The radiation-free condition $u_d(\mathbf{x}) \rightarrow 0$ as $|\mathbf{x}|\rightarrow\infty$ is thus satisfied strongly in that in fact $u_d(\mathbf{x})=0$ when $\mathbf{x}\in \mathbb{R}^3 \setminus \left(C \cup D \right)$.

\begin{center}
	\begin{tikzpicture}
	\draw[step=0.2, white] (-4,-3) grid (4,2.6);
	\draw [black,thick] plot [smooth cycle] coordinates {(3,0) (2,1.3) (0,2) (-1.5,1) (-3,0) (-2.7,-1.8) (0,-2.3) (2.5,-1)};
	\draw (0.8,1.4) node {\large $\widetilde{C}$};
	\begin{scope}
	\clip (1.6,-0.9) rectangle (3.2,1.3);
	\draw [green!70!black,ultra thick] plot [smooth cycle] coordinates {(3,0) (2,1.3) (0,2) (-1.5,1) (-3,0) (-2.7,-1.8) (0,-2.3) (2.5,-1)};
	\end{scope}
	\draw[green!50!black] (3.4,-0.7) node {\large $\partial \widetilde{C}_\ell$};
	\filldraw[fill=black!20!white, draw=black!20!white] plot [smooth cycle] coordinates {(1.2,0) (0,1.3) (-2.2,0.2) (-2.5,-1.5) (0,-2)}; 
	\draw (-1.5,-1.1) node {\large $C$};
	\draw (0.1,0.1) -- (-0.1,-0.1);
	\draw (-0.1,0.1) -- (0.1,-0.1);
	\draw (-0.3,-0.3) node {\large $\mathbf{O}$};
	\draw[thick] (0,1.8) -- (0,2.2);
	\draw[thick] (-2.8,-0.2) -- (-3.2,0);
	\draw[thick] (-2.6,-1.6) -- (-2.9,-1.9);
	\draw[thick] (0,-2.1) -- (0,-2.5);
	\draw[thick] (2.8,-1.1) -- (2.4,-0.8);
	\draw[thick] (1.75,1.15) -- (2.15,1.45);
	\draw [-{Stealth[scale=2.0]},thick,dashed,red] (3.6,0.4) -- (-0.4,-1.5);
	\draw[red] (1.6,-1.05) node {\large $\mathbf{x}-\mathbf{x}_\ell$};
	\draw [-{Stealth[scale=2.0]},thick,dashed,blue] (3.6,0.4) -- (2.2,1.15);
	\draw[blue] (3.4,1.1) node {\large $\mathbf{y}-\mathbf{x}_\ell$};
	\draw [-{Stealth[scale=2.0]},thick] (2.2,1.15) -- (-0.4,-1.5);
	\draw (0.9,0.5) node {\large $\mathbf{x}-\mathbf{y}$};
	\filldraw[fill=black] (1.3,2.2) circle [radius=0.08];
	\filldraw[fill=black] (-2.4,1.8) circle [radius=0.08];
	\filldraw[fill=black] (-3.5,-1) circle [radius=0.08];
	\filldraw[fill=black] (-2,-2.5) circle [radius=0.08];
	\filldraw[fill=black] (2,-2) circle [radius=0.08];
	\filldraw[fill=green!70!black,draw=green!70!black] (3.6,0.4) circle [radius=0.08];
	\end{tikzpicture}
	
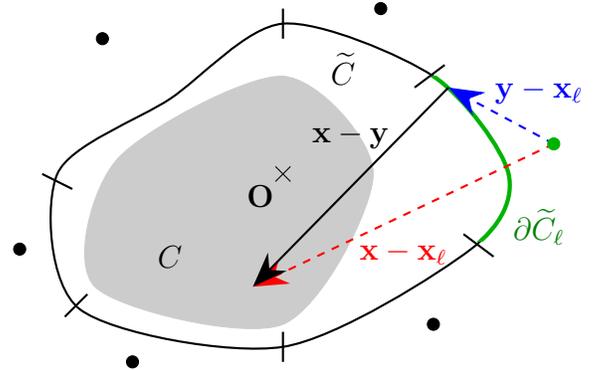
\captionof{figure}{A two-dimensional depiction of the domain $\widetilde{C}$ where the first case of \eqref{kh_int} holds and the cloaked region $C \subseteq \widetilde{C}$ (the grey area). The continuous source distribution on the bounding surface $\partial \widetilde{C}$ is replaced by the point active sources (the small dots) at $\mathbf{x}_\ell$ where $\ell=1,2,\cdots,L$. The face $\partial \widetilde{C}$ is divided into different sections and particularly for $\mathbf{y} \in \partial \widetilde{C}_\ell$ (the green curve), we decompose the position vector $\mathbf{x}-\mathbf{y}$ (the black arrow) into $\mathbf{x}-\mathbf{x}_\ell$ (the red dashed arrow) and $\mathbf{y}-\mathbf{x}_\ell$ (the blue dashed arrow) and apply the addition formula \eqref{add} to expand the wavefunctions as one centred at the source at $\mathbf{x}_\ell$ (the green dot).}
	\label{C_tilde}
\end{center}

\subsection{2. Geometry of the source distribution and the cloaked region $C$}
\begin{figure}[h!]  
	\centering
	\begin{tabular}{cc}
		\multicolumn{2}{c}{
			(a)\subf{\includegraphics[trim={1cm 1.5cm 0cm 1.2cm},clip,width=0.45\textwidth]{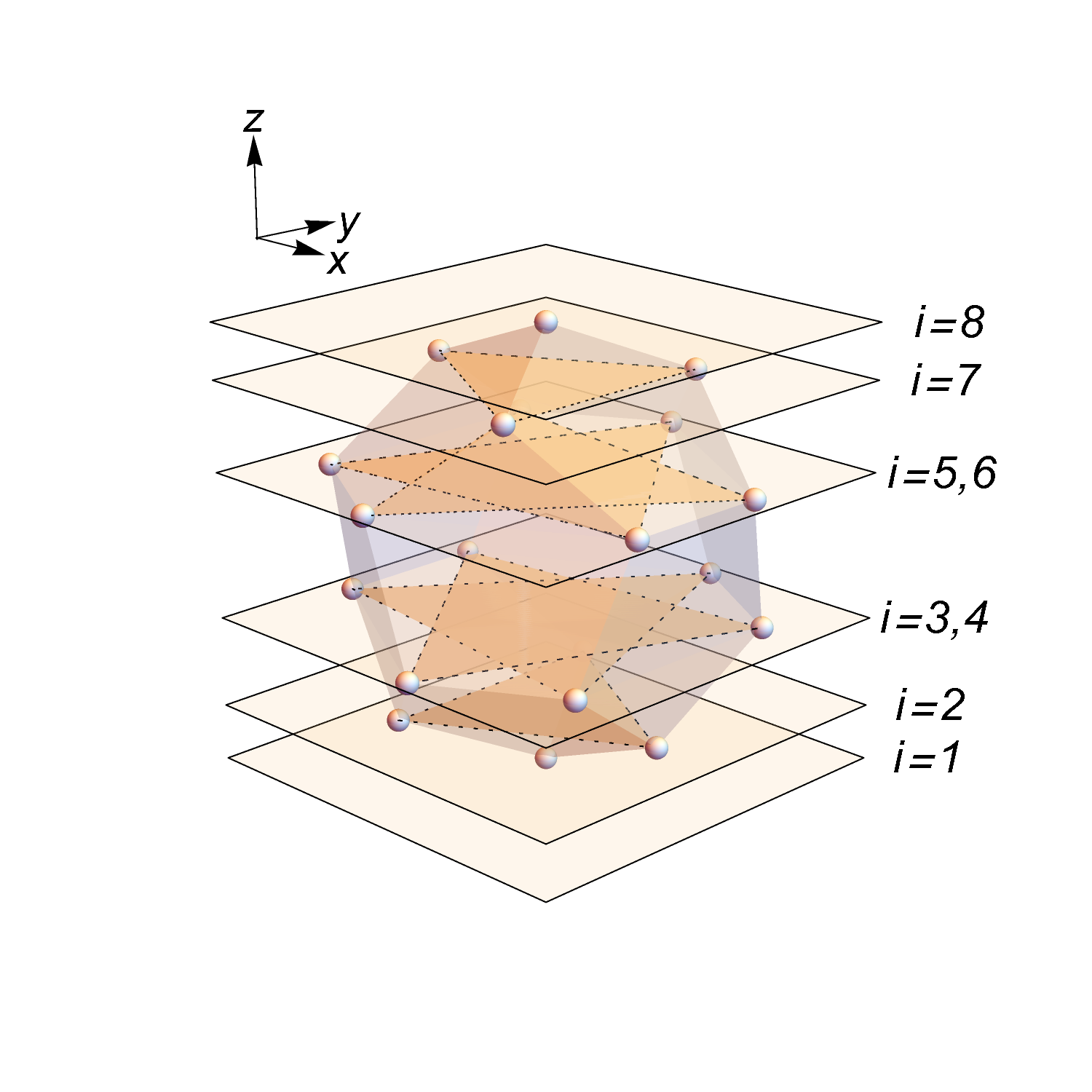}}} 
		\\
		(b)\subf{\includegraphics[clip,width=3cm]{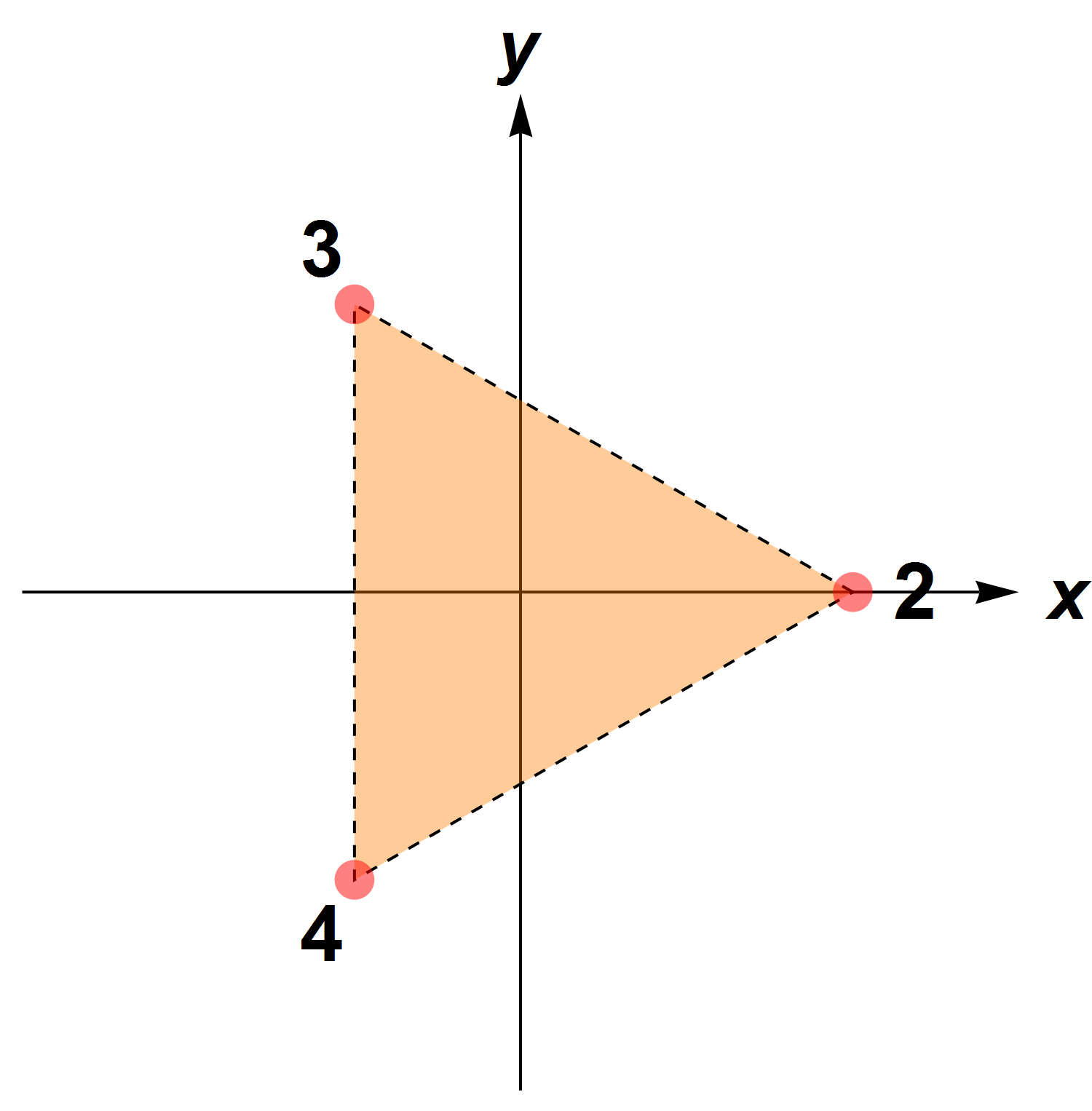}}
		&
		(c)\subf{\includegraphics[clip,width=3cm]{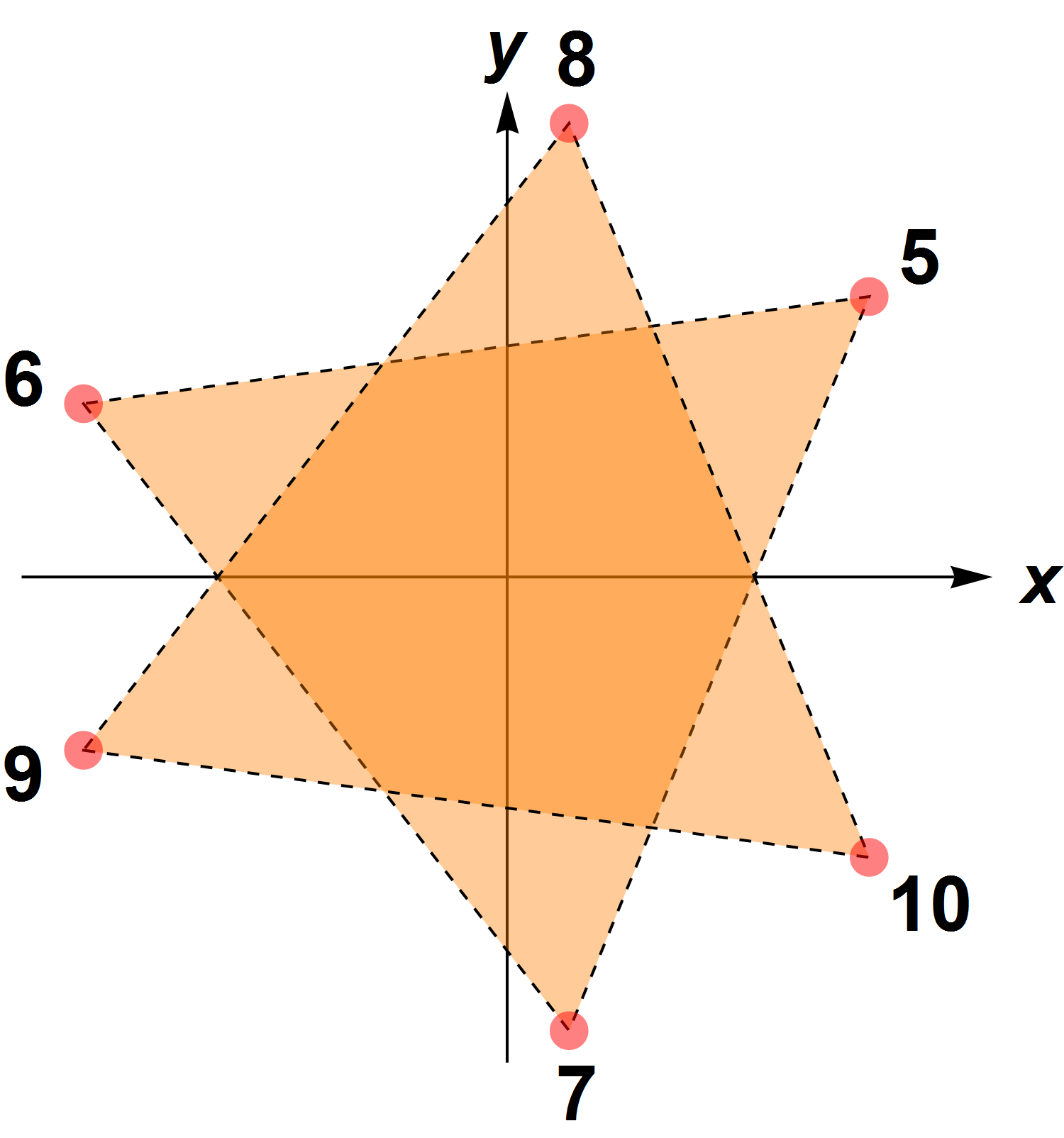}}
		\\
		(d)\subf{\includegraphics[clip,width=3cm]{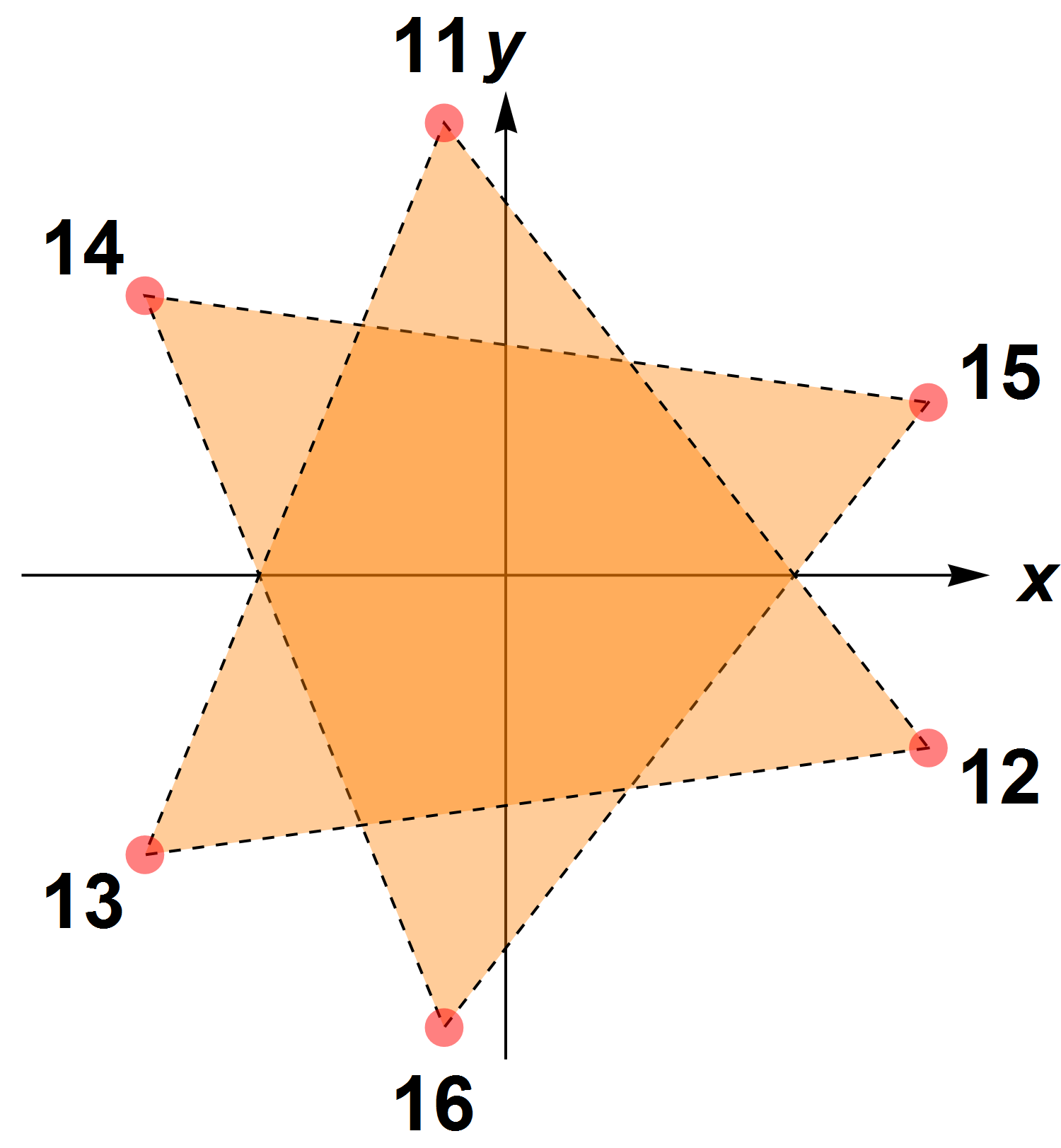}}
		&
		(e)\subf{\includegraphics[clip,width=3cm]{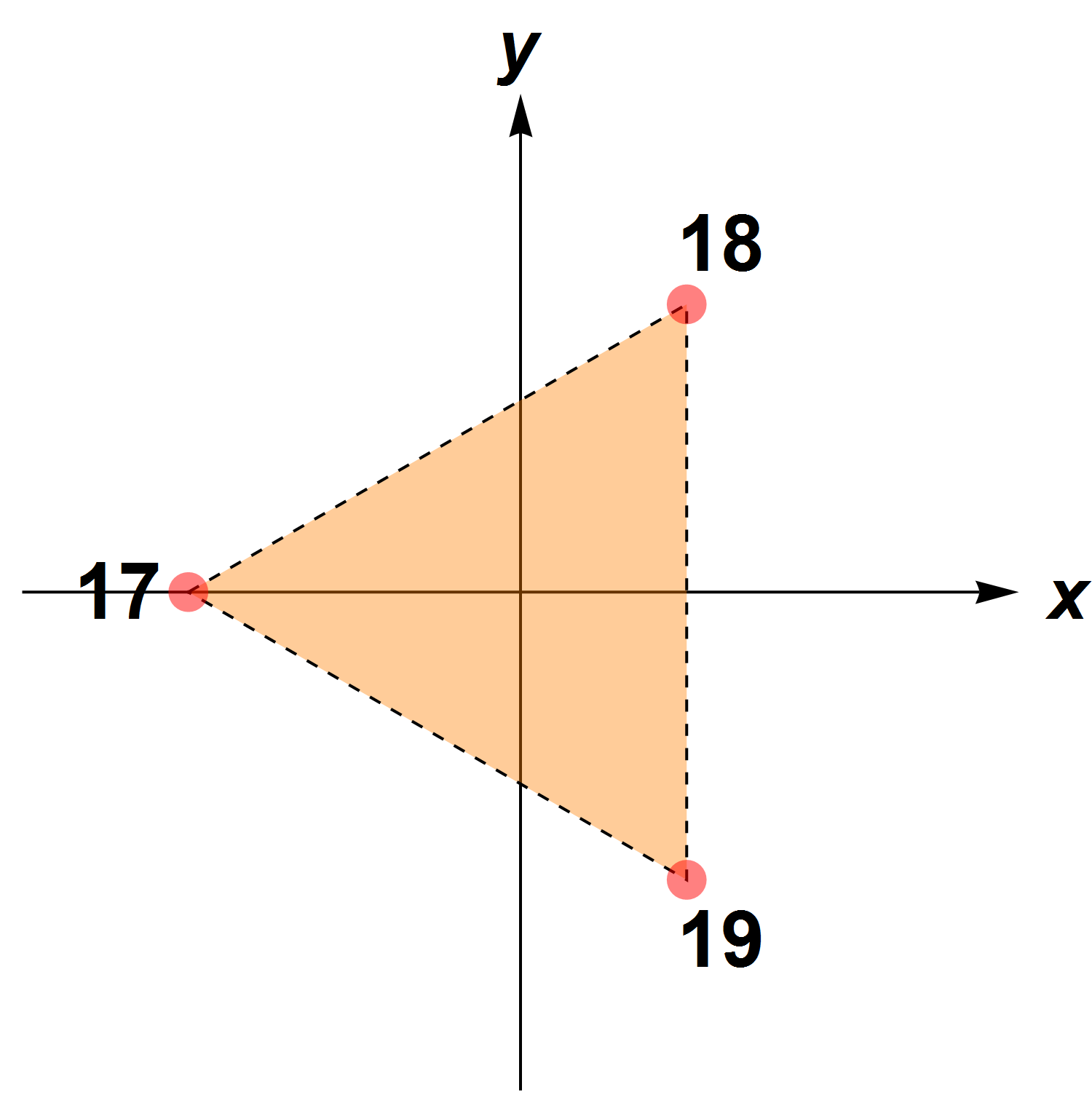}}
	\end{tabular}
	\caption{
		Illustration of the numbering of $L=20$ active sources (a regular dodecahedron) based on their geometry described by \eqref{x_ell} -- \eqref{3d_r_p_varphi}. In (a), the sources are located on $N_L=8$ horizontal planes, with the cross section shown for (b) $i=2$, (c) $i=3,4$, (d) $i=5,6$ and (e) $i=7$. } \label{L20}
\end{figure}

Given the characteristic indices $(p,q)$ of a Platonic solid, the length of its side $sx_0$ and the position vector of the bottommost source $\mathbf{x}_1=(0,0,-x_0)$, we aim to determine the locations of the remaining sources. Since the source distribution has a $q$-fold rotational symmetry about the $z$ axis by the definition of $q$, the sources must be distributed on a number of horizontal planes sliced through the polyhedron and the number of sources residing on each plane must be an integral multiple of $q$ (except when the source is located on the $z$ axis like the $\ell=1$ one). We can assign the indices $1\leq \ell \leq L$ to the active sources based on two rules: 1) the plane it is located on starting from the bottom; 
2) in either the clockwise or counterclockwise direction for sources on the same plane. Any sources can thus be located by specifying a set of two integers $(i,j)$. The first index $i=\lceil (\ell-1)/q \rceil +1$ represents the order of the plane and ranges from $i=1$ to $i=N_L= \lceil (L-1)/q \rceil +1$, where $\lceil .\rceil$ is the ceiling function. The second index $j=\text{mod}(\ell-2,q)$ denotes the order of the source on a particular plane and has the range $0\leq j \leq q-1$, where $\text{mod}(.,.)$ is the modulo function. By simple geometric arguments we can show for all cases of $L$ that 
\begin{align}
\mathbf{x}_\ell = \mathbf{x}_{i,j} = x_0(n_i r_i \cos \varphi_{i,j}, r_i \sin \varphi_{i,j}, z_i), \label{x_ell}
\end{align}
where
\begin{align}
n_i &=
\begin{dcases}
1, \qquad \qquad  &i \leq \max\left(\frac{N_L+1}{2},2 \right) \\
-1, \qquad \qquad  &i>\max\left(\frac{N_L+1}{2},2 \right)
\end{dcases}
, \label{n} \\
z_i &= 
\begin{dcases}
-1, \   &i=1 \\
\frac{s^2-2}{2}, \   &i=2 \\
s^2 \left [1-\cos (3\pi/5) \right ]-1, \   &2<i< \frac{N_L+1}{2}, \\
& L=20 \\
-z_{N_L+1-i}, \  & i> \frac{N_L+1}{2},L\neq 4
\end{dcases}
, \label{3d_r_p_z} \\
r_i &= \sqrt{1-z_i^2}, \label{3d_r_p_r} \\ 
\varphi_{i,j} &= 
\varphi_{i,0} + \frac{2 j\pi}{q}, \label{3d_r_p_phi} \\
\varphi_{i,0} &=
\begin{dcases}
0,    &i = 2 \\
\arcsin \left\{\frac{\sqrt{2s^2 [1-\cos (3\pi/5)] }}{2r_3} \right\},   &i=3, L=20 \\
\varphi_{3,0} + \arccos \left( 1-\frac{s^2}{2r_4^2} \right),   &i =4, L=20 \\
\varphi_{N_L+1-i,0},  &i> \frac{N_L+1}{2}, \\
&L \neq 4
\end{dcases}
. \label{3d_r_p_varphi} 
\end{align}
Note that $j$ is not needed when $\ell=1 \ \text{or} \ L$ provided that $L \neq 4$ as these two sources are positioned on the $z$ axis. For convenience, we set $j=0$ With $\mathbf{x}_\ell$ in the form \eqref{x_ell} -- \eqref{3d_r_p_varphi}, the numbering of sources on a plane goes counterclockwise starting from Quadrant I for $i\leq \max\left((N_L+1)/2,2 \right)$ and clockwise starting from Quadrant II for $i>\max\left((N_L+1)/2,2 \right)$. 
Given the two indices $i,j$, we can retrieve $\ell$ as
\begin{align}
\ell=
\begin{cases}
1, & i=1 \\
(i-2)q+j+2, & 1<i \leq N_L \\
L, & i=N_L, L \neq 4
\end{cases}
. \label{l_i_j}
\end{align}
In Fig. \ref{L20} we illustrate the labelling of $L=20$ active sources located at the vertices of an imaginary regular dodecahedron with $p=5, q=3$ based on the principles discussed above. Fig. \ref{L20}(a) shows that when the position vector of the source $\ell=1$ 
is taken as $\mathbf{x}_1=-x_0\widehat{\mathbf{e}}_z$, the sources are distributed on a total of $N_L=8$ horizontal planes, forming an equilateral triangle on each cross section except for the sources $\ell=1,20$ which reside at the  bottommost and topmost position respectively. (Note that $z_3=z_4$ and $z_5=z_6$ by \eqref{3d_r_p_z}.) In Fig. \ref{L20}(b) -- (e) each cross section for $i=2,3,4,5,6,7$ and the corresponding order $\ell$ for each active source are shown. 

As we have demonstrated in Part 1, the cloaked region $C$ is the domain enclosed by the imaginary spheres $S_\ell$, each centred at $\mathbf{x}_\ell$ with radius $a_\ell x_0$. We will keep $a_\ell$ constant at $a$ for all $\ell$ so that all spherical faces $\partial C_\ell$ have an identical geometric shape. For our formulation of the source amplitudes to hold, $a$ must be within the range
\begin{align}
\frac{s}{2\sin(\pi/p)} \leq a < 1. \label{range}
\end{align}
Note that we have $a<1$ as otherwise all the space interior to the source distribution would be occupied by the spheres and $C$ would not exist.
As for the lower bound of \eqref{range}, we need to ensure that adjacent spheres intersect with each other so that $C$ remains completely bounded by the spheres. 
The possible intersecting point furthest away from the origin is the centroid of each flat polygonal face of the Platonic solid, which suggests that $a$ must be at least as large as $s/[2\sin(\pi/p)]$ since this is the distance between the centroid and each vertex of a $p$-sided regular polygon with side $s$.

\begin{figure}[H]
	\centering 
	\includegraphics[trim={1.5cm 2.5cm 1.5cm 1cm},clip,width=0.25\textwidth]{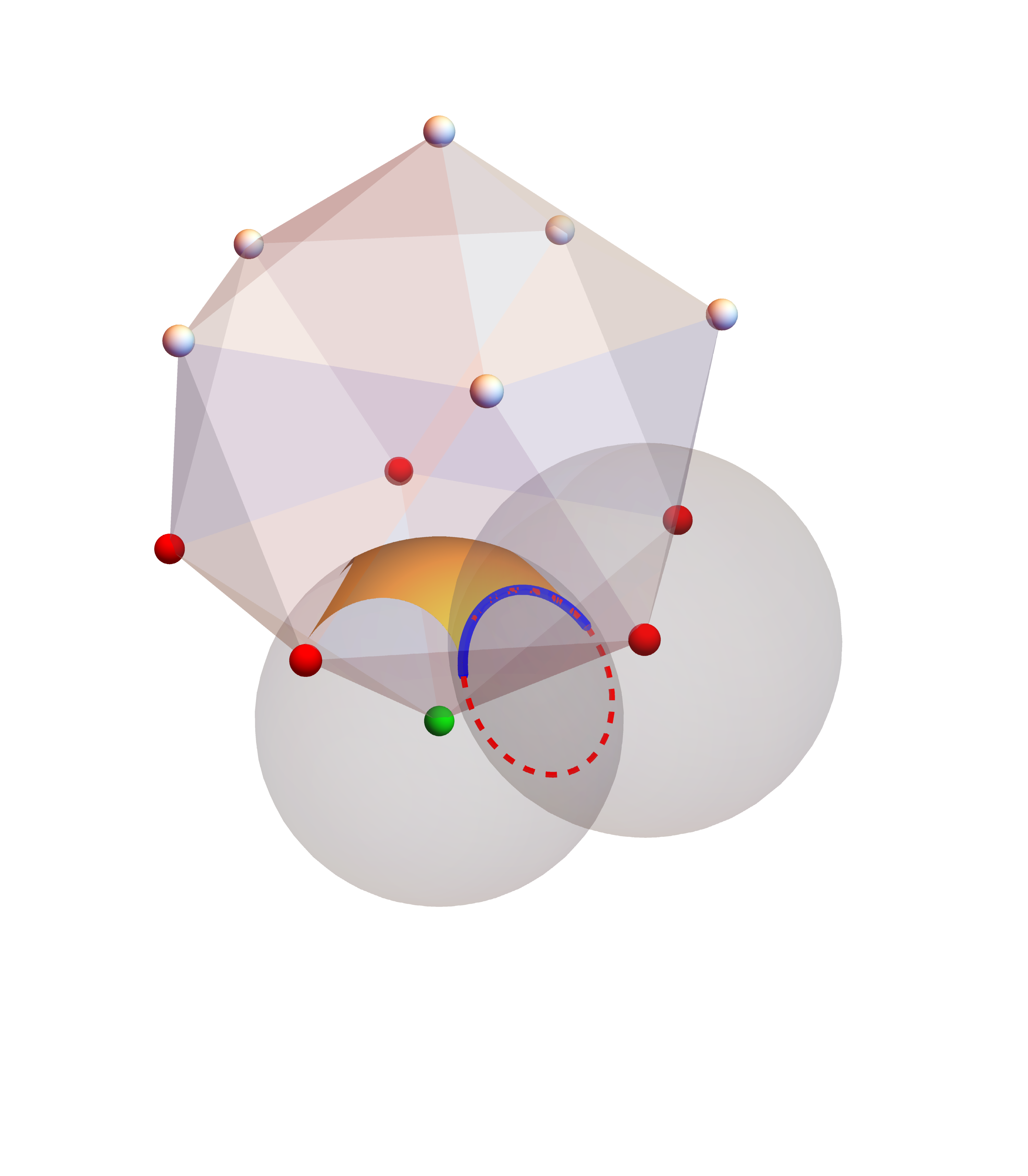}
	\caption{Illustration of the geometry of the face $\partial C_1$, which is part of the surface of the sphere $S_1$ centred at the source $\ell=1$ (the green dot), when $a$ takes the minimum value in \eqref{range}. Its boundary is formed by the intersection of $S_1$ with the spheres $S_\ell$, each centred at the source $\ell =2,\cdots,  q+1$ (the red dots). The blue solid part of the intersecting circle is represented by the plus form of \eqref{cos_theta} while the red dashed part is by the minus one. The two circular arcs meet with each other at the centroids of two adjacent polygonal faces.}
	\label{Cos}
\end{figure}

When we evaluate the coefficient $q_{1,nm}$ of the source $\ell=1$, we need to first perform the surface integral $\mathcal{I}_{n\nu}^{m\mu}$ over the spherical face $\partial C_1$. To obtain a parametric form of $\partial C_1$ with respect to the point $\mathbf{x}_1$, we see from Fig.\ \ref{Cos} that $\partial C_1$ is bounded by a set of circular arcs, each being part of the intersecting circle (the blue solid and red dashed curve) between $S_1$ and 
$S_\ell$ for $\ell = 2, \cdots, q+1$, where the spheres $S_\ell$ centre at the sources on the plane $i=2$. The equations for $S_1$ and $S_\ell$ are thus
\begin{align}
\left|(x,y,z) - x_0(0,0,-1) \right| &= ax_0,  \label{S_1} \\
\left |(x,y,z)-x_0(r_2 \cos\varphi_\ell, r_2 \sin\varphi_\ell, z_2) \right| &=ax_0, \label{S_2}
\end{align}
respectively, with $z_2,r_2$ given by \eqref{3d_r_p_z}, \eqref{3d_r_p_r} and  $\varphi_\ell = \varphi_{2,j}= 2\text{mod}(\ell-2,q)\pi/q = 2(\ell-2)\pi/q$ by \eqref{3d_r_p_phi}, \eqref{3d_r_p_varphi} and the fact that $\ell-2\leq q-1<q$.  
Eliminating the squared terms from \eqref{S_1} and \eqref{S_2}, we have
\begin{align}
(r_{2}\cos\varphi_{\ell}) x + (r_{2}\sin\varphi_{\ell}) y + (1+z_{2})z = 0, \label{3d_sc_p_plane}
\end{align}
which gives the plane that the intersecting circle between $S_1$ and $S_\ell$ lies on. If we introduce the parameterization $(x,y,z) = x_0 (a\sin\theta \cos \varphi, a\sin\theta \sin\varphi, -1+a\cos \theta)$ and apply the identity $\sin^2\theta + \cos^2\theta =1$, then \eqref{3d_sc_p_plane} becomes
\begin{gather}
a^2 \big[1+h \cos^2(\varphi-\varphi_{\ell}) \big]\cos^2\theta - 2a \cos\theta \nonumber \\
+ 1-a^2h \cos^2(\varphi-\varphi_{\ell}) = 0, \label{3d_sc_p_Z_2}
\end{gather}
where $h=(1-z_2)/(1+z_2)$. Solving for $\cos\theta$ in terms of $\varphi$ yields
\begin{align}
\cos \theta &= \Big \{1 \pm [a^2h^2\cos^4 (\varphi-\varphi_\ell) - h(1-a^2) \nonumber \\
& \ \ \ \times \cos^2 (\varphi-\varphi_\ell) ]^{1/2} \Big \}  / \left\{ a \left[1+h \cos^2(\varphi-\varphi_\ell) \right] \right \}. \label{cos_theta}
\end{align}
Note that \eqref{cos_theta} represents the entire circumference of the intersecting circle between $S_1$ and $S_\ell$. 
The plus and minus form equal each other when the square root term vanishes, which gives
\begin{align}
	\cos(\varphi-\varphi_\ell) &= \sqrt{\frac{1-a^2}{a^2h}}, \label{phi_+_-}\\
	\cos\theta &= a. \label{theta_+_-}
\end{align}
When $a$ takes its minimum value as stated in \eqref{range}, we can show that indeed $\varphi = \varphi_\ell \pm \pi/q$ in \eqref{phi_+_-} by substituting the values of $h$ and lower bound of $a$ corresponding to the five different source distributions. Considering the regularity of the Platonic solids, we can deduce that the values of $\varphi$ and $\theta$ defined by \eqref{phi_+_-} and \eqref{theta_+_-} give the centroids of the two adjacent polygonal faces that meet at both source 1 and $\ell$. The intersecting circle between $S_1$ and $S_\ell$ thus passes through these two centroids and they are where the two cases of \eqref{cos_theta} coincide with one another.
As Fig.\ \ref{Cos} shows for the minimum case of $a$, only the circular arc interior to the Platonic solid (the blue solid curve) forms part of the boundary of $\partial C_1$ and it is defined by smaller values of $\theta$ (as viewed from $\mathbf{x}_1$) than the exterior one (the red dashed curve) is. Since $\cos \theta$ is monotonically decreasing for $\theta \in [0,\pi]$, only the plus form of \eqref{cos_theta} represents the boundary of $\partial C_1$.
For values of $a$ greater than the lower bound of \eqref{range}, the intersecting circle no longer passes through the centroids. 
It can be proved that the two points where the plus and minus part of the circumference meet each other occur at values of $\varphi$ such that $|\varphi-\varphi_\ell|>\pi/q$ 
since $\cos(\varphi-\varphi_\ell)$ can be shown to be monotonically decreasing with $a$.
Nevertheless, 
taking into account the $q$-fold rotational symmetry of $\partial C_1$ about the $z$ axis, we can show that each of the $q$ circular arcs bounding the face $\partial C_1$ spans only the range $|\varphi-\varphi_\ell|\leq \pi/q$ and the plus form of \eqref{cos_theta} suffices to fully describe the boundary of $\partial C_1$. 

The volume $\mathcal{V}=|C|$ of the cloaked region $C$ can be found using the divergence theorem \cite{weber2005mathematical}, which states that for any vector field $\mathbf{f}$,
\begin{align}
	\int_C \nabla \cdot \mathbf{f}(\mathbf{x}) \ dV(\mathbf{x}) = \int_{\partial C} \mathbf{f}(\mathbf{x}) \cdot \mathbf{n} \ dS(\mathbf{x}), \label{div}
\end{align}
with $\mathbf{n}$ the outward unit normal emanated from the area element $dS(\mathbf{x})$. Taking $\mathbf{f}(\mathbf{x})=\mathbf{x}=(x,y,z)$, we have
\begin{align}
	\mathcal{V} = \frac{1}{3} \int_{\partial C} \mathbf{x} \cdot \mathbf{n} \ dS(\mathbf{x}) = \frac{L}{3} \int_{\partial C_1} \mathbf{x} \cdot \mathbf{n} \ dS(\mathbf{x}) \label{V}
\end{align}
as each $\partial C_\ell$ is identical to $\partial C_1$ in terms of its geometric shape.
If we shift the reference point from the origin to $\mathbf{x}_1$ by replacing $\mathbf{x}-\mathbf{x}_1$ with $\mathbf{x}$, then \eqref{V} becomes
\begin{align}
	\mathcal{V} &= \frac{L}{3} \int_{\partial C_1} (\mathbf{x}+\mathbf{x}_1 ) \cdot \mathbf{n} \ dS(\mathbf{x}) \nonumber \\
	&= \frac{Lq}{3}  \int_{-\pi/q}^{\pi/q} \int_0^{\arccos [g_\ell(\varphi)]} (x,y,z-x_0) \cdot \left[-\frac{(x,y,z)}{a_0}\right] \nonumber \\
	& \ \ \ \times (a_0)^2 \sin\theta \ d\theta d\varphi \nonumber \\
	&= \frac{1}{3} Lq a_0^2 x_0 \int_{-\pi/q}^{\pi/q} \int_0^{\arccos [g_\ell(\varphi)]} (\cos\theta - a) \sin\theta \ d\theta d\varphi,
	\label{V_2}
\end{align}
where $a_0=ax_0$ and $g_\ell(\varphi)$ is the plus form of \eqref{cos_theta}. Note that \eqref{V_2} is applicable to the whole range of $a$ in \eqref{range}.

\subsection{3. Truncation error of the source coefficients $q_{\ell,nm}$}

In \eqref{q_plane_wave}, we have expressed the source amplitude $q_{\ell,nm}$ as an infinite series in terms of the incident wave coefficient $Q_{\nu \mu}$. To numerically implement the cloaking approach, only a finite number of terms from the series can be retained which leads to truncation error. We will show that this error can be bounded by choosing an appropriate value of truncation order.

Recall from \eqref{dC_l} in Part 1 the integral form of the active field $u_d$ with 
\begin{align}
	u_d(\mathbf{x}) &= \frac{ik}{\sqrt{4\pi}} \sum_{\ell=1}^{L} \int_{\partial \widetilde{C}_\ell} [u_i(\mathbf{y}) \nabla_{\mathbf{y}} V_0^0(\mathbf{x}-\mathbf{y}) \cdot \mathbf{n} \nonumber \\
	& \ \ \ - V_0^0(\mathbf{x}-\mathbf{y}) \nabla_\mathbf{y} u_i(\mathbf{y}) \cdot \mathbf{n} ] \ dS(\mathbf{y}). \label{dC_l_te}
\end{align} 
If we substitute the expansion formula of $V_0^0(\mathbf{x}-\mathbf{y})$ in \eqref{V_0^0} into \eqref{dC_l_te} and compare the expression with the ansatz of $u_d$ in \eqref{u_d}, then we can write the source amplitude $q_{\ell,nm}$ as
\begin{align}
	q_{\ell,nm} &= ik \int_{\partial C_\ell} \Big[u_i(\mathbf{y})\nabla_\mathbf{y} \overline{U^{m}_n (\mathbf{y}-\mathbf{x}_\ell) } \cdot \mathbf{n} \nonumber \\
	& \ \ \ -\overline{U^{m}_n (\mathbf{y}-\mathbf{x}_\ell) } \nabla_\mathbf{y} u_i(\mathbf{y})\cdot \mathbf{n} \Big] \ dS(\mathbf{y}). \label{q_te}
\end{align} 
For a plane wave incidence, we have by \eqref{e_iky-x_l} that 
\begin{align}
	u_i(\mathbf{y}) = e^{i\mathbf{k}\cdot \mathbf{y}}  = e^{i\mathbf{k} \cdot \mathbf{x}_\ell} \sum_{\nu=0}^\infty \sum_{\mu=-\nu}^\nu Q_{\nu \mu} U^\mu_\nu (\mathbf{y}-\mathbf{x}_\ell), \label{u_i_y}
\end{align}
where $Q_{\nu \mu} = 4\pi i^\nu \overline{Y^\mu_\nu(\widehat{\mathbf{k}})}$. From \eqref{q_te} and \eqref{u_i_y} we observe that the truncation error of $q_{\ell,nm}$ originates from the truncation of the expansion of $u_i$. If we discard the terms with $\nu > V$ in \eqref{u_i_y} and denote the truncated source amplitude and incident plane wave as $\widetilde{q}_{\ell,nm}$ and $\widetilde{u}_i$ respectively, then the truncation error is
\begin{align}
	&\ \ \  |q_{\ell,nm} - \widetilde{q}_{\ell,nm}| \nonumber \\
	&\leq k \int_{\partial C_\ell} \Big\{ |u_i(\mathbf{y}) - \widetilde{u}_i(\mathbf{y})| |\nabla_\mathbf{y} \overline{U^{m}_n (\mathbf{y}-\mathbf{x}_\ell) }\cdot \mathbf{n}| \nonumber \\
	& \ \ \ + |\overline{U^{m}_n (\mathbf{y}-\mathbf{x}_\ell) }| |\nabla_\mathbf{y} [u_i(\mathbf{y})- \widetilde{u}_i(\mathbf{y})] \cdot \mathbf{n}| \Big\} \ dS(\mathbf{y}). \label{|q-q|}
\end{align} 
It has already been shown in \cite{kennedy2007intrinsic} and \cite{abhayapala2003characterization} that
\begin{align}
	|u_i(\mathbf{y}) - \widetilde{u}_i(\mathbf{y})| &= \left| \sum_{\nu=0}^\infty \sum_{\mu=-\nu}^\nu Q_{\nu \mu} U^\mu_\nu (\mathbf{y}-\mathbf{x}_\ell) \right| \nonumber \\
	&\leq \sum_{\nu>V}^\infty (2\nu+1) |j_\nu(k|\mathbf{y}-\mathbf{x}_\ell|)| \label{u-u}
\end{align}
and when $V=\lceil eka_0/2 \rceil+\Delta$ (as $|\mathbf{y}-\mathbf{x}_\ell|=a_0$ for $\mathbf{y}\in \partial C_\ell$) with $\Delta$ a nonnegative integer, \eqref{u-u} is bounded as 
\begin{align}
	|u_i(\mathbf{y}) - \widetilde{u}_i(\mathbf{y})| \leq \frac{\sqrt{e/2}}{e-1} e^{-\Delta} \approx 0.6785 e^{-\Delta}. \label{|u-u|}
\end{align}
The result \eqref{|u-u|} states that the truncation error in $u_i$ decays exponentially for truncation order greater than the critical value $\lceil eka_0/2 \rceil$.
As for the truncation error of $\nabla u_i$, we can rewrite it in a form similar to \eqref{u-u}:
\begin{align}
	& \ \ \ |\nabla_\mathbf{y} [u_i(\mathbf{y})- \widetilde{u}_i(\mathbf{y})] \cdot \mathbf{n}| \nonumber \\
	&\leq k\sum_{\nu>V}^\infty (2\nu+1) |j_\nu'(ka_0)| \nonumber \\
	&= k\sum_{\nu>V}^\infty (2\nu+1) \left|\frac{\nu}{2\nu+1} j_{\nu-1} (ka_0) - \frac{\nu+1}{2\nu+1} j_{\nu+1} (ka_0)  \right| \nonumber \\
	&\leq  k\sum_{\nu>V}^\infty \left[\nu \left| j_{\nu-1} (ka_0)\right| + (\nu+1) \left| j_{\nu+1} (ka_0)  \right| \right] \nonumber \\
	&\leq  k\sum_{\nu>V}^\infty \left[(2\nu-1) \left| j_{\nu-1} (ka_0)\right| + (2\nu+3) \left| j_{\nu+1} (ka_0)  \right| \right] \nonumber \\
	&\leq k \left[\sum_{\nu>V-1}^\infty (2\nu+1) |j_\nu(ka_0)| + \sum_{\nu>V+1}^\infty (2\nu+1) |j_\nu(ka_0)| \right] \label{nabla_u-u}
\end{align}
where $j_\nu'(ka_0)$ is expressed as a linear combination of 
$j_{\nu-1}(ka_0)$ and $j_{\nu+1}(ka_0)$ in the third line \cite{abramowitz1988handbook}.
Taking $V=\lceil eka_0/2 \rceil +\Delta+1$ and referring to \eqref{|u-u|}, we 
have
\begin{align}
	|\nabla_\mathbf{y} [u_i(\mathbf{y})- \widetilde{u}_i(\mathbf{y})] \cdot \mathbf{n}| &\leq k\frac{\sqrt{e/2}}{e-1} (1+e^{-2}) e^{-\Delta} \nonumber \\
	&\approx 0.7703k e^{-\Delta}. \label{|nabla_u-u|}
\end{align}
Note that at this value of $V$, \eqref{|u-u|} becomes
\begin{align}
	|u_i(\mathbf{y}) - \widetilde{u}_i(\mathbf{y})| \leq \frac{\sqrt{e/2}}{e^2-e} e^{-\Delta} \approx 0.2496 e^{-\Delta}. \label{|u-u|_+1}
\end{align}
By \cite{lohofer1998inequalities}, the term $\overline{U^m_n(\mathbf{y}-\mathbf{x}_\ell)}$ can be bounded as
\begin{align}
	|\overline{U^m_n(\mathbf{y}-\mathbf{x}_\ell)}| &= \left| A^m_n j_n(ka_0) P^m_n(\cos \theta) e^{-im\varphi} \right| \nonumber \\
	&\leq \sqrt{\frac{2n+1}{4\pi \rho_m}} |j_n(ka_0)| , \label{|U|}
\end{align}
with
\begin{align}
	\rho_m =
	\begin{cases}
		1, \ \ \ &m=0\\
		2, \ \ \ &m\neq 0
	\end{cases}
	. \label{rho}
\end{align}
Due to the rotational symmetry of the source configuration, the surface area of each piece of spherical face $dC_\ell$ is same as that of $dC_1$. The equation \eqref{theta_+_-} defines the polar angle of the two end points of the circular arc $\cos \theta = g_\ell(\varphi)$ with respect to $\mathbf{x}_1$ for $\ell=2,\cdots,q+1$. For $-\pi/q \leq \varphi \leq \pi/q$, we thus have $g_\ell(\varphi) \geq a$. Using these arguments, we can show that
\begin{align}
	\int_{\partial C_\ell} \ dS(\mathbf{y}) &= \int_{\partial C_1} \ dS(\mathbf{y}) \nonumber \\
	&= q\int_{-\pi/q}^{\pi/q} \int_0^{\arccos[g_\ell (\varphi)]} a_0^2 \sin \theta \ d\theta d\varphi \nonumber \\
	&= qa_0^2 \int_{-\pi/q}^{\pi/q} [1-g_\ell(\varphi)] \ d\varphi \nonumber \\
	&\leq 2\pi a_0^2(1-a). \label{area}  
\end{align}
Substituting \eqref{|nabla_u-u|} -- 
\eqref{area} into \eqref{|q-q|} gives
\begin{align}
	& \ \ \ |q_{\ell,nm} - \widetilde{q}_{\ell,nm}| \nonumber \\
	&\leq k^2 \sqrt{\frac{2n+1}{4\pi \rho_m}} \frac{\sqrt{e/2}}{e-1} \left[e^{-1}|j_n'(ka_0)| + (1+e^{-2})|j_n(ka_0)| \right] \nonumber \\
	& \ \ \ \times \int_{\partial C_\ell} \ dS(\mathbf{y}) \ e^{-\Delta} \nonumber \\
	&\leq 2\pi (ka_0)^2 (1-a) \sqrt{\frac{2n+1}{4\pi \rho_m}} \frac{\sqrt{e/2}}{e-1} \nonumber \\
	& \ \ \ \times  \left[e^{-1}|j_n'(ka_0)| + (1+e^{-2})|j_n(ka_0)| \right] e^{-\Delta}.   \label{|q-q|_2}
\end{align}

While it has been shown that there exists a critical value for the truncation order $V$ such that the truncation error of the source amplitude $q_{\ell,nm}$ approaches zero for $V$ sufficiently larger than it, the exact minimum value of $V$ required to achieve a certain level of accuracy in computation remains to be determined as the upper bound \eqref{|q-q|_2} depends on the arbitrary parameter $\Delta$. We are concerned about how an error in $q_{\ell,nm}$ will lead to discrepancy in the active field $u_d|_\ell$ generated by the $\ell$th source and thus degrade the performance of the cloaking system. 
Define
\begin{align}
	\widetilde{u}_d|_\ell = \sum_{n=0}^\infty \sum_{m=-n}^n \widetilde{q}_{\ell,nm} V^m_n(\mathbf{x} - \mathbf{x}_\ell). \label{u_d_l}
\end{align} 
A measure to quantify this effect is the relative squared truncation error averaged over the surface of a unit sphere $\Omega$ centred at $\mathbf{x}_\ell$, denoted by $\varepsilon$, where
\begin{align}
	\varepsilon &= \frac{ \displaystyle  \int_\Omega \left| u_d|_\ell(\mathbf{x}-\mathbf{x}_\ell) - \widetilde{u}_d|_\ell (\mathbf{x}-\mathbf{x}_\ell) \right|^2  \ dS(\mathbf{x}-\mathbf{x}_\ell)}{  \displaystyle  \int_\Omega \left| u_d|_\ell (\mathbf{x}-\mathbf{x}_\ell) \right|^2 \  dS(\mathbf{x}-\mathbf{x}_\ell)} \nonumber \\
	&= \frac{\displaystyle \sum_{n=0}^\infty \sum_{m=-n}^n \left| q_{\ell,nm} - \widetilde{q}_{\ell,nm} \right|^2 |h_n^{(1)}(k|\mathbf{x}-\mathbf{x}_\ell|)|^2}{\displaystyle \sum_{n=0}^\infty \sum_{m=-n}^n |q_{\ell,nm}|^2 |h_n^{(1)}(k|\mathbf{x}-\mathbf{x}_\ell|)|^2}. \label{epsilon}
\end{align}
To establish an upper bound on $\varepsilon$, we can use the maximum value of $|q_{\ell,nm}|$ as a proxy for the benchmark in \eqref{epsilon}. Note that with
\begin{gather}
	|u_i(\mathbf{y})| = |e^{i\mathbf{k}\cdot \mathbf{y}}|=1, \label{|u|} \\
	|\nabla_\mathbf{y} u_i(\mathbf{y})\cdot \mathbf{n}| = 
	|ik \widehat{\mathbf{k}} \cdot (\widehat{\mathbf{y}-\mathbf{x}_\ell})  e^{i\mathbf{k}\cdot (\mathbf{y}-\mathbf{x}_\ell)} e^{i\mathbf{k}\cdot \mathbf{x}_\ell}|
	\leq k \label{|nabla_u|}
\end{gather}
and \eqref{|U|}, we have
\begin{align}
	|q_{\ell,nm}| &\leq k \int_{\partial C_\ell} \Big[|u_i(\mathbf{y})| |\nabla_\mathbf{y} \overline{U^{m}_n (\mathbf{y}-\mathbf{x}_\ell) } \cdot \mathbf{n}| \nonumber \\
	& \ \ \ + |\overline{U^{m}_n (\mathbf{y}-\mathbf{x}_\ell) }| |\nabla_\mathbf{y} u_i(\mathbf{y})\cdot \mathbf{n}| \Big] \ dS(\mathbf{y}) \nonumber\\
	&\leq 2\pi (ka_0)^2 (1-a) \sqrt{\frac{2n+1}{4\pi \rho_m}} \nonumber \\
	& \ \ \ \times \left[|j_n'(ka_0)| + |j_n(ka_0)| \right]. \label{|q|}   
\end{align}
Substituting \eqref{|q-q|_2} and \eqref{|q|} into \eqref{epsilon} and noting that $N$ is the multipole order for each active source, we can obtain an upper bound for $\epsilon$ in the form
\begin{align}
	\varepsilon
	\leq \frac{e}{2(e-1)^2} F(\mathbf{x}) e^{-2\Delta} \label{epsilon_bound}
\end{align}
with
\begin{align}
	F(\mathbf{x}) &=
	\frac{\displaystyle \sum_{n=0}^N \left[e^{-1}|j_n'(ka_0)| + (1+e^{-2})|j_n(ka_0)|  \right]^2 H_n(\mathbf{x}) }{\displaystyle \sum_{n=0}^N \left[|j_n'(ka_0)| + |j_n(ka_0)| \right]^2 H_n(\mathbf{x})}, \label{F} \\
	H_n(\mathbf{x}) &= (2n+1)(n+1) |h_n^{(1)}(k|\mathbf{x}-\mathbf{x}_\ell|)|^2. \label{H}
\end{align}
Given the tolerance level $E$, the minimum truncation order $V$ required for the source amplitude $q_{\ell,nm}$ such that $\epsilon <E$ is thus $V=\lceil eka_0/2 \rceil +\Delta+1$, where
\begin{align}
	\Delta = \left\lceil \frac{1}{2}\ln \left[\frac{e}{2E(e-1)^2}\max F(\mathbf{x})\right] \right\rceil. \label{Delta}
\end{align}
Applying this expression of $V$ onto the settings of Fig. \ref{fig3} with $ka_0= 5\pi \cdot \displaystyle \frac{s}{2\sin(\pi/5)}, N=10 \ \text{and} \ E=1\%$ where $s$ is the length of the side of a regular dodecahedron with $x_0=1$, we truncate the series expansion of the source coefficient in \eqref{q_plane_wave} at $V=16$.

\subsection{4. Computation of the surface integral $\mathcal{I}^{m\mu}_{n\nu}(\mathbf{x}_1,\partial C_1)$}

To implement the cloaking strategy discussed, we need to evaluate the surface integral
\begin{align}
	\mathcal{I}_{n\nu}^{m\mu}(\mathbf{x}_1, \partial C_1) = \int_{\partial C_1} \overline{Y^m_n(\widehat{\mathbf{y}-\mathbf{x}_1})} Y^\mu_\nu(\widehat{\mathbf{y}-\mathbf{x}_1}) \ dS(\mathbf{y}-\mathbf{x}_1), \label{I_com}
\end{align}
which will be applied to determine the strengths of all active sources given that they are equidistant from the origin and all spherical faces $\partial C_\ell$ have an identical geometric shape in a Platonic source distribution. 
By the definition of the normalized spherical harmonic function and the parametric form of $\partial C_1$ derived in Part 2 (and stated in \eqref{cos} -- \eqref{h}), we can write the integral as
\begin{align}
	& \ \ \ \mathcal{I}^{m\mu}_{n\nu}(\mathbf{x}_1,\partial C_1)  \nonumber \\
	&= a_{0}^2 A^m_n A^\mu_\nu \sum_{\ell=2}^{q+1} \int_{\varphi_{\ell} - \frac{\pi}{q}}^{\varphi_{\ell} + \frac{\pi}{q}}  \bigg[\int^{1}_{ g_{\ell}(\varphi)} P^m_n (w) P^\mu_\nu (w) \ dw  \bigg] \nonumber \\
	& \ \ \ \times e^{i(\mu-m)\varphi} \ d\varphi. 	\label{3d_i_I_expand}
\end{align}
Now replace the azimuthal angle $\varphi$ by $\varphi+\varphi_\ell$ in the integration with respect to  $\varphi$. This substitution eliminates the dependence of  $g_\ell(\varphi)$ on the index $\ell$ as can be seen from \eqref{g} and thus we may denote $g(\varphi) = g_\ell(\varphi+\varphi_\ell)$. Note that $g(\varphi)$ is even about $\varphi=0$. The surface integral then becomes
\begin{align}
	\mathcal{I}^{m\mu}_{n\nu}(\mathbf{x}_1,\partial C_1) 
	&= 2a_{0}^2 A^m_n A^\mu_\nu 
	\bigg(\sum_{\ell=2}^{1+q} e^{i(\mu- m) \varphi_{\ell} } \bigg) \mathcal{J}^{m\mu}_{n\nu}, \label{3d_a_e_o_2}
\end{align}
where
\begin{align}
	\mathcal{J}^{m\mu}_{n\nu} &= \int_0^{\pi/q} \mathcal{K}^{m\mu}_{n\nu}(\varphi) \cos \big[(\mu-m)\varphi \big] \ d\varphi, \label{J_phi} \\
	\mathcal{K}^{m\mu}_{n\nu}(\varphi) &= \int^{1}_{g(\varphi)} P^m_n (w) P^\mu_\nu (w) \ dw . \label{3d_i_I_theta}
\end{align}
Up to this point it appears that simplifying the surface integral into its analytic form is not trivial. Nevertheless, we may still look for ways to speed up the computation of the expressions \eqref{3d_a_e_o_2} -- \eqref{3d_i_I_theta}. Note that the most computationally-expensive part is the double integral $\mathcal{J}^{m\mu}_{n\nu}$ and given the ranges of the indices $n,m,\nu,\mu$, a direct evaluation of \eqref{3d_a_e_o_2} -- \eqref{3d_i_I_theta} over these ranges would entail performing the double integral for a total of
\begin{align}
	\sum_{n=0}^N \sum_{\nu=0}^V (2n+1)(2\nu+1) = (N+1)^2(V+1)^2 \label{times}
\end{align}   
times, where $N$ is the multiple order of each active source and $V$ is the truncation order of the source amplitude discussed in Part 3. The computation will be costly especially when higher order sources are used in the cloaking system or when the frequency of the incident wave is high which requires a larger truncation order. We are therefore seeking a simpler form of $\mathcal{J}^{m\mu}_{n\nu}$ and thus $\mathcal{I}^{m\mu}_{n\nu}(\mathbf{x}_1,\partial C_1)$ that can facilitate the computation.

By \cite{gradshteyn2014table}, the associated Legendre functions in \eqref{3d_i_I_theta} can be written in an alternative form
\begin{align}
	P^m_n(w) = 
	\bigg(\frac{1+w}{1-w} \bigg)^{m/2} \sum_{\tau=0}^{\infty} c_{n,\tau}^m \bigg(\frac{1-w}{2} \bigg)^\tau \label{3d_int_cir_p_hg}
\end{align}
with
\begin{align}
	c_{n,\tau}^m &= \frac{(-n)_\tau (n+1)_\tau}{\Gamma(1-m)(1-m)_\tau \tau!}, \label{c_gamma} \\
	(p)_\tau &=
	\begin{cases}
		1, \ \ \ &\tau=0 \\ p(p+1)(p+2)\cdots (p+\tau-1). \ \ \ &\tau> 0
	\end{cases}
	. \label{3d_int_cir_poch}
\end{align}
In \eqref{c_gamma}, $\Gamma(1-m)$ is the gamma function and is convergent provided that $1-m$ is not a nonpositive integer. To satisfy this condition we can restrict $m$ to the nonpositive integers only. For positive $m$, we can apply the identity
\begin{align}
	P^m_n(w) = (-1)^m \frac{(n+m)!}{(n-m)!} P_n^{-m}(w), \ \ \ m> 0 \label{3d_i_P}
\end{align} 
so that the azimuthal order of the associated Legendre function becomes nonpositive. The integral $\mathcal{K}^{m\mu}_{n\nu}(\varphi)$ can now be rewritten as
\begin{align}
	\mathcal{K}^{m\mu}_{n\nu}(\varphi) = a^{m\mu}_{n\nu} \int^{1}_{g(\varphi)} P^{-|m|}_n (w) P^{-|\mu|}_\nu (w) \ dw , \label{K}
\end{align}
where
\begin{align}
	a^{m\mu}_{n\nu} =
	\begin{dcases}
		1,  \   &m \leq 0 \ \text{and} \ \mu \leq 0 \\
		(-1)^m \frac{(n+m)!}{(n-m)!}, \   &m>0 \ \text{and} \ \mu \leq 0 \\
		(-1)^{m+\mu} \frac{(n+m)!}{(n-m)!} \frac{(\nu+\mu)!}{(\nu-\mu)!}, \   &m>0 \ \text{and} \ \mu>0
	\end{dcases}
.\label{a}
\end{align}
In \eqref{3d_int_cir_poch},
$(p)_\tau$ is called the Pochhammer symbol \cite{gradshteyn2014table} and vanishes for $\tau \geq 1-p$ if $p$ is a nonpositive integer. Referring back to \eqref{c_gamma}, we observe that $(1-m)_\tau > 0$ for all $\tau$ since $1-m$ must be positive by our choice of $m$ and $c^m_{n,\tau}$ is hence well-defined for all $\tau$.  Note also that $(-n)_\tau$ and thus $c^m_{n,\tau}$ vanish identically for $\tau\geq n+1$, which means that the series expansion for $P^m_n(w)$ in \eqref{3d_int_cir_p_hg} is finite and terminates at $\tau=n$. 
Substituting \eqref{3d_int_cir_p_hg} into \eqref{K} and replacing $w$ by $1-2w$ inside the integral, we have 
\begin{align}
	& \ \ \ \mathcal{K}^{m\mu}_{n\nu}(\varphi)/a^{m\mu}_{n\nu} \nonumber \\
	&= \sum_{\tau=0}^{n} \sum_{\sigma=0}^{\nu} 
	\frac{C_{n\nu,\tau\sigma}^{m\mu}}{2^{\tau+\sigma+1}}
	\int_{g}^1 (1+w)^{-\frac{|m|+|\mu|}{2}} (1-w)^{\tau+\sigma+\frac{|m|+|\mu|}{2}}  dw \nonumber \\
	&= \sum_{\tau=0}^{n} \sum_{\sigma=0}^{\nu} C_{n\nu,\tau\sigma}^{m\mu} \int_{0}^{\frac{1-g}{2}} (1-w)^{-\frac{|m|+|\mu|}{2}} w^{\tau+\sigma+\frac{|m|+|\mu|}{2}} \  dw \nonumber \\
	&=  \sum_{\tau=0}^{n} \sum_{\sigma=0}^{\nu} C_{n\nu,\tau\sigma}^{m\mu} B_{\frac{1-g}{2}} \Big(\tau+\sigma +\frac{|m|+|\mu|}{2}+1, 1-\frac{|m|+|\mu|}{2}  \Big),
	\label{3d_i_I_theta_2}
\end{align}
where the dependence of $g$ on $\varphi$ is understood and $C_{n\nu,\tau\sigma}^{m\mu} = 2 c_{n,\tau}^{-|m|} c_{\nu,\sigma}^{-|\mu|}$. 
In \eqref{3d_i_I_theta_2}, $B_x(a,b)$ is the incomplete beta function \cite{gradshteyn2014table} defined by
\begin{align}
	B_x(a,b) = \int_0^x t^{a-1} (1-t)^{b-1} \ dt, \label{3d_i_beta}
\end{align}
which is convergent when $a$ is positive \cite{srivastava2012zeta}. This condition is again satisfied since in \eqref{3d_i_I_theta_2}, 
$\displaystyle \tau+\sigma +\frac{|m|+|\mu|}{2}+1 \geq 1$. 
Now substituting \eqref{3d_i_I_theta_2} into \eqref{J_phi} gives
\begin{align}
	& \ \ \ \mathcal{J}^{m\mu}_{n\nu} \nonumber \\ 
	&= a_{n\nu}^{m\mu} \sum_{\tau=0}^{n} \sum_{\sigma=0}^{\nu} C_{n\nu,\tau\sigma}^{m\mu} \int_0^{\pi/q} \cos \big(|\mu-m|\varphi \big)  \nonumber \\
	& \ \ \ \times  B_{\frac{1-g}{2}} \Big(\tau+\sigma +\frac{|m|+|\mu|}{2}+1, 1-\frac{|m|+|\mu|}{2} \Big) \ d\varphi. \label{3d_i_I_phi}
\end{align}
With \eqref{3d_i_I_phi}, we have expressed $\mathcal{J}^{m\mu}_{n\nu}(\varphi)$ as a linear combination of integrals in terms of the incomplete beta function, which can be readily evaluated using the corresponding built-in function of some common numerical software and thus reduces a fair portion of the computational cost.
 

While the sum of integrals $\mathcal{J}^{m\mu}_{n\nu}$ now takes a simpler form which is more convenient to evaluate, the integral still needs to be performed for many times when the multipole order $N$ or the truncation parameter $V$ becomes large. In the following we outline how we can make use of the properties of $\mathcal{J}^{m\mu}_{n\nu}$ to significantly reduce the number of integrations required. Inspection of its form in \eqref{3d_i_I_phi} and the definition of $C^{m\mu}_{n\nu,\tau \sigma}, a_{n\nu}^{m\mu}$ in \eqref{c_gamma}, \eqref{a} respectively shows that $\mathcal{J}^{m\mu}_{n\nu}$ is symmetric with respect to the two groups of indices $(n,m)$ and $(\nu,\mu)$ such that $\mathcal{J}^{m\mu}_{n\nu} = \mathcal{J}^{\mu m}_{\nu n}$. Without loss of generality, we may consider only the case $n\leq \nu$ where
$0\leq n \leq \min (N,V)$ and $n\leq \nu \leq \max (N,V)$. 
The symmetry of $\mathcal{J}^{m\mu}_{n\nu}$ means that
\begin{align}
	& \ \ \ \mathcal{I}^{m\mu}_{n\nu}(\mathbf{x}_1,\partial C_1) \nonumber \\
	&= 2a_{0}^2 A^m_n A^\mu_\nu 
	\bigg(\sum_{\ell=2}^{1+q} e^{i(\mu- m) \varphi_{\ell} } \bigg)
	\begin{dcases}
		\mathcal{J}^{m\mu}_{n\nu}, \ \ n\leq \nu \\
		\mathcal{J}^{\mu m}_{\nu n}, \ \ n> \nu
	\end{dcases}  
	.\label{swap}
\end{align}
A further inspection of the integrand in \eqref{3d_i_I_phi} reveals that there exists different combinations of the indices $\tau,\sigma,m,\mu$ that indeed give the same value of the integral. Take
$i=\tau+\sigma, j=|m|+|\mu|, k=|\mu-m|$ and rewrite \eqref{3d_i_I_phi} as  
\begin{align}
	\mathcal{J}_{n\nu}^{m\mu} &= a_{n\nu}^{m\mu}  \sum_{i=0}^{n+\nu} \left(\sum_{\substack{\tau+\sigma=i, \\ 0\leq \tau\leq n, 0\leq \sigma\leq \nu }} C_{n\nu,\tau\sigma}^{m\mu} \right)\mathcal{B}_{ijk},  \label{J=aCB} \\
	\mathcal{B}_{ijk} &=
	\int_0^{\pi/q}  B_{\frac{1-g}{2}} \Big(i +\frac{j}{2}+1, 1-\frac{j}{2} \Big)  \cos k\varphi  \ d\varphi. \label{B}
\end{align}
Note that $i,j,k$ remain invariant under the swap of the indices in \eqref{swap}. 
The strategy is to group the summands in \eqref{3d_i_I_phi} by identifying the integrals that have the same value of $i=\tau+\sigma$ and collecting the coefficients $C^{m\mu}_{n\nu,\tau\sigma}$ multiplied to each of them.
With this approach, we can avoid repeatedly evaluating integrals that give the same result.  
To determine the permutations of $(\tau,\sigma)$, we note that $n\leq\nu$ is already assumed and there are only three possible cases for the value of $i$:\\
\textcolor{white}{h}\\
Case 1: $i\leq n$ 
\begin{flalign}
	(\tau,\sigma) = (0,i),(1,i-1),(2,i-2),\cdots, (i-1,1),(i,0). \textcolor{white}{abc} \label{case_1}
\end{flalign}
Case 2: $n<i\leq \nu$
\begin{flalign}
	\nonumber (\tau,\sigma)
	&= (0,i),(1,i-1),(2,i-2),\cdots,(n-1,i-n+1),&& \\ & \ \ \ \  (n,i-n). \label{case_2}
\end{flalign}
Case 3: $i>\nu$
\begin{flalign}
	\nonumber (\tau,\sigma) &= (i-\nu,\nu),(i-\nu+1,\nu-1),(i-\nu+2,\nu-2), && \\
	& \ \ \ \ \cdots, (n-1,i-n+1),(n,i-n). \label{case_3}
\end{flalign}
Note that $i,j \leq N+V$ since
\begin{gather}
	0 \leq \tau + \sigma \leq n+\nu \leq N+V, \label{i_range} \\
	0 \leq |m| + |\mu| \leq n+\nu \leq N+V. \label{j_range}
\end{gather}
As for $k$, we have $|\mu-m|\leq|\mu| + |m|=j$. The lower bound of $k$ depends on whether $j$ is odd or even because if the sum of two integers is odd (even), then their difference must also be odd (even). The possible values taken by $k$ are thus
\begin{align}
	k=
	\begin{dcases}
		1, 3, 5, \cdots, j-2, j, \ \ & j \ \text{is odd} \\
		0, 2, 4, \cdots, j-2, j, \ \ & j \ \text{is even}
	\end{dcases} \label{k_o_e}
\end{align}
or more compactly,
$k= \text{mod}(j,2), \text{mod}(j,2) +2, \cdots, j-2, j$. Therefore under the change of indices $(n,\nu,m,\mu) \to (i,j,k)$, the integral $\mathcal{B}_{ijk}$ is now evaluated for a total of  
\begin{align}
	&\ \ \ (N+V+1)\sum_{j=0}^{N+V}  \left[ \frac{j-\text{mod}(j,2)}{2}+1 \right] \nonumber \\
	&= (N+V+1)
	\begin{dcases}
		\left( M +1 \right) \left( M +2 \right),  &N+V \ \text{is odd} \\
		\left( M +1 \right)^2,  &N+V \ \text{is even}
	\end{dcases}
	 \label{times_2}
\end{align}
times, where  $M= \left\lfloor (N+V)/2 \right\rfloor$ and $\lfloor . \rfloor$ is the floor function. Putting \eqref{times_2} in the context of Fig. \ref{fig3}, we have $N=10$ and $V=16$ and thus 5292 integrations are run in total. Compare it with the original form of the integral where 34969 integrations have to be performed by \eqref{times}. Our method has led to around $85\%$ decrease in the number of integrations done (along with an integrand less costly to evaluate). The reduction is more significant for larger $N$ and $V$ as the leading order of \eqref{times} is quartic but only cubic in \eqref{times_2}. 




\subsection{5. Transformation approach with rotation of the coordinate system }
Consider a rotation in three-dimensional space $\mathbf{R}(\widehat{\mathbf{v}},\Theta)$ which maps the coordinate system $\mathbf{x}=(x,y,z)$ to $\mathbf{x}'=(x',y',z')=\mathbf{R}\mathbf{x}$ with $\widehat{\mathbf{v}}=(v_x,v_y,v_z)$ the unit rotation axis vector and $\Theta$ the rotation angle (in a counterclockwise sense). We require that in the rotated frame $\mathbf{x}'$, 1) the $\ell$th active source occupies the bottommost location with position vector $\mathbf{x}_\ell'=(0,0,-x_0)=\mathbf{x}_1$; 2) the rotated spherical face $\partial C_\ell'$ 
is aligned in the same orientation as $\partial C_1$ is in the original frame $\mathbf{x}$ such that $\partial C_\ell' = \partial C_1$.
We are seeking the form of $\widehat{\mathbf{v}}_\ell, \Theta_\ell$ and thus the rotation matrix $\mathbf{R}_\ell$ for $\ell=1,2,\cdots,L$. 


We first consider the active sources located on the layers with $2 \leq i \leq \max\left((N_L+1)/2,2 \right)$. The rotation matrix $\mathbf{R}$ can be found using the Rodrigues' rotation formula \cite{rodriguez1840lois}:
\begin{align}
\mathbf{R}(\widehat{\mathbf{v}},\Theta) = \mathbf{I}\cos\Theta -\mathbf{V}\sin\Theta + \widehat{\mathbf{v}} \widehat{\mathbf{v}}^T (1-\cos\Theta), \label{3d_o_R}
\end{align}
where $\mathbf{I}$ is the $3\times 3$ identity matrix; $\widehat{\mathbf{v}}^T$ denotes the transpose of $\widehat{\mathbf{v}}$ and $\mathbf{V}$ is defined by
\begin{align}
\mathbf{V}=
\begin{pmatrix}
0 & -v_z & v_y \\ v_z &0&-v_x\\-v_y&v_x&0 
\end{pmatrix}
. \label{3d_a_V}
\end{align}
To find the rotation axis $\widehat{\mathbf{v}}$, we recall from Part 2 that with $a$ taking the minimum value in \eqref{range} the vertices of the cloaked region $C$ coincide with the centroid of each $p$-sided polygonal face of the Platonic solid. By the definition of $p$, both the polyhedron and $C$ have a $p$-fold rotational symmetry about an axis through the origin and any vertex of $C$. Since the source $\ell=1$ is the common vertex of $q$ congruent polygonal faces, these faces give a total of $q$ centroids which the rotation axes may pass through. We can take $\mathbf{v}_\ell = |\mathbf{v}_\ell|  \widehat{\mathbf{v}}_\ell = |\mathbf{v}_\ell| (v_{\ell,x}, v_{\ell,y}, v_{\ell,z})$ to be the position vector of the centroid associated with the rotation $\mathbf{R}_\ell$ that 
satisfies the two conditions stipulated above. By the regularity and symmetry of the source distribution, $\mathbf{v}_\ell$ is perpendicular to the polygonal face on which the centroid is located. 
It is hence straightforward to show that in a Cartesian coordinate system centred at the origin,
\begin{align}
\mathbf{v}_\ell = x_0 \sqrt{1-\overline{a}^2} \left( \overline{a} \cos \psi_{2,j}, \overline{a}\sin \psi_{2,j}, -\sqrt{1-\overline{a}^2} \right), \label{v_l}
\end{align}
where $\overline{a}$ is the minimum radius in \eqref{range} and $\psi_{2,j} = \varphi_{2,j}+ \pi/q$ with $\varphi_{2,j}$ and $j$ defined as in Part 2. The corresponding unit vector $\widehat{\mathbf{v}}_\ell$ is thus
\begin{align}
\widehat{\mathbf{v}}_\ell = \left( \overline{a} \cos \psi_{2,j}, \overline{a} \sin \psi_{2,j}, -\sqrt{1-\overline{a}^2} \right). \label{v_2}
\end{align}
When $a$ takes values greater than $\overline{a}$, the vertices of $C$ no longer coincide with the centroid of each polygonal face of the Platonic solid. With \eqref{cos_theta} the polar angle of the $q$ vertices immediately above the source $\ell=1$ is given by 
\begin{align}
	\cos \theta &= \Big \{1 + [a^2h^2\cos^4 (\pi/q) - h(1-a^2) \nonumber \\
	& \ \ \ \times \cos^2 (\pi/q) ]^{1/2} \Big \}  / \left\{ a \left[1+h \cos^2(\pi/q) \right] \right\} \label{cos_theta_c_q}
\end{align}
in a spherical coordinate system centred at $\mathbf{x}_1$. In Part 2 it has already been shown in \eqref{phi_+_-} that
\begin{align}
	\cos \frac{\pi}{q} = \sqrt{\frac{1-\overline{a}^2}{\overline{a}^2h}}. \label{cos_pi_q}
\end{align}
Substituting \eqref{cos_pi_q} into \eqref{cos_theta_c_q}, we have 
\begin{align}
	\cos\theta = \frac{\overline{a}^2+\sqrt{(1-\overline{a}^2)(a^2-\overline{a}^2)}}{a}. \label{cos_theta_a}
\end{align}
Note that at $a=\overline{a}$, we recover the relation $\cos\theta =\overline{a}$ in \eqref{theta_+_-} which defines the centroid of each polygonal face that meets at the source $\ell=1$. If we again let the rotation axis $\mathbf{v}_\ell$ be the position vector of the vertex of $C$ that is associated with the rotation of the source $\ell$ for $\ell=2,\cdots,q+1$, then 
with some algebra and the expression \eqref{cos_theta_a} we can prove that in a Cartesian coordinate system centred at the origin, 
\begin{align}
	\mathbf{v}_\ell  &= x_0 \Big(\sqrt{1-\overline{a}^2} - \sqrt{a^2-\overline{a}^2} \Big) \nonumber \\
	& \ \ \ \times \Big( \overline{a} \cos \psi_{2,j}, \overline{a} \sin\psi_{2,j}, -\sqrt{1-\overline{a}^2} \Big). \label{v_l_3} 
\end{align}
Comparing \eqref{v_l_3} with \eqref{v_l}, we can deduce that the vertex of $C$, the centroid of the polygonal face and the origin are collinear with one another, which implies that the unit vector $\widehat{\mathbf{v}}_\ell$ in the form \eqref{v_2} indeed applies to the whole range of $a$.

\begin{figure}[h!] 
	\centering
	\begin{tabular}{c}
		(a)\subf{\includegraphics[trim={1cm 0.5cm 0.3cm 0.5cm},clip,width=6cm]{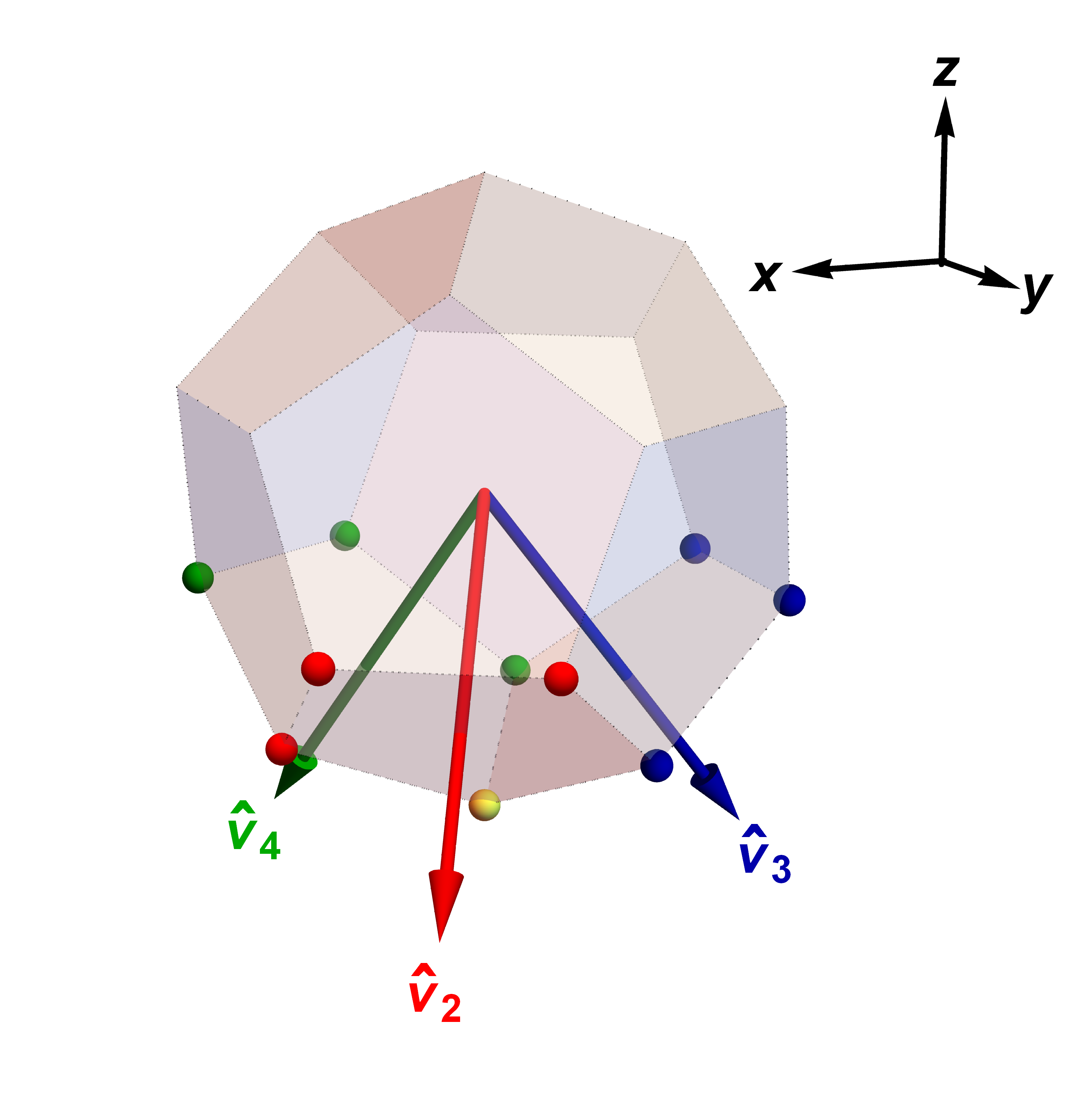}}
		\\
		(b)\subf{\includegraphics[trim={1cm 0.5cm 0.3cm 0.5cm},clip,width=6cm]{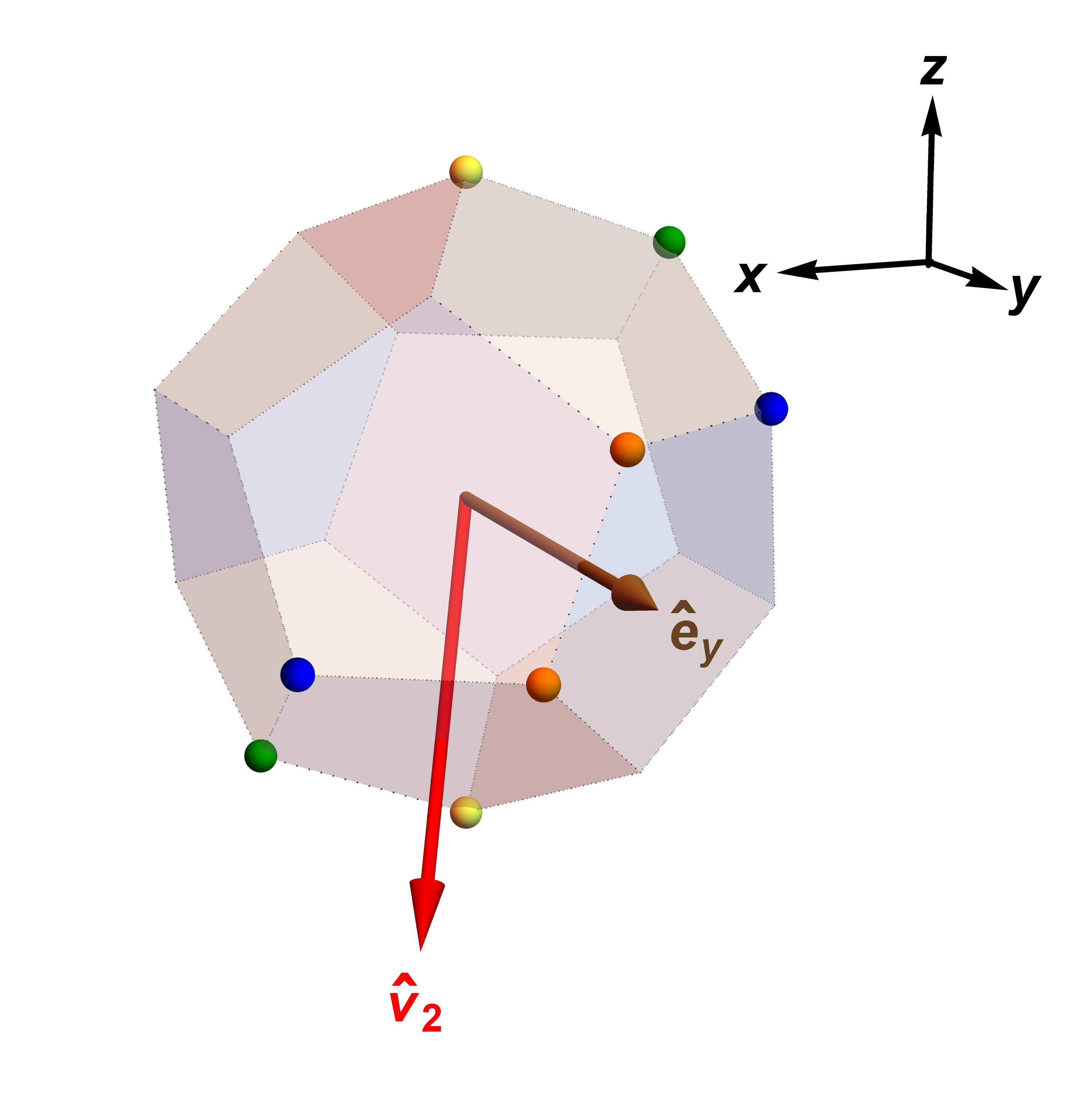}}
	\end{tabular}
	\caption{Visualization of the rotational approach in the original frame $\mathbf{x}=(x,y,z)$ for $L=20$ when the active sources are located on planes with (a) $i \leq \max\left((N_L+1)/2,2\right)=4.5$ (which gives $i=1,2,3,4$) and (b) $i>4.5$ (which gives $i=5,6,7,8$). The source $\ell=1$ is depicted by the bottommost yellow dot in both subfigures. In (a), sources that possess the common rotation axis $\widehat{\mathbf{v}}_\ell$ are represented by dots of the same colour. In (b), for the case $j=0$, sources that are rotationally symmetric about the unit $y$ axis $\widehat{\mathbf{e}}_y$ (the brown arrow) are shown in the same colour.}\label{R_a_b}
\end{figure}

Since each polygonal face consists of $p$ sides and both the first and $\ell$th source are located at the vertices of this polygon, the rotation angle $\Theta_\ell$ must be an integral multiple of $2\pi/p$. 
Given the form of $\widehat{\mathbf{v}}_\ell$ in \eqref{v_2}, a clockwise rotation of the original frame $\mathbf{x}$ by an angle of $2(i-1)\pi/p$ about the axis $\widehat{\mathbf{v}}_\ell$ will `apparently' move the source located on the $i$th layer 
to the bottommost location. (A concrete example for the case $L=20$ will be illustrated later in Fig. \ref{R_a_b}.) Provided the convention that positive $\Theta_\ell$ denotes a counterclockwise rotation of the frame $\mathbf{x}$, we take $\Theta_\ell=-2(i-1)\pi/p$.
Substituting \eqref{v_2} and this value of $\Theta_\ell$ into \eqref{3d_o_R}, we have the rotation matrix $\mathbf{R}_\ell=\{R_{\ell,st}\}_{s,t=1}^3$ given by
\begin{align}
	& \ \ \ R_{\ell, st} \nonumber \\
	&=
	\begin{dcases}
	\cos \frac{2(i-1)\pi}{p}+ v_{\ell,s}^2  \left[1-\cos \frac{2(i-1)\pi}{p} \right], \ \ \ &s=t\\
	-\epsilon_{stu} v_{\ell,u} \sin \frac{2(i-1)\pi}{p} \\
	 \ \ \ +  v_{\ell,s} v_{\ell,t} \left[1-\cos \frac{2(i-1)\pi}{p} \right], \ \ \ &s\neq t
	\end{dcases}
	,\label{3d_a_R_ij}
\end{align}
where $\epsilon_{stu}$ is the Levi-Civita symbol \cite{weber2005mathematical} defined by 
\begin{align}
\epsilon_{stu} =
\begin{cases}
1, \ &\textnormal{if} \ (s,t,u)=(1,2,3),(2,3,1)\ \textnormal{or} \ (3,1,2)\\
-1, \ &\textnormal{if} \ (s,t,u)=(2,1,3),(3,2,1)\ \textnormal{or} \ (1,3,2)
\end{cases}
.\label{3d_a_epsilon_ijk}
\end{align}

As for the active sources located on the layers $i>\max\left((N_L+1)/2,2 \right)$, we note that with the orientation of each Platonic solid specified by \eqref{x_ell} -- \eqref{3d_r_p_varphi} in Part 2, they possess a two-fold rotational symmetry about the $y$ axis, except for the case $L=4$. In mathematical terms, this rotation can be expressed by the matrix $\mathbf{R}(\widehat{\mathbf{e}}_y,t)$ where
\begin{align}
\mathbf{R}(\widehat{\mathbf{e}}_y,t)=
\begin{pmatrix}
\cos t & 0 & -\sin t \\
0 & 1 &0 \\
\sin t & 0 & \cos t
\end{pmatrix}
\label{R_y}
\end{align}
with $\widehat{\mathbf{e}}_y$ the unit vector pointing in the positive $y$ direction and $t$ the angle of rotation. By setting $t=\pi$ we can show that for $i>\max\left((N_L+1)/2,2 \right)$,
\begin{align}
\mathbf{x}_{i,j} = \mathbf{R}(\widehat{\mathbf{e}}_y,\pi) \mathbf{x}_{N_L+1-i,j}. \label{x=x} 
\end{align}
The relation \eqref{x=x} means that 
we can always match a source located on the plane $i>\max\left((N_L+1)/2,2 \right)$ with one on a lower layer $i\leq \max\left((N_L+1)/2,2 \right)$ and their position vectors are related to each other by \eqref{x=x}. We can make use of the results derived in the previous case 
and express the rotation matrix $\mathbf{R}_\ell$ for $i>\max\left((N_L+1)/2,2 \right)$ as
\begin{align}
\mathbf{R}_{\ell} = \mathbf{R}_{\ell'}(\widehat{\mathbf{v}}_{\ell'},\Theta_{\ell'}) \mathbf{R}(\widehat{\mathbf{e}}_y,\pi) \label{R=RR}
\end{align}
where
\begin{align}
\ell'=
\begin{cases}
(N_L-i-1)q+j+2, & \max\left(\frac{N_L+1}{2},2 \right)<i <N_L \\
1, & i=N_L, L \neq 4
\end{cases} \label{l'_i_j}
\end{align}
and $\mathbf{R}_{\ell'}(\widehat{\mathbf{v}}_{\ell'},\Theta_{\ell'})$ is in the same form as \eqref{3d_a_R_ij} with $\widehat{\mathbf{v}}_{\ell'}$ defined by \eqref{v_2} and $ -2(i-1)\pi/p$ replaced by $-2(N_L-i)\pi/p$.


To illustrate the method outlined above, we look at the case $L=20$ in Fig.\ \ref{R_a_b}, which shows the different rotation axes associated with the rotation that transforms the position vector $\mathbf{x}_\ell$ to $\mathbf{x}_1=(0,0,-x_0)$ (the yellow dot at the bottom in both subfigures). Fig.\ \ref{R_a_b}(a) shows the case for the layers $2 \leq i \leq \max\left((N_L+1)/2,2\right)$. For $L=20$, it involves those sources with $i=2,3,4$ (or equivalently $2\leq \ell \leq 10$). Since $(p,q)=(5,3)$ by Table \ref{geo}, they form a total of three regular pentagons that meet at the first source. These sources can be classified into three groups with components of a group having the same value of $j$ and represented by dots of the same colour. 
The three groups are $j=0$ coloured in red, $j=1$ in blue and $j=2$ in green. By the numbering of the sources on the layers $i=2,3,4$ in Fig. \ref{L20}(b) -- (c) in Part 2, the three groups consist of sources with $\ell=\{2,5,8\}, \{3,6,9\} \ \text{and} \ \{4,7,10\}$ respectively. 
Sources in the same group share the common rotation axis $\widehat{\mathbf{v}}_\ell$, where $\ell$ is the index of any source in the group since the form of $\widehat{\mathbf{v}}_\ell$ in \eqref{v_2} depends only on $j$ and is the same for all components of a group. (Note that in Fig.\ \ref{R_a_b}(a), the minimum value of $\ell$ in the group is taken.)
Each of these rotation axes points in the direction \eqref{v_2} towards the centroid of the pentagon that the group is located on. The angle of rotation for each active source in the group is given by $\Theta_{\ell}=-2(i-1)\pi/5$ with $i$ the order of the layer the source resides on. For example, for the group $\ell= \{2,5,8\}$, we have $i=\{2,3,4\}$. A clockwise rotation by $2\pi/5,4\pi/5,6\pi/5$ of the frame $\mathbf{x}$ about the axis $\widehat{\mathbf{v}}_2$ will `bring' these sources onto the lowest position respectively. In Fig.\ \ref{R_a_b}(b), the case with $i>\max\left((N_L+1)/2,2\right)$ is illustrated. The two-fold rotational symmetry of the source distribution about the unit vector in the $y$ direction $\widehat{\mathbf{e}}_y$ (the brown arrow) means that we can match any source with $i=5,6,7,8$ to another source with $i=4,3,2,1$ which has the same value of $j$ by the relation \eqref{x=x}. In the figure we focus on the case with $j=0$ and depict the sources which form a pair according to \eqref{x=x} in the same colour. Again by the labelling of sources in Fig. \ref{L20} in Part 2, the pairs are $\ell=\{1,20\}(\text{yellow}), \{2,17\}(\text{green}), \{5,14\}(\text{blue})$ and $\{8,11\}(\text{orange})$. The rotation matrices for the upper four sources can be found using \eqref{R=RR} and the knowledge of the corresponding matrices for the lower four.

The rotation matrix $\mathbf{R}$ is related to how the form of the active field $u_d$ transforms between the original space $\mathbf{x}$ and the rotated one $\mathbf{x}'$. (The subscript of $\mathbf{R}_\ell$ is suppressed for convenience as its dependence on $\ell$ is understood.) With the incident wave propagating in the direction $\widehat{\mathbf{k}}$, the active field $u_d|_\ell$ radiated by the $\ell$th source can be posed in the original frame $\mathbf{x}$ as
\begin{align}
u_d|_\ell (\mathbf{x})&= \sum_{n=0}^\infty \sum_{m=-n}^n q_{\ell, nm}(\widehat{\mathbf{k}},\mathbf{x}_\ell, \partial C_\ell) V^m_n (\mathbf{x}-\mathbf{x}_\ell) \label{3d_a_phi_d_ell}
\end{align}
where $q_{\ell, nm}(\widehat{\mathbf{k}},\mathbf{x}_\ell, \partial C_\ell)$ is the source coefficient that we are seeking. Alternatively, if we switch to the rotated frame $\mathbf{x}'$, then $u_d|_\ell$ is in the form
\begin{align}
u_d|_\ell (\mathbf{x}')
&= \sum_{n=0}^\infty \sum_{m=-n}^n q_{\ell,nm}'(\widehat{\mathbf{k}}',\mathbf{x}_\ell', \partial C_\ell') V^m_n (\mathbf{x}'-\mathbf{x}_\ell') \nonumber \\
&= \sum_{n=0}^\infty \sum_{m=-n}^n  q_{1,nm}(\widehat{\mathbf{k}}',\mathbf{x}_1, \partial C_1) 
V^m_n(\mathbf{x}'-\mathbf{x}_\ell')
\label{3d_a_phi_d_ell_'}
\end{align}
where $q_{\ell,nm}'$ is the source amplitude in the rotated space with $\widehat{\mathbf{k}}' = \mathbf{R} \widehat{\mathbf{k}}$. 
The equality between $q_{\ell,nm}'(\widehat{\mathbf{k}}', \mathbf{x}_\ell', \partial C_\ell')$ and $q_{1,nm}(\widehat{\mathbf{k}}', \mathbf{x}_1, \partial C_1)$ holds since
\begin{align}
& \ \ \ \ q_{\ell,nm}'(\widehat{\mathbf{k}}', \mathbf{x}_\ell', \partial C_\ell') \nonumber \\ 
&= -ik^2 \sum_{t=0}^\infty \sum_{s=-t}^t Q_{ts}(\widehat{\mathbf{k}}') q_{\ell,nm,ts}(\mathbf{x}_\ell',\partial C_\ell') \nonumber \\
&= -ik^2 \sum_{t=0}^\infty \sum_{s=-t}^t Q_{ts}(\widehat{\mathbf{k}}') q_{1,nm,ts}(\mathbf{x}_1,\partial C_1) \nonumber \\
&= q_{1,nm}(\widehat{\mathbf{k}}', \mathbf{x}_1, \partial C_1). \label{q} 
\end{align}
Note that $q_{1,nm}(\widehat{\mathbf{k}}', \mathbf{x}_1, \partial C_1)$ can be evaluated using the explicit formula \eqref{q_l} -- \eqref{Range} with the parametric form of $\partial C_1$ in \eqref{cos} -- \eqref{h} and $\widehat{\mathbf{k}}'$ in place of $\widehat{\mathbf{k}}$. (In \eqref{q}, the form of the source coefficient with a general incident wave is used as the case of plane wave incidence should follow immediately.) The forms of representation \eqref{3d_a_phi_d_ell} and \eqref{3d_a_phi_d_ell_'} are equivalent with each other as they describe the same quantity $u_d|_\ell$ but only in different basis functions.
The correspondence between the two expressions can be established by considering the transformation of the spherical harmonic function $Y^m_n(\widehat{\mathbf{x}-\mathbf{x}_\ell})$ between the two frames \cite{man2016wigner}:
\begin{align}
Y^m_n(\widehat{\mathbf{x}-\mathbf{x}_\ell}) 
= \sum_{m'=-n}^{n}  Y^{m'}_n( \widehat{\mathbf{x}'-\mathbf{x}_\ell'}) D^{n}_{m'm}(\gamma, \beta, \alpha), \label{3d_o_Y}
\end{align}
where $D^{n}_{m'm}(\gamma, \beta, \alpha)$ is the $(m',m)$th entry of the Wigner D-matrix \cite{wigner2012group} with degree $n$ defined by
\begin{align}
D^{n}_{m'm}(\gamma, \beta, \alpha)=e^{im' \gamma} d^{n}_{m'm}(\beta) e^{im \alpha} \label{3d_o_D}
\end{align}
with
\begin{align}
&\ \ \ d^{n}_{m'm}(\beta) \nonumber \\
&=  \sum_j C^n_{m'm,j} \bigg(\cos \frac{\beta}{2} \bigg)^{2n+m-m'-2j} \bigg(\sin \frac{\beta}{2} \bigg)^{m'-m+2j}, \label{3d_o_d} \\
&C^n_{m'm,j} = \frac{(-1)^{j}[(n+m')!(n-m')!(n+m)!(n-m)!]^{1/2}}{(n+m-j)!j!(m'-m+j)!(n-m'-j)!}. \label{C}
\end{align}
In \eqref{3d_o_d} -- \eqref{C}, the summation over $j$ is such that the factorial terms are nonnegative. The arguments $(\gamma,\beta,\alpha)$ are the Euler angles \cite{varshalovich1988description} of the rotation matrix $\mathbf{R}(\widehat{\mathbf{v}},\Theta)$ when it is decomposed into three separate rotations about the $z,y,z$ axes of the original space $\mathbf{x}$ such that
\begin{align}
\mathbf{R}(\widehat{\mathbf{v}},\Theta)= \mathbf{R} (\widehat{\mathbf{e}}_z,\gamma) \mathbf{R} (\widehat{\mathbf{e}}_y,\beta) \mathbf{R} (\widehat{\mathbf{e}}_z,\alpha), \label{3d_o_euler}
\end{align}
where $\mathbf{R}(\widehat{\mathbf{e}}_y,\beta)$ is already defined in \eqref{R_y} and  
\begin{align}
\mathbf{R}(\widehat{\mathbf{e}}_z,\alpha)=
\begin{pmatrix}
\cos \alpha & \sin \alpha & 0 \\
-\sin \alpha & \cos \alpha &0 \\
0 & 0 & 1
\end{pmatrix}
.\label{R_z}
\end{align}
Comparing elementwise between the left and right side of \eqref{3d_o_euler}, we have
\begin{align}
\alpha &= -\arctan \bigg(\frac{R_{23}} {R_{13}} \bigg ),\label{3d_a_alpha} \\
\beta &= \arccos (R_{33}),\label{3d_a_beta} \\
\gamma &= \arctan \bigg(\frac{R_{32}} {R_{31}} \bigg ),\label{3d_a_gamma} 
\end{align}
with the entries of $\mathbf{R}$ described by either \eqref{3d_a_R_ij} or \eqref{R=RR}, depending on the value of $\ell$.
If we substitute \eqref{3d_o_Y} into \eqref{3d_a_phi_d_ell} and compare with \eqref{3d_a_phi_d_ell_'}, then for $m'=-n,\cdots,n$, we have
\begin{align}
\sum_{m=-n}^n  D^{n}_{m'm}(\gamma, \beta, \alpha) q_{\ell, nm}(\widehat{\mathbf{k}}, \mathbf{x}_\ell, \partial C_\ell) = q_{1,nm'}(\widehat{\mathbf{k}}', \mathbf{x}_1, \partial C_1). \label{3d_a_lin_eq}
\end{align}
Note that \eqref{3d_a_lin_eq} can be written more compactly in the matrix form \eqref{Dq=q} if we set $\mathbf{D}^n=\{ D^n_{m'm}\}_{m',m=-n}^n$ and $\mathbf{q}_{\ell,n}=\{ q_{\ell,nm} \}_{m=-n}^n$.

\subsection{6. Wave scattering and parameter study}
Consider a smooth object subject to an incident scalar wave $\mathfrak{u}_i$ and denote the ensuing scattered wave as $\mathfrak{u}_s$. The two wave fields can be expanded in terms of the basis functions as  
\begin{align}
	\mathfrak{u}_i(\mathbf{x}) &= \sum_{n=0}^\infty \sum_{m=-n}^n \mathfrak{Q}_{nm} U^m_n(\mathbf{x}), \label{u_i_6} \\
	\mathfrak{u}_s (\mathbf{x}) &= \sum_{n=0}^{\infty} \sum_{m=-n}^n \mathfrak{a}_{nm} V^m_n(\mathbf{x}), \label{u_s_6} 
\end{align}
with $\mathfrak{Q}_{nm}$ and $\mathfrak{a}_{nm}$ the incident and scattering coefficient respectively. For simplicity, we only look at the case of acoustic wave but our arguments can be easily generalized to other cases like electromagnetic or elastic wave. It has been derived by \cite{waterman1969new} in the acoustic context that
\begin{align}
	\mathfrak{a}_{nm} = \sum^\infty_{j=0}\sum_{i=-j}^j T_{nm,ji} \mathfrak{Q}_{ji}, \label{a=TQ}
\end{align}
where $T_{nm,ji}$ is a transition matrix depending on the geometry of the scattering object and the boundary condition imposed on its surface. For a perfect sphere of radius $A$, $T_{nm,ji}$ takes the form 
\begin{numcases}{T_{nm,ji}= -\delta_{nj}\delta_{mi}}
	\frac{j_n'(kA)}{h_n'(kA)} & $\text{for sound-hard case}$ \label{T_hard}
	\\
	\frac{j_n(kA)}{h_n(kA)} & $\text{for sound-soft case}$ \label{T_soft}
\end{numcases}
with $\delta_{hk}$ the Kronecker delta function. 

By adopting procedures similar to what is done in \cite{shearer2015antiplane}, it can be shown that $W$, the acoustic power scattered to the far field by the object in three dimensions averaged over a wave period, is given by
\begin{align}
	W=\frac{\omega\rho}{2k}\sum_{n=0}^\infty \sum_{m=-n}^n |\mathfrak{a}_{nm}|^2, \label{W}
\end{align}
with $\omega$ the angular frequency and $\rho$ the density of the medium. 
We can apply \eqref{W} as a measure to assess the effectiveness of our cloaking approach discussed before by defining the quantity $\sigma$ in the form
\begin{align}
	\sigma &= \frac{W(u_d\neq 0)}{W(u_d=0)} \nonumber \\ 
	&= \frac{\displaystyle \sum_{n=0}^{\infty} \sum_{m=-n}^n \left|a_{nm}(u_d\neq 0) + f_{nm} \right|^2}{\displaystyle \sum_{n=0}^\infty \sum_{m=-n}^n \left | a_{nm}(u_d=0) \right|^2}, \label{sigma_6}
\end{align}
where $u_d$ and $a_{nm}$, defined in \eqref{u_d} and \eqref{u_s}, are the active field produced by the sources and the scattering coefficient of the sphere inside the cloaked region. The term $f_{nm}$ is the far-field amplitude of $u_d$ such that
\begin{align}
	u_d(\mathbf{x}) &= \sum_{n=0}^\infty \sum_{m=-n}^n f_{nm} V^m_n(\mathbf{x}) 
	. \label{u_d_6}
\end{align}
In \eqref{sigma_6}, for the case without any control, the coefficient $\mathfrak{a}_{nm}$ depends solely on the scattered field $u_s$ and is thus equivalent to $a_{nm}(u_d=0)$. However, when the cloaking devices are activated ($u_d \neq 0$), 
$u_s$ is influenced also by $u_d$ and 
$\mathfrak{a}_{nm}$ becomes $a_{nm}(u_d \neq 0)$. 
Note that the contribution from $f_{nm}$ should also be included in $W(u_d\neq 0)$ since we require not only a reduction in the scattered field but also minimal radiation from the active sources to the far field and \eqref{u_d_6} has the same basis function as \eqref{u_s_6} does. 
The quantity $\sigma$ is therefore the ratio of the total power 
radiated by the whole system before and after control. To ensure that the cloaking method is effective, we need $\sigma<1$. 

In the following we will find the expressions for 
the three coefficients in \eqref{sigma_6}.
Substituting $\mathfrak{Q}_{nm} = Q_{nm}$ and \eqref{T_soft} into \eqref{a=TQ} gives $a_{nm}(u_d=0)$ in the form
\begin{align}
	a_{nm}(u_d=0) = -\frac{j_n(kA)}{h_n(kA)} Q_{nm} \label{a_nm_u_d=0}
\end{align}
when the scattering object is a sound-soft sphere. As for $a_{nm}(u_d\neq 0)$, we recall from Part 1 that we apply the addition theorem for spherical wavefunctions in \eqref{add} -- \eqref{cal_S} to determine the explicit form of the source amplitudes. Similarly, to obtain the form of $a_{nm}(u_d\neq 0)$, we need to first invoke the addition formula to write the active field $u_d$ in terms of the incoming wavefunction $U^m_n(\mathbf{x})$. Since $|\mathbf{x}|<|\mathbf{x}_\ell|$ on the surface of the scattering sphere,
by putting $z_n(k|\mathbf{r}-\mathbf{r}_0|)=h_n^{(1)}(k|\mathbf{r}-\mathbf{r}_0|), \mathbf{r}=\mathbf{x}, \mathbf{r}_0=\mathbf{x}_\ell \ \text{and} \ \mathbf{r}_0'=\mathbf{0}$ in the second case of \eqref{add} for $\ell=1,2,\cdots,L$, we can rewrite the active field from every source in a form similar to \eqref{u_i_6}. However, given the complexity of the expressions in \eqref{add} -- \eqref{cal_S} and the fact that the evaluation has to be repeated for $L$ times, each with a different position vector $\mathbf{x}_\ell$, the computation of the coefficients related to the active field will be expensive. 
A way to circumvent this problem is to again make use of the regularity and rotational symmetry of the Platonic source distribution like what we have done in Part 5 when we determine the amplitudes of all active sources using only knowledge of that of the bottommost source and the geometry of Platonic solids.  
To distinguish between the contributions from different active sources, we now denote the rotated space for each source as $\mathbf{x}^{(\ell)} = (x^{(\ell)},y^{(\ell)},z^{(\ell)})=\mathbf{R}_\ell \mathbf{x}$ (which is different from the notation in Part 5). In the rotated frame $\mathbf{x}^{(\ell)}$, we have the position vector     
$\mathbf{x}_\ell^{(\ell)}=\mathbf{x}_1=(0,0,-x_0)$, the spherical face
$\partial C_\ell^{(\ell)}= \partial C_1$ and the propagating vector of the incident wave
$\widehat{\mathbf{k}}^{(\ell)} = \mathbf{R}_\ell \widehat{\mathbf{k}}$. Note that $\mathbf{x}^{(1)} = \mathbf{x}$ and $\widehat{\mathbf{k}}^{(1)} = \widehat{\mathbf{k}}$. In the new notation, the equation of transformation between the original and rotated spaces for the spherical harmonic function \eqref{3d_o_Y} becomes
\begin{align}
	Y^m_n(\widehat{\mathbf{x}-\mathbf{x}_\ell}) = \sum_{m'=-n}^{n}  Y^{m'}_n( \widehat{\mathbf{x}^{(\ell)}-\mathbf{x}_1}) D^{n}_{m'm}(\gamma_\ell, \beta_\ell, \alpha_\ell) \label{3d_o_Y_6}
\end{align}
for $\ell=1,2,\cdots,L$, where $\gamma_\ell, \beta_\ell, \alpha_\ell$ are the Euler angles of the rotation matrix $\mathbf{R}_\ell$. The system of linear equations for the source amplitudes in \eqref{3d_a_lin_eq} is now
\begin{align}
	\sum_{m=-n}^n  D^{n}_{m'm}(\gamma_\ell, \beta_\ell, \alpha_\ell) q_{\ell, nm}(\widehat{\mathbf{k}}, \mathbf{x}_\ell, \partial C_\ell) \nonumber \\
	= q_{1,nm'}(\widehat{\mathbf{k}}^{(\ell)}, \mathbf{x}_1, \partial C_1). \label{3d_a_lin_eq_6}
\end{align}
Substituting \eqref{3d_o_Y_6} and \eqref{3d_a_lin_eq} into the expression of $u_d|_\ell$ in \eqref{3d_a_phi_d_ell} for $\ell=1,2,\cdots,L$ and summing the active fields over all sources, we have 
the total active field $u_d$ in the form
\begin{align}
	u_d = \sum_{\ell=1}^L \sum_{n=0}^\infty \sum_{m=-n}^n q_{1,nm}(\widehat{\mathbf{k}}^{(\ell)}) 
	V^m_n(\mathbf{x}^{(\ell)}-\mathbf{x}_1), \label{u_d_D}
\end{align}
where the dependence of $q_{1,nm}$ on the geometric parameters $\mathbf{x}_1$ and $\partial C_1$ is already understood. An explicit formula for $q_{1,nm}$ is already available in \eqref{q_l} -- \eqref{Range} along with the parametric form for the surface of integration $\partial C_1$ in \eqref{cos} -- \eqref{h}.
Note that the expression of $S^{m\mu}_{n\nu} (\mathbf{r}_0'-\mathbf{r}_0)$ in \eqref{S} can be
simplified for $\mathbf{r}_0'-\mathbf{r}_0 = \mathbf{0}-\mathbf{x}_\ell^{(\ell)} = -\mathbf{x}_1 = x_0 \widehat{\mathbf{e}}_z$, which is the pointing in the positive $z$ direction. The spherical harmonic function  in \eqref{S} now becomes  
\begin{align}
	Y_{\mathfrak{q}}^{\mu-m} (\widehat{\mathbf{e}}_z) &= A^{0}_{\mathfrak{q}} \delta_{\mu m} P^{0}_{\mathfrak{q}}(\cos 0) e^{i(0)\varphi} \nonumber \\
	&= \sqrt{\frac{2\mathfrak{q}+1}{4\pi}} \delta_{\mu m} P^0_\mathfrak{q} (1)	 
	, \label{3d_s_Y}
\end{align}
where $\mathfrak{q}=q_0+2q$ and the exponent $\mu-m$ must vanish since the azimuthal angle $\varphi$ is undefined for $\theta=0$. By \cite{gradshteyn2014table}, the associated Legendre polynomial $P^\mathfrak{p}_\mathfrak{q}(x)$ can be written as 
\begin{align}
	P^\mathfrak{p}_\mathfrak{q}(x) &= \frac{(-1)^\mathfrak{p}}{2^\mathfrak{q} \mathfrak{q}!} (1-x^2)^{\mathfrak{p}/2} \frac{d^{\mathfrak{p}+\mathfrak{q}}}{dx^{\mathfrak{p}+\mathfrak{q}}} (x^2-1)^\mathfrak{q}. \label{s_P}
\end{align}
With $\mathfrak{p}=0 \ \text{and} \ x=1$, we apply the general Leibniz rule \cite{olver2000applications} on the derivative such that
\begin{align}
	& \ \ \ \frac{d^\mathfrak{q}}{dx^\mathfrak{q}} (x^2-1)^\mathfrak{q} \bigg|_{x=1} \nonumber \\
	&= \frac{d^\mathfrak{q}}{dx^\mathfrak{q}} (x+1)^\mathfrak{q} (x-1)^\mathfrak{q} \bigg|_{x=-1} \nonumber \\
	&= \sum^\mathfrak{q}_{j=0} \binom{\mathfrak{q}}{j} \bigg[\frac{\mathfrak{q}!}{(\mathfrak{q}-j)!}(x+1)^{\mathfrak{q}-j} \bigg] \bigg[\frac{\mathfrak{q}!}{j!}(x-1)^{j} \bigg] \bigg|_{x=1} \nonumber \\
	&=\delta_{0j} \binom{\mathfrak{q}}{j} \bigg[\frac{\mathfrak{q}!}{(\mathfrak{q}-j)!}(x+1)^{\mathfrak{q}-j} \bigg] \bigg[\frac{\mathfrak{q}!}{j!}(x-1)^{j} \bigg] \bigg|_{x=1} \nonumber \\
	&= 2^\mathfrak{q} \mathfrak{q}!. \label{3d_s_leibniz}
\end{align}
By \eqref{s_P} and \eqref{3d_s_leibniz}, we have
$Y_{q_0+2q}^{\mu-m} (\widehat{\mathbf{e}}_z) = \sqrt{[2(q_0+2q)+1]/(4\pi)} \delta_{\mu m} $ and thus
\begin{align}
	&\ \ \ S^{m\mu}_{n\nu} (-\mathbf{x}_1) \nonumber \\
	&= 2\sqrt{\pi} \delta_{m\mu} (-1)^m \sum_{q=0}^{(n+\nu-q_0)/2} i^{\nu-n+q_0+2q}  \sqrt{2(q_0+2q)+1} \nonumber \\
	& \ \ \ \times z_{q_0+2q} (kx_0)
	 \mathcal{G} (n,m,\nu,-\mu,q_0+2q), \label{S_x_1}
\end{align}
where $q_0 \equiv|n-\nu|$ since in \eqref{q_0}, only the first case is possible with $\mu=m$.

Now we can finally apply the second case of the addition formula \eqref{add} with $\mathbf{r}=\mathbf{x}^{(\ell)}, \mathbf{r}_0=\mathbf{x}_\ell^{(\ell)}=\mathbf{x}_1 \ \text{and} \ \mathbf{r}_0'=\mathbf{0}$ to expand the total active field in \eqref{u_d_D} as
\begin{align}
	u_d &= \sum_{\ell=1}^L \sum_{n=0}^\infty \sum_{m=-n}^n  q_{1,nm}(\widehat{\mathbf{k}}_\ell^{(\ell)}) \sum_{\nu=0}^\infty S^{mm}_{n\nu}(-\mathbf{x}_1) U^m_\nu (\mathbf{x}^{(\ell)}),
 	\label{u_d_mm}
\end{align}
where $|\mathbf{x}^{(\ell)}| < |\mathbf{x}_1| = x_0$ on the surface of the scattering sphere. Note that the term $S^{mm}_{n\nu}(-\mathbf{x}_1)$ is now independent of $\ell$ and needs to be evaluated only for $-\mathbf{x}_1$ using the simplified form in \eqref{S_x_1}. The Kronecker delta function in \eqref{S_x_1} also filters out the azimuthal modes with $\mu \neq m$. The next step is to revert \eqref{u_d_mm} to the originial space $\mathbf{x}$ so that it has the same basis function as the incident field $\mathfrak{u}_i$ does in \eqref{u_i_6}. Referring back to how the spherical harmonic function is transformed from $\mathbf{x}$ to $\mathbf{x}^{(\ell)}$ in \eqref{3d_o_Y_6} and the order of the Euler angles in \eqref{3d_o_euler} in Part 5, we can simply reverse the rotation 
and write
\begin{align}
	Y^{m}_n(\widehat{\mathbf{x}}^{(\ell)}) 
	= \sum_{m'=-n}^{n}  Y^{m'}_n( \widehat{\mathbf{x}}) D^{n}_{m'm}(-\alpha_\ell, -\beta_\ell, -\gamma_\ell), \label{3d_o_Y_inv}
\end{align}
where $D^{n}_{m'm}(-\alpha_\ell, -\beta_\ell, -\gamma_\ell)$ is by construction the inverse of $D^{n}_{mm'}(\gamma_\ell, \beta_\ell, \alpha_\ell)$ which is defined in \eqref{3d_o_D} -- \eqref{C}. Substitution of \eqref{3d_o_Y_inv} into \eqref{u_d_mm} yields
\begin{align}
	u_d(\mathbf{x}) &=  \sum_{\nu=0}^\infty \sum_{m'=-\nu}^\nu  \Bigg[\sum_{\ell=1}^L \sum_{n=0}^\infty \sum_{m=-n}^n  q_{1,nm}(\widehat{\mathbf{k}}^{(\ell)})  S^{mm}_{n\nu}(-\mathbf{x}_1) \nonumber \\
	& \ \ \ \times D^{\nu}_{m'm}(-\alpha_\ell, -\beta_\ell, -\gamma_\ell) \Bigg] U_\nu^{m'} (\mathbf{x}), \ \ |\mathbf{x}|<x_0. 
	\label{u_d_mm_D}
\end{align}
Now comparing \eqref{u_d_mm_D} with the expansion of $\mathfrak{u}_i$ in \eqref{u_i_6} and including the contribution of the external incident wave $u_i$, we obtain the scattering coefficient $a_{\nu m'}(u_d \neq0)$ as
\begin{align}
	& \ \ \ a_{\nu m'}(u_d\neq 0) \nonumber \\
	&= -\frac{j_\nu(kA)}{h_\nu(kA)} \Bigg[Q_{\nu m'} + \sum_{\ell=1}^L \sum_{n=0}^\infty \sum_{m=-n}^n  q_{1,nm}(\widehat{\mathbf{k}}^{(\ell)})  S^{mm}_{n\nu}(-\mathbf{x}_1) \nonumber \\
	& \ \ \ \times D^{\nu}_{m'm}(-\alpha_\ell, -\beta_\ell, -\gamma_\ell)  \Bigg] \label{a_nm_neq_0}
\end{align}
in the sound-soft case. Similarly, if we apply the addition formula for $|\mathbf{r}-\mathbf{r}_0'| > |\mathbf{r}_0'-\mathbf{r}_0|$ in \eqref{add} -- \eqref{cal_S} on $u_d$, we can show that for $|\mathbf{x}|>x_0$,
\begin{align}
	u_d(\mathbf{x}) &=  \sum_{\nu=0}^\infty \sum_{m'=-\nu}^\nu  \Bigg[\sum_{\ell=1}^L \sum_{n=0}^\infty \sum_{m=-n}^n  q_{1,nm}(\widehat{\mathbf{k}}^{(\ell)})  \widehat{S}^{mm}_{n\nu}(-\mathbf{x}_1) \nonumber \\
	& \ \ \ \times D^{\nu}_{m'm}(-\alpha_\ell, -\beta_\ell, -\gamma_\ell) \Bigg] V_\nu^{m'} (\mathbf{x}), \label{u_d_mm_D_far}
\end{align}
where
\begin{align}
	&\ \ \ \widehat{S}^{m\mu}_{n\nu} (-\mathbf{x}_1) \nonumber \\
	&= 2\sqrt{\pi} (-1)^m \delta_{m\mu} \sum_{q=0}^{(n+\nu-q_0)/2} i^{\nu-n+q_0+2q} \sqrt{2(q_0+2q)+1} \nonumber \\
	& \ \ \ \times j_{q_0+2q} (kx_0)
	\mathcal{G} (n,m,\nu,-\mu,q_0+2q), \label{S_x_1_far}
\end{align}
and thus
\begin{align}
	&\ \ \ f_{\nu m'} \nonumber \\
	&= \sum_{\ell=1}^L \sum_{n=0}^\infty \sum_{m=-n}^n  q_{1,nm}(\widehat{\mathbf{k}}^{(\ell)})  \widehat{S}^{mm}_{n\nu}(-\mathbf{x}_1) D^{\nu}_{m'm}(-\alpha_\ell, -\beta_\ell, -\gamma_\ell). \label{q_nm_neq_0}
\end{align}
upon comparing \eqref{u_d_mm_D_far} with \eqref{u_d_6}.

The method illustrated above where we make use of the property that the spherical harmonic function vanishes for nonzero azimuthal order on the $z$ axis can also be applied to simplify the expression of 
the source coefficient $q_{\ell,nm}$ in \eqref{q_l} -- \eqref{Range} under a general wave incidence. As we evaluate the surface integral $\mathcal{I}_{n\nu}^{m\mu}(\mathbf{x}_1,\partial C_1)$ and adopt the rotational approach in Part 5 to determine all source amplitudes, the term   
$\widehat{S}^{s\mu}_{t\nu}(\mathbf{x}_\ell)$ needs to be computed only for $\mathbf{x}_1=-x_0\widehat{\mathbf{e}}_z$. Similar to what is done in \eqref{3d_s_Y} -- \eqref{S_x_1}, we have $Y_{q_0+2q}^{\mu-s} (-\widehat{\mathbf{e}}_z) = (-1)^{q_0+2q}\sqrt{[2(q_0+2q)+1]/(4\pi)} \delta_{\mu s}$ and thus  
\begin{align}
	&\ \ \ \widehat{S}^{s\mu}_{t\nu} (\mathbf{x}_1) \nonumber \\
	&= 2\sqrt{\pi} \delta_{s\mu} (-1)^s \sum_{q=0}^{(t+\nu-q_0)/2} i^{\nu-t-(q_0+2q)}  \sqrt{2(q_0+2q)+1} \nonumber \\
	& \ \ \ \times j_{q_0+2q} (kx_0)
	\mathcal{G} (t,s,\nu,-\mu,q_0+2q), \label{S_x_1_ts}
\end{align}
where $q_0 \equiv|t-\nu|$. The term $q_{1,nm,ts}$ from \eqref{q_2} now becomes 
\begin{align}
	q_{1, nm, ts} &= \sum_{\nu=0}^\infty  \widehat{S}^{ss}_{t\nu} (\mathbf{x}_1) D_{\nu n} \mathcal{I}_{n\nu}^{ms}(\mathbf{x}_1, \partial C_1). \label{q_2_ts}
\end{align}

\begin{figure}[H]
	\centering
	\includegraphics[trim={0cm 0cm 0cm 0cm},clip,width=8.5cm]{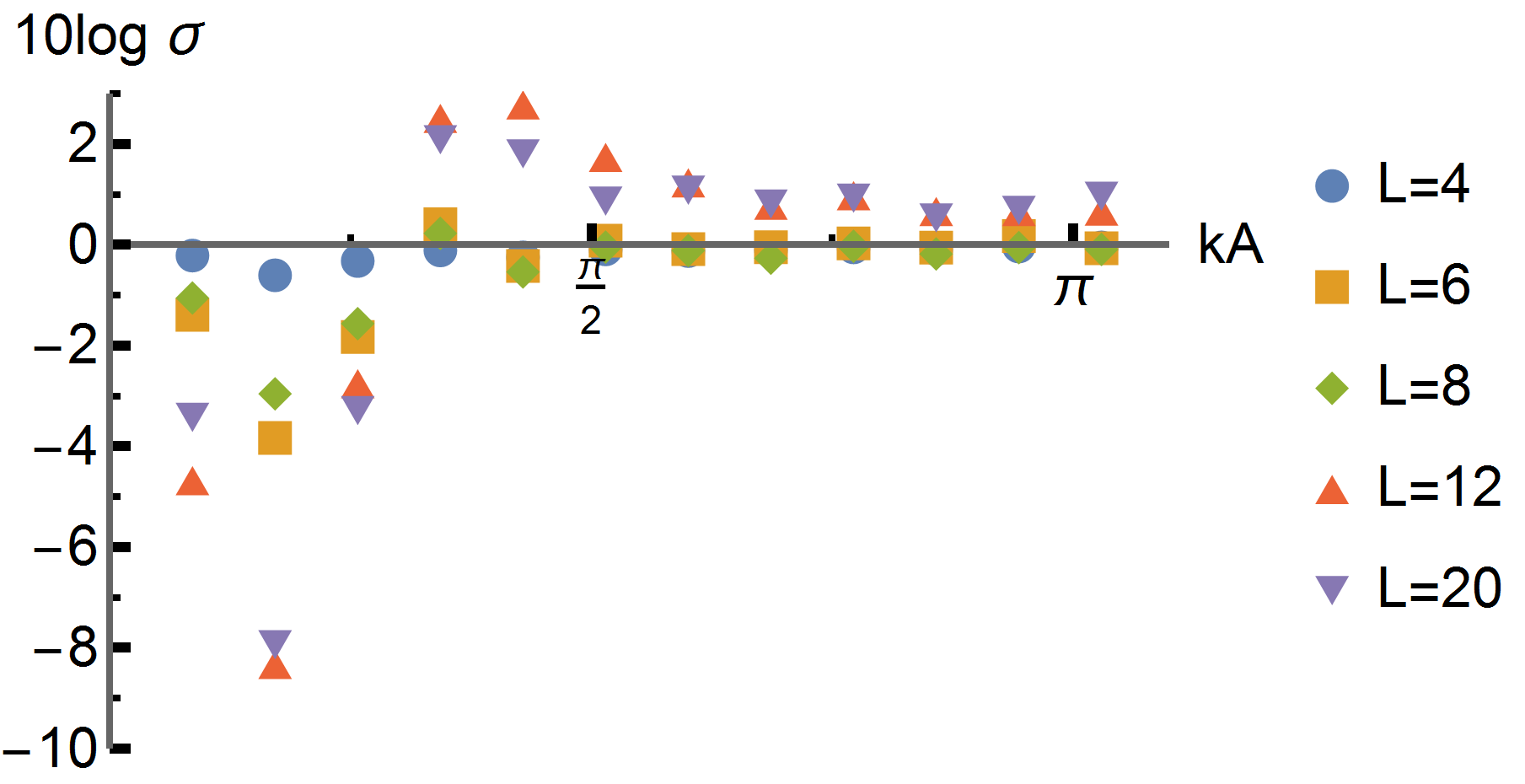}
	\caption{The sound power level $\text{SWL}=10\log \sigma$ versus the nondimensionalized radius $kA$ of the sound-soft sphere inside the cloaked region for five different number of sources $L$ and $\pi/2 \leq kx_0 \leq 6\pi$ at an interval of $\pi/2$ (or $0.0858\pi \leq kA \leq 1.0294\pi$ at an interval of around 0.0858$\pi$). 
	Here a multipole order of $N=1$ (i.e., monopole and dipole only) is used.  
	The other parameters are the same as those in Fig. \ref{sc_plot}.} 
	\label{N_1}
\end{figure}

In Fig. \ref{N_1}, we plot the sound power level $\text{SWL}=10\log \sigma$ as a function of $kA$, which is the dimensionless radius of the scattering sphere inside the silent region. The quantity $\sigma$ is in the form \eqref{sigma_6} with expressions for $a_{nm}(u_d=0),a_{nm}(u_d \neq 0) \ \text{and} \ f_{nm}$ given by \eqref{a_nm_u_d=0}, \eqref{a_nm_neq_0} and \eqref{q_nm_neq_0} respectively. The active sources consist of monopoles and dipoles only with multipole order $N=1$. The settings of the plot are otherwise identical to those of Fig. \ref{sc_plot}. Comparing Fig. \ref{N_1} with Fig. \ref{sc_plot}, we observe that the range of $kA$ where there is a substantial power reduction is mostly confined to $kA <\pi/4$ in all cases of $L$. The reduction attained is also less than that in Fig. \ref{sc_plot}. 
For $kA >\pi/4$ the power is even amplified slightly when $L=12,20$. Nonetheless, the plot shows that in more practical scenarios where only monopole and dipole sources are realizable, our cloaking approach is still capable of considerably reducing the power radiated to far field for lower frequencies. Another point to note from both Fig. \ref{sc_plot} and \ref{N_1} is that while the cloaking effect gets better in general for more larger value of $L$, increasing the number of sources does not necessarily increase the reduction at some frequencies. This phenomenon is especially obvious when we increase from $L=6$ to 8 or from 12 to 20. A possible reason is that the volume of the cloaked region indeed becomes smaller in these two cases as can be seen in Table \ref{geo}.

\begin{figure}[H]
	\centering
	\includegraphics[trim={0cm 0cm 0cm 0cm},clip,width=8.5cm]{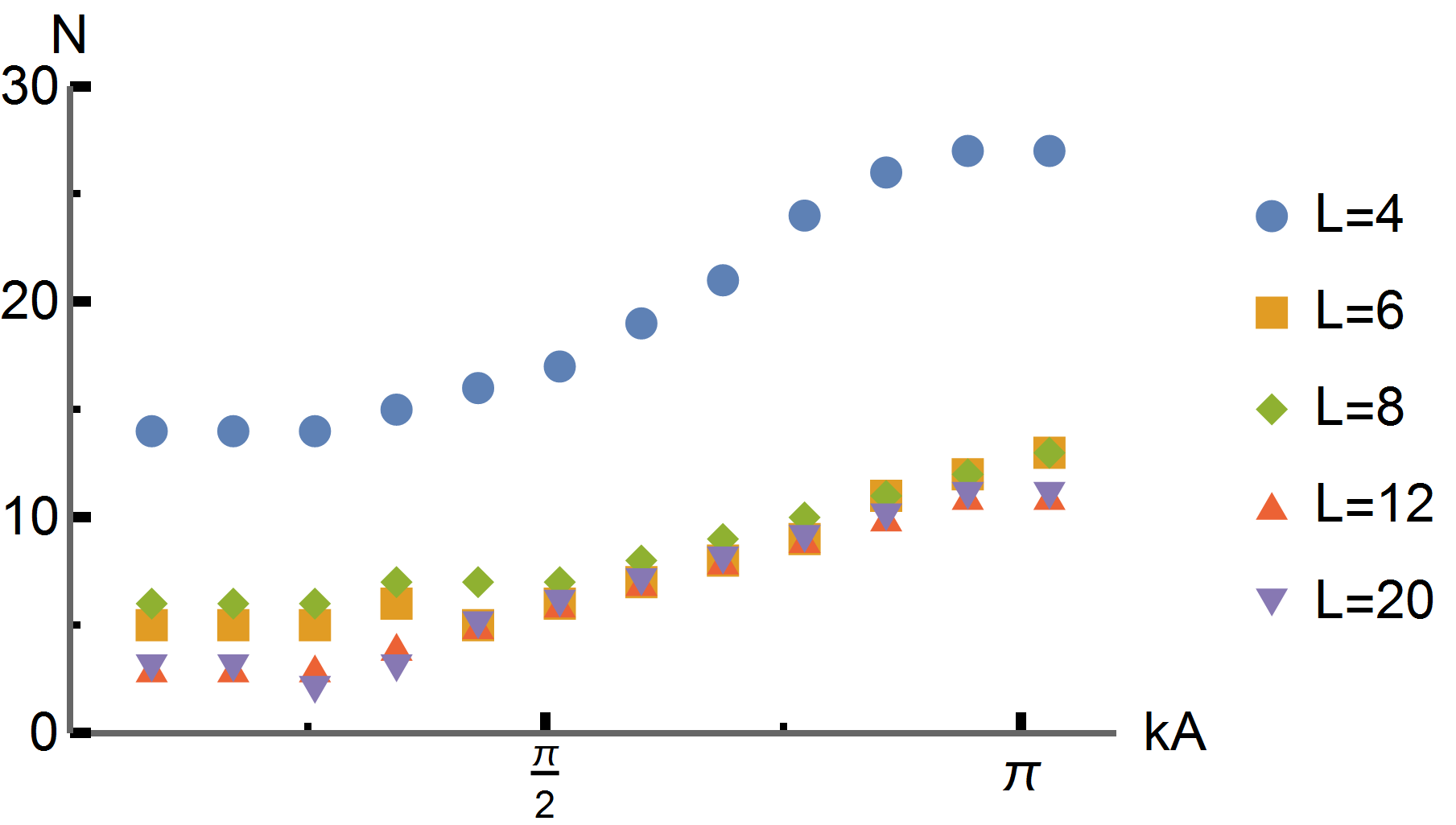}
	\caption{
		The order of multipole required for the five source configurations to attain a minimum reduction of 10 decibels over the range $\pi/2 \leq kx_0 \leq 6\pi$ at an interval of $\pi/2$ (or $0.0858\pi \leq kA \leq 1.0294\pi$ at an interval of around 0.0858$\pi$).  	
		Note that when $L=4$, the SWL fails to reach $-10$ decibels at $kA \approx 0.8579\pi, 0.9437\pi, 1.0294\pi$ in the range $N\leq30$. Here the values of $N$ which can achieve a reduction closest to $10$ decibels are chosen for these three cases instead.} 
	\label{10Log_sigma_10}
\end{figure}

In Fig. \ref{10Log_sigma_10}, we show the multipole order $N$ required for each active source in order to achieve a minimum reduction of 10 decibels for each observation point in the wavenumber ranged $\pi/2 \leq kx_0 \leq 6\pi$ (or $0.0858\pi \leq kA \leq 1.0294\pi$) for the five different source distributions. It appears that the more sources employed, the less multipoles required to reach the same level of reduction in general. The result also seems to agree with what is previously suggested by Fig. \ref{sc_plot} and \ref{N_1} as the order of multipoles needed to attain a certain degree of reduction becomes higher when the frequency increases.  

%
%

\end{document}